\newfam\msbfam
\font\twlmsb=msbm10 at 12pt
\font\eightmsb=msbm10 at 8pt
\font\sixmsb=msbm10 at 6pt
\textfont\msbfam=\twlmsb
\scriptfont\msbfam=\eightmsb
\scriptscriptfont\msbfam=\sixmsb
\def\cj{\fam\msbfam}
\def\A{{\cj A}}

\def\C{{\cj C}}

\def\R{{\cj R}}

\centerline{\bf FIBRE BUNDLES, CONNECTIONS, GENERAL RELATIVITY,} 

\

\centerline{\bf AND EINSTEIN-CARTAN THEORY}

\

\

\centerline{\bf M. Socolovsky}

\

\centerline{\it  Instituto de Ciencias Nucleares, Universidad Nacional Aut\'onoma de M\'exico}
\centerline{\it Circuito Exterior, Ciudad Universitaria, 04510, M\'exico D. F., M\'exico} 

\

\

\

{\bf Introduction}

\

The main purpose of this article is to present in the most natural way, that is, in the context of the theory of vector and principal bundles and connections in them, fundamental geometrical concepts related to general relativity (GR) and one of its extensions, the Einstein-Cartan theory (EC). Central concepts are the curvature tensor ${R^\mu}_{\nu\rho\sigma}$, the torsion tensor ${T^\mu}_{\nu\rho}$ and the non-metricity tensor $Q_{\mu\nu\rho}=-D_\mu g_{\nu\rho}$ as properties of connections in a Riemannian or pseudo-Riemannian manifold, with metric $g_{\mu\nu}$ and affine connection ${\Gamma^\mu}_{\nu\rho}$. ($D_\mu$ is the covariant derivative with respect to ${\Gamma^\mu}_{\nu\rho}$.) GR has to do with a metric symmetric connection, the Levi-Civita connection, that only allows for ${R^\mu}_{\nu\rho\sigma}$; EC theory involves a metric but not necessarily symmetric connection, that allows also for ${T^\mu}_{\nu\rho}$; while the theory of Weylian manifolds involves a non necessarily metric ($Q_{\mu\nu\rho}\neq 0$) and non necessarily symmetric (${T^\mu}_{\nu\rho}\neq 0$) connection. (In units of length $[L]$, $[{R^\mu}_{\nu\rho\sigma}]=[L]^{-2}$, $[{T^\mu}_{\nu\rho}]=[Q_{\mu\nu\rho}]=[{\Gamma^\mu}_{\nu\rho}]=[L]^{-1}$, while $[g_{\mu\nu}]=[L]^0$.) 

\

One of the most beautiful equations of Physics is the equality to zero of the Einstein tensor, that is, the Einstein's equations in {\it vacuum}: $$G_{\mu\nu}=0,$$ where $$G_{\mu\nu}=R_{\mu\nu}-{{1}\over{2}}g_{\mu\nu}R,$$ with $R=g^{\nu\sigma}R_{\nu\sigma}=g^{\nu\sigma}g^{\mu\rho}R_{\mu\nu\rho\sigma}=g^{\nu\sigma}{R^\rho}_{\nu\rho\sigma}$. $G_{\mu\nu}=0$ is equivalent to Ricci flatness: $$R_{\mu\nu}=0.$$ This however does not imply vanishing curvature; so, in GR, {\it empty space-time can be curved}. Instead, in EC theory, torsion must be zero in vacuum.

\

It is remarkable that $G_{\mu\nu}$ appears naturally when the Bianchi equations for the Levi-Civita connection are expressed in terms of the Ricci tensor $R_{\mu\nu}$ and the scalar curvature $R$. Then, $G_{\mu\nu}$ is a purely geometric object.

\

The use of the tetrads ($e_c$) formalism along with their duals, the coframes or anholonomic coordinates ($e^a$), allows us to discover how GR and EC theory have an internal or gauge symmetry (Utiyama, 1956), implemented by a connection that takes values in the Lie algebra of the Lorentz group ${\cal L}_4$: the spin connection $\omega_{ab}$. The variations of the Einstein-Hilbert action of pure gravity or gravity coupled to Dirac matter with respect to $\omega_{ab}$ and $e_c$ lead, respectively, to the Cartan and Einstein equations, the former involving torsion and the spin of matter, and the latter involving curvature and the energy-momentum of matter.

\

Later, through a shift of the $e^c$'s one finds the translation gauge potential $B^a$; together, $B^a$ and $\omega_{bc}$, define a Poincar\'e connection, extending the symmetry group of GR and EC theory to the semidirect product ${\cal P}_4\odot{\cal D}$, where ${\cal P}_4={\cal T}_4\odot{\cal L}_4$ is the Poincar\'e group, with ${\cal T}_4$ the translation group, and ${\cal D}$ the group of general coordinate transformations.

\

Finally, in the last section, we discuss the problem of defining a gauge invariant field strength for the Maxwell field coupled to gravity, and the subsisting problem of the $U(1)$-gauge dependence of torsion in the solution of the Cartan equation.

\

\centerline{\it CONTENT}

\

1. Connections in smooth real vector bundles $\xi$

\

2. Linear connection in a differentiable manifold $M$

\

3. Total covariant derivative of a section in $\xi$

\

4. Local expressions for the connection (covariant derivative) and the total covariant derivative

\

5. Local expressions for the covariant derivatives of vector fields and 1-differential forms in a manifold

\

6. Example: Trivial connection in $\R ^n$

\

7. Transformation of the connection coefficients in a manifold (local concept). Provisional definition of 

\ \ \ \ tensors in a manifold. Tensors in arbitrary vector bundles.

\

8. Directional covariant derivative and parallel transport of tensors; geodesics.

\

9. Curvature and torsion of a connection in $\xi$

\

10. Geometric interpretation of curvature and torsion

\

11. Exterior covariant derivative and curvature 2-form

\

12. Bianchi equation and Bianchi identities

\

13. The Levi-Civita connection

\

14. Physics 1: Equivalence principle in GR

\

15. Covariant components of the curvature tensor

\

16. Ricci tensor for the Levi-Civita connection

\

17. Physics 2: (Local) Einstein equations in empty space-time

\

18. Ricci (or curvature) scalar (with Levi-Civita connection)

\

19. Einstein tensor

\

20. Physics 2$^\prime$: (Local) Einstein equations in empty space-time

\

21. Examples in $m=D=1,2,3$. Generalization to $m\geq 4$ and Weyl tensor

\

22. For the Levi-Civita connection, $G^{\mu\nu};_\nu=0$

\

23. Physics 3: (Local) Einstein equations in the presence of matter

\

24. Tensor bundles as associated bundles to the bundle of frames of $M^n$, ${\cal F}_{M^n}$

\

25. Vertical bundle of a principal fibre bundle

\

26. Soldering form on ${\cal F}_{M^n}$

\

27. Linear connection in a manifold $M^n$ on ${\cal F}_{M^n}$

\

28. Tetrads and spin connection

\

29. Curvature and torsion in terms of spin connection and tetrads. Cartan structure equations;  

\ \ \ \ \ Bianchi identities

\

30. Spin connection in non-coordinate basis

\

31. Locally inertial coordinates

\

32. Einstein-Cartan equations

\

33. Lorentz gauge invariance of Einstein and Einstein-Cartan theories

\

34. Poincar\'e gauge invariance of Einstein and Einstein-Cartan theories

\

35. Torsion and gauge invariance

\

Acknowledgements

\

References

\

Appendix A

\

Appendix B

\

Appendix C

\

Appendix D

\

Appendix E

\

Appendix F

\

{\bf 1. Connections in smooth real vector bundles}

\

Let $\xi:\R^m-E\buildrel{\pi}\over \longrightarrow M^n$ be a smooth $m$ dimensional real vector bundle over $M^n\equiv M $, a differentiable manifold of dimension $n$. Let $\Gamma(TM)$ denote the sections of the tangent bundle of $M$ and $\Gamma(E)$ denote the sections of $E$. $E$ is an $m+n$ dimensional differentiable manifold; this can be easily shown from the local triviality condition.

\

{\it A connection} in $\xi$ is a function $$\nabla:\Gamma(TM)\times \Gamma(E) \to \Gamma(E),$$ $$(X,s)\mapsto \nabla(X,s)\equiv\nabla_Xs$$ which has the following properties:

\

i) $\nabla_{X+X^\prime}s=\nabla_Xs+\nabla_{X^\prime}s$

\

ii) $\nabla_{fX}s=f\nabla_Xs$, where $f\in C^\infty (M,\R)$: smooth real valued functions on $M$

\

iii) $\nabla_X(s+s^\prime)=\nabla_Xs+\nabla_Xs^\prime$ 

\

iv) $\nabla_X(fs)=X(f)s+f\nabla_Xs$ (Leibnitz rule)

\

$\Gamma(TM)$ and $\Gamma(E)$ are infinite dimensional vector spaces over $\R$, but modules over $C^\infty (M,\R)$ as a ring, with dimensions $n$ and $m$ respectively. The Leibnitz rule shows that $\nabla$ is not $C^\infty (M,\R)$-linear in the second entry. As will be shown below, this will be reflected in the fact that under a change of local coordinates, the set of connection coefficients (Christoffel symbols) is not a tensor.

\

The value of the connection at $(X,s)$ is called the {\it covariant (or invariant) derivative of} $s$ {\it in the direction of} $X$. $\nabla_Xs:M\to E$, $x\mapsto \nabla_Xs(x)=(x,(\nabla_Xs)_x)$, with $(\nabla_Xs)_x\in E_x$: the fibre in $E$ over $x$; $E_x$ is a real $m$ dimensional vector space. 

\

Notice that we can define the operator $$\nabla_X:\Gamma(E)\to\Gamma(E), \ \nabla_X(s)=\nabla_Xs$$ $\nabla_X\in Lin_\R(\Gamma(E))$ and obeys the Leibnitz rule. 

\

One summarizes these concepts in the following diagram: $$\matrix{& & & \R^m \cr & & & \vert \cr & & &  E \cr}$$ $$\matrix{& & & s\uparrow \downarrow \pi\uparrow \nabla_Xs \cr}$$ $$\ \ \ \ \ \ \ \ \R^n-TM\buildrel{\pi_M}\over \rightarrow M$$  $$\ \ \ \ \ \ \ \ \ \ \ \ \ \ \ \ \ \ \buildrel{X}\over \leftarrow$$

\

$E=\bigcup_{x\in M}\{x\}\times E_x\equiv \coprod_{x\in M}E_x$.

\

{\it Note}: $M$ is a differentiable manifold; as such is a topological space. This global structure is defined in $M$ prior to any connection on $\xi$.

\

{\bf 2. Linear connection in a differentiable manifold $M$}

\

A {\it linear connection} on $M$ is a connection in its tangent bundle. With $E=TM$ we have: $$\nabla: \Gamma(TM)\times \Gamma(TM)\to \Gamma(TM),$$ $$(X,Y)\mapsto \nabla(X,Y)\equiv \nabla_XY$$ with 

\

i$^\prime$) $\nabla_{X+X^\prime}(Y)=\nabla_XY+\nabla_{X^\prime}Y$

\

ii$^\prime$) $\nabla_{fX}(Y)=f\nabla_XY$

\

iii$^\prime$) $\nabla_X(Y+Y^\prime)=\nabla_XY+\nabla_XY^\prime$

\

iv$^\prime$) $\nabla_X(fY)=X(f)Y+f\nabla_XY$

\

We shall denote by {\it Conn}($\xi$) the set of connections in the vector (or principal, where appropiate) bundle $\xi$.

\

Again, $\nabla_X\in Lin_\R(\Gamma(TM))$ and obeys the Leibnitz rule.

\

{\bf 3. Total covariant derivative of a section in $\xi$}

\

Let $\Gamma(T^*M\otimes E)$ be the set of differential 1-forms in $M$ with values in $E$. $T^*M\otimes E$ is a vector bundle on $M$: $$\R^{n\times m}-T^*M\otimes E\to M, \ \ \ with \ \ \ T^*M\otimes E=\coprod_{x\in M}T_{x}^*M\otimes E_x.$$  The section $\nabla s\in \Gamma(T^*M\otimes E)$ is defined by $$\nabla s:\Gamma(TM)\to \Gamma(E), \ \ X\mapsto \nabla s(X):= \nabla_Xs.$$ As for any differential form on $M$, $\nabla s$ is $\C^\infty (M,\R)$-linear i.e. $\nabla s(fX)=f\nabla s(X)$; however, $$\nabla (fs)=sdf+f\nabla s.$$ In fact, $\nabla (fs)(X)=\nabla_X(fs)=X(f)s+f\nabla_Xs=f\nabla s(X)+sX(f)=f\nabla s(X)+sdf(X)=(f\nabla s+sdf)(X).$ 

\

$\nabla s$ is called the {\it total covariant derivative of the section $s$}. In detail, 

\

$\nabla s:M\to T^*M\otimes E$, $x\mapsto (\nabla s)(x)=(x,(\nabla s)_x)$, $(\nabla s)_x=\alpha_x\otimes v_x:T_xM\to E_x$, $X_x\mapsto (\nabla s)_x(X_x)=\alpha_x(X_x)v_x=\lambda_xv_x$ where $\alpha_x\in T_{x}^*M$, $v_x\in E_x$ and $\lambda_x\in \R$.

\ 

For a linear connection on $M$, $$\nabla Y:\Gamma(TM)\to \Gamma(TM), \ \ X\mapsto (\nabla Y)(X)=\nabla_XY.$$

\

{\bf 4. Local expressions for $\nabla _Xs$ and $\nabla s$}

\

Let $(U_\alpha, \sigma_i)_{\alpha\in J, \ i\in \{1,\dots,m\}}$ be a basis of local sections of $E$ i.e. $\sigma_i:U_{\alpha}\to E_\alpha=\pi^{-1}(U_{\alpha})$, $x\mapsto \sigma_i(x)=(x,\sigma_{ix})\in \{x\}\times E_x$, with  $\pi \circ\sigma_i=Id_{U_\alpha}$, and such that if $s\in \Gamma(E_\alpha)$, then $s=\Sigma_{i=1}^m s^i\sigma_i$ with $s^i\in C^\infty (U_\alpha,\R)$. Let ${{\partial}\over{\partial x_{\alpha}^\mu}}\equiv \partial_\mu$ be a local coordinate basis of $\Gamma(TU_\alpha)$ i.e. if $X\in \Gamma(TU)$ then $X=X^\mu \partial_\mu$ with $X^\mu \in C^\infty (U_\alpha,\R)$. (If the domains of the local trivializations of the bundle $E$ do not coincide with the domains of the atlas $\cal U$ of the manifold $M$, one can always consider their intersections.) Then, locally, $$\nabla_Xs=\nabla_{X^\mu \partial_\mu}(s^i\sigma_i)=X^\mu \nabla_{\partial_{\mu}}(s^i\sigma_i)=X^\mu((\partial_\mu s^i)\sigma_i+s^i \nabla_{\partial_{\mu}} \sigma_i);$$ and since $\nabla_{\partial_{\mu}}\sigma_i\in \Gamma(E_\alpha)$ then $$\nabla_{\partial_{\mu}}\sigma_i:=\Gamma^j_{\mu i}\sigma_j$$ for a unique set of $n\times m^2$ functions $\Gamma^j_{\mu i}:U_\alpha\to \R$, called the {\it Christoffel symbols} of the connection $\nabla$ in the atlas $\cal U$. For a linear vector bundle, $m=1$, and then there are only $n$ symbols: $\Gamma^j_{\mu i}\equiv\Gamma_\mu$. 

\

We then write $$\nabla_Xs=X^\mu((\partial_\mu s^i)\sigma_i+s^i\Gamma^j_{\mu i}\sigma_j)=X^\mu(\partial_\mu s^j +s^i\Gamma^j_{\mu i})\sigma_j=X^\mu(\partial_\mu \delta^j_i+\Gamma^j_{\mu i})s^i\sigma_j=X^\mu D_{\mu i}^j s^i \sigma_j$$ where we have defined the {\it local covariant derivative operator} $$D_{\mu i}^j:=\partial_\mu \delta^j_i+\Gamma^j_{\mu i}.$$ 

\

(Units: $[\Gamma^j_{\mu i}]=[x^\mu]^{-1}$; in natural units $[\Gamma^j_{\mu i}]=[mass]$ if $[x^\mu]=[length]$.)

\

We can also write $$\nabla_Xs=X^\mu s^j_{;\mu}\sigma_j \ \ with \ \ D_\mu s^j\equiv s^j_{;\mu}=D^j_{\mu i}s^i=s^j_{,\mu} +\Gamma_{\mu i}^j s^i \ \ and \ \ s^j_{,\mu}=\partial_\mu s^j.$$ Notice that the ordinary derivative term $s^j,_\mu$ is due to the Leibnitz rule.

\

Locally, we can write $$\nabla s=dx^\mu\otimes \nabla_{\partial_{\mu}}s.$$ In fact, $\nabla s(X)=(dx^\mu\otimes\nabla_{\partial_{\mu}}s)X=dx^\mu(X)\nabla_{\partial{\mu}}s=dx^\mu(X^\nu \partial_\nu)\nabla_{\partial_{\mu}}s=X^\nu dx^\mu(\partial_\nu)\nabla_{\partial{\mu}}s=X^\nu\delta^\mu_\nu\nabla_{\partial_{\mu}}s=X^\mu\nabla_{\partial_{\mu}}s=\nabla_Xs$. Then $$\nabla s=dx^\mu\otimes(\partial_\mu\delta^j_i+\Gamma^j_{\mu i})s^i\sigma_j=dx^\mu\otimes(\partial_\mu s^j\sigma_j+\Gamma^j_{\mu i}s^i\sigma_j)=dx^\mu\otimes \partial_\mu s^j\sigma_j+dx^\mu\otimes \Gamma^j_{\mu i}s^i \sigma_j=ds^j\otimes \sigma_j+\Gamma^j_i \otimes s^i\sigma_j$$ where $\Gamma^j_i$ is an $m\times m$ matrix of 1-forms on $U_\alpha$ given by $$\Gamma^j_i=dx^\mu \Gamma^j_{\mu i}.$$ We can write $\Gamma^j_i\otimes s^i\sigma_j=\Gamma^j_is^i\otimes\sigma_j=(\Gamma s)^j\otimes\sigma_j$ and then $$\nabla s=ds^j\otimes \sigma_j+(\Gamma s)^j\otimes \sigma_j=(ds^j+(\Gamma s)^j)\otimes \sigma_j=((d+\Gamma)s)^j\otimes \sigma_j$$ i.e. $$\nabla s=(d+\Gamma)s\in \Gamma(T^*U\otimes E).$$ Then, locally, $$\nabla=d+\Gamma.$$

\

Let $s^i_{;\mu}(x)$=0 i.e. $$s^i_{,\mu}(x)+\Gamma^i_{\mu j}(x)s^j(x)=0.$$ Multiplying by $dx^\mu |_x$ we obtain $$s^i_{,\mu}(x)dx^\mu |_x=-\Gamma^i_{\mu j}(x)s^j(x)dx^\mu |_x \in T^*_xU.$$ The r.h.s. 
$$-\Gamma ^i_{\mu j}(x)s^j(x)dx^\mu |_x  \equiv (\delta_{\vert\vert}s)^i|_x,$$ is called the {\it infinitesimal parallel transport (transfer) (Schroedinger, 1950) of the section $s^i$ by the connection $\nabla$ along the 1-form} $dx^i |_x $ (see section {\bf 8}), and we see that for a covariantly constant section, it coincides with the differential of $s^i$ at $x$, $ds^i\vert_x$. The transfer of the section, proportional to $dx^\lambda \vert_x$ and to the section itself, just follows the values of $s^i$ along $dx^\lambda$ (when $s^i_{;\mu}=0$). When $s^i;_\mu \neq 0$, $(\delta_{\vert\vert}s)^i\vert _x$ still is the parallel transfer of $s^i$ through $dx^\lambda$, but it fails to follow the value of the section. (A more detailed discussion can be found in Cheng, 2010.)

\

{\bf 5. Local expression for $\nabla_XY$} 

\

When $E=TM$, $\sigma_i=\partial_\nu$ and $$\nabla_{\partial_{\mu}}\partial_\nu=\Gamma^\rho_{\mu\nu}\partial_\rho$$ for a unique set of $n^3$ smooth functions  $\Gamma^\rho_{\mu\nu}:U_\alpha\to \R$. ($\Gamma^\rho_{\mu\nu}$ is not necessarily symmetric or antisymmetric in $\mu$ and $\nu$.) Then, $$\nabla_XY=X^\mu D_{\mu\nu}^\rho Y^\nu\partial_\rho, \ \ D^\rho_{\mu\nu}=\partial_\mu\delta^\rho_\nu+\Gamma^\rho_{\mu\nu},$$ $$D^\rho_{\mu\nu}Y^\nu=\partial_\mu Y^\rho+\Gamma^\rho_{\mu\nu}Y^\nu, \ \ \nabla_XY=(\nabla_XY)^\rho\partial_\rho,$$ $$(\nabla_XY)^\rho=X^\mu(\partial_\mu\delta^\rho_\nu+\Gamma^\rho_{\mu\nu})Y^\nu.$$ The quantity $$D_{\mu\nu}^\rho Y^\nu=\partial_\mu Y^\rho +\Gamma^\rho_{\mu\nu}Y^\nu\equiv Y^\rho_{;\mu}\equiv D_\mu Y^\rho$$ is the {\it covariant derivative of the local vector field $Y^\rho$ in the direction of ${{\partial}\over {\partial x^\mu}}$}. Then, $$(\nabla_XY)^\rho=X^\mu D_\mu Y^\rho.$$ 

\

If $V^\mu$ is a local vector field and $A_\mu$ is a local differential 1-form in $M^n$, then $\varphi=V^\mu A_\mu$ is a scalar (0-rank tensor) i.e. $\varphi^\prime=\varphi$ or $V^{\prime\mu}A^\prime_\mu=V^\mu A_\mu$ under $x^\nu\to x^{\prime\nu}$. The covariant derivative of a scalar is naturally defined as $$\varphi_{;\nu}=\varphi_{,\nu}$$ and the Leibnitz rule is assumed for the covariant derivative of the product of arbitrary tensors $T$ and $S$:$$(TS)_{;\mu}=T_{;\mu}S+TS_{;\mu}$$. Then, $$\varphi_{;\nu}=(V^\mu A_\mu)_{,\nu}=(\partial_\nu V^\mu)A_\mu +V^\mu\partial_\nu A_\mu=V^\mu_{;\nu}A_\mu+V^\mu A_{\mu;\nu}$$ and so $$V^\mu A_{\mu;\nu}=V^\mu _{,\nu}A_\mu+V^\mu A_{\mu,\nu}-V^\mu _{;\nu}A_\mu=V^\mu_{,\nu} A_\mu+V^\mu A_{\mu,\nu}-V^\mu_{,\nu}A_\mu-\Gamma^\mu_{\nu\rho}V^\rho A_\mu=V^\mu A_{\mu,\nu}-\Gamma^\mu_{\nu\rho}V^\rho A_\mu=V^\rho A_{\rho;\nu}$$ where $$A_{\rho;\nu}=A_{\rho,\nu}-\Gamma^\mu_{\nu\rho}A_\mu\equiv D_\nu A_\rho$$ is the {\it covariant derivative of the local 1-form $A_\rho$ in the direction of ${{\partial}\over {\partial x^\nu}}$.}

\

{\bf 6. Example: the trivial connection in $\R^n$}

\

Consider $M=\R^n$; then with $X,Y\in\Gamma(TM)$ we define $$\nabla^0_XY:=X(Y^\nu)\partial_\nu=X^\rho \partial_\rho(Y^\nu)\partial_\nu \ .$$ Additivity in both $X$ and $Y$, and $C^\infty(M,\R)$-linearity in $X$ are trivial; finally, $\nabla^0_X(fY)=X^\rho\partial_\rho(fY^\nu)\partial_\nu=X(f)Y+fX(Y^\nu)\partial_\nu=X(f)Y+f\nabla^0_XY$.

\

In particular, $$\nabla^0_{\partial_\mu}(\partial_\nu)=\Gamma^{0\rho}_{\mu\nu}\partial_\rho=\nabla^0_{\partial_\mu}(\delta^\rho_\nu)\partial_\rho=\partial_\mu(\delta^\rho_\nu)\partial_\rho=0$$ and therefore $$\Gamma^{0\rho}_{\mu\nu}=0.$$ Note. We can compare with the Lie derivative ${\cal L} _XY$: $${\cal L} _XY=[X,Y]=X(Y^\nu)\partial_\nu-Y(X^\nu)\partial_\nu=[X,Y]^\nu\partial_\nu=\nabla^0_X Y-\nabla^0_Y X;$$ then $$\nabla^0_X Y-\nabla^0_Y X-[X,Y]=0.$$ As we shall see later, this means that $\nabla^0$ is torsion free. 

\

For $\nabla^0Y$ we have: $$\nabla^0Y=d(Y^\rho)\partial_\rho.$$ In fact, $(\nabla^0Y)(X)=dx^\nu \partial_\nu(Y^\rho)\partial_\rho(X)=dx^\nu(X)\partial_\nu(Y^\rho)\partial_\rho=X(x^\nu)\partial_\nu(Y^\rho)\partial_\rho=X^\nu \partial_\nu(Y^\rho)\partial_\rho=X(Y^\rho)\partial_\rho$

$=\nabla^0_X(Y)$.

\

{\bf 7. Transformation of $\Gamma^\mu_{\nu\rho}$ under a change of local coordinates (charts) in $M$}

\

From the definition of the $\Gamma^\rho_{\mu\nu}$-functions and using $$\partial_\lambda={{\partial}\over{\partial x^\lambda}}={{\partial}\over{\partial x^{\prime \sigma}}}{{\partial x^{\prime \sigma}}\over{\partial x^\lambda}}$$ one obtains: $$\nabla_{{{\partial x^{\prime\sigma}}\over{\partial x^\mu}}{{\partial}\over{\partial x^{\prime\sigma}}}}({{\partial}\over{\partial x^{\prime\lambda}}}{{\partial x^{\prime\lambda}}\over{\partial x^\nu}})={{\partial x^{\prime\sigma}}\over{\partial x^\mu}}\nabla_{{{\partial}\over{\partial x^{\prime\sigma}}}}({{\partial x^{\prime\lambda}}\over{\partial x^\nu}}{{\partial}\over{\partial x^{\prime\lambda}}})
={{\partial x^{\prime\sigma}}\over{\partial x^\mu}}({{\partial}\over{\partial x^{\prime\sigma}}}({{\partial x^{\prime\lambda}}\over{\partial x^\nu}}){{\partial}\over{\partial x^{\prime\lambda}}}+{{\partial x^{\prime\lambda}}\over{\partial x^\nu}}(\nabla_{{{\partial}\over{\partial x^{\prime\sigma}}}}{{\partial}\over{\partial x^{\prime\lambda}}}))$$ $$={{\partial x^{\prime\sigma}}\over{\partial x^\mu}}({{\partial x^\alpha}\over{\partial x^{\prime\sigma}}}{{\partial^2x^{\prime\lambda}}\over{\partial x^\alpha \partial x^\nu}}{{\partial}\over{\partial x^{\prime\lambda}}}+{{\partial x^{\prime\lambda}}\over{\partial x^\nu}}\Gamma^{\prime\rho}_{\sigma\lambda}{{\partial}\over{\partial x^{\prime\rho}}})={{\partial x^{\prime\sigma}}\over{\partial x^\mu}}{{\partial x^\alpha}\over{\partial x^{\prime\sigma}}}{{\partial^2x^{\prime\lambda}}\over{\partial x^\alpha \partial x^\nu}}{{\partial}\over{\partial x^{\prime\lambda}}}+{{\partial x^{\prime\sigma}}\over{\partial x^\mu}}{{\partial x^{\prime\beta}}\over{\partial x^\nu}}\Gamma^{\prime\lambda}_{\sigma\beta}{{\partial}\over{\partial x^{\prime\lambda}}}$$ $$={{\partial^2x^{\prime\lambda}}\over{\partial x^\mu \partial x^\nu}}{{\partial}\over{\partial x^{\prime\lambda}}}+{{\partial x^{\prime\alpha}}\over{\partial x^\mu}}{{\partial x^{\prime\beta}}\over{\partial x^\nu}}\Gamma^{\prime\lambda}_{\alpha\beta}{{\partial}\over{\partial x^{\prime\lambda}}}=\Gamma^\rho_{\mu\nu}{{\partial x^{\prime\lambda}}\over{\partial x^\rho}}{{\partial}\over{\partial x^{\prime\lambda}}}$$ which implies, by the linear independence of the coordinate tangent fields ${{\partial}\over{\partial x^{\prime\lambda}}}$, $$\Gamma^\rho_{\mu\nu}{{\partial x^{\prime\lambda}}\over{\partial x^\rho}}={{\partial^2x^{\prime\lambda}}\over{\partial x^\mu \partial x^\nu}}+{{\partial x^{\prime\alpha}}\over{\partial x^\mu}}{{\partial x^{\prime\beta}}\over{\partial x^\nu}}\Gamma^{\prime\lambda}_{\alpha\beta};$$finally, multiplying by the inverse transformation ${{\partial x^\gamma}\over{\partial x^{\prime\lambda}}}$ and using ${{\partial x^\gamma}\over{\partial x^{\prime\lambda}}}{{\partial x^{\prime\lambda}}\over{\partial x^\rho}}=\delta^\gamma_\rho$ one obtains $$\Gamma^\gamma_{\mu\nu}={{\partial x^\gamma}\over{\partial x^{\prime\lambda}}}{{\partial x^{\prime\alpha}}\over{\partial x^\mu}}{{\partial x^{\prime\beta}}\over{\partial x^\nu}}\Gamma^{\prime\lambda}_{\alpha\beta}+{{\partial x^\gamma}\over{\partial x^{\prime\lambda}}}{{\partial^2x^{\prime\lambda}}\over{\partial x^\mu \partial x^\nu}}.$$ The second term on the r.h.s., which comes from the fact that $\nabla$ obeys the Leibnitz rule, shows that the $\Gamma^\mu_{\nu\rho}$ functions do not transform as tensors.

\

{\bf Provisional definition of a tensor} (previous to the existence of any metric in the manifold)

\

In a chart $U_\alpha(x^\mu_\alpha\equiv  x^\mu)$ of $M^n$, a {\it tensor} with $r$ ``contravariant'' and $s$ ``covariant'' indices is a set of $n^{r+s}$ functions on $U_\alpha$ with values in $\R$, $$\{T^{\mu_1 \dots \mu_r}_{\nu_1 \dots \nu_s}, \ \ \mu_k,\nu_l=1,\dots,n, \ \ k=1,\dots,r, \ \ l=1,\dots,s\}$$ such that in a chart $U_\beta(x^{\prime\nu}_\beta \equiv x^{\prime\nu})$ which overlaps with $U_\alpha(x^\mu_\alpha)$ becomes the set $$T^{\prime \alpha_1\dots\alpha_r}_{\beta_1\dots\beta_s}={{\partial x^{\prime \alpha_1}}\over{\partial x^{\mu_1}}}\dots{{\partial x^{\prime\alpha_r}}\over{\partial x^{\mu_r}}}{{\partial x^{\nu_1}}\over{\partial x^{\prime\beta_1}}}\dots{{\partial x^{\nu_s}}\over{\partial x^{\prime\beta_s}}}T^{\mu_1\dots\mu_r}_{\nu_1\dots\nu_s}$$ for $x^\prime=x^\prime(x)$ in the overlap $U_\alpha\cap U_\beta$.

\

{\it Remark}: The concept of covariant and contravariant indices has sense only if there exists a metric in the manifold.
   
\

An $r$-contravariant and $s$-covariant tensor can be considered a $C^\infty (M,\R)$-multilinear map from the tensor product of $s$ factors of $\Gamma(TM)$ and $r$ factors of $\Gamma(T^*M)$ with values in $C^\infty(M,\R)$. On a chart $U_\alpha$, $$T=T^{\mu_1\dots\mu_r}_{\nu_1\dots\nu_s}dx_\alpha^{\nu_1}\otimes\dots \otimes dx_\alpha^{\nu_s}\otimes{{\partial}\over\ {\partial x_\alpha^{\mu_1}}}\otimes\dots \otimes{{\partial}\over{\partial x_\alpha^{\mu_r}}}:(\otimes_s(\Gamma (TU_\alpha))\otimes (\otimes_r\Gamma(T^*U_\alpha))\to C^\infty (U_\alpha,\R),$$  $$V_1^{\rho_1}{{\partial}\over{\partial x_\alpha^{\rho_1}}}\otimes\dots \otimes V_s^{\rho_s}{{\partial}\over{\partial x_\alpha^{\rho_s}}}\otimes A_{1\sigma_1}dx_\alpha^{\sigma_1}\otimes\dots \otimes A_{r\sigma_r}dx_\alpha^{\sigma_r}\mapsto T^{\mu_1\dots \mu_r}_{\nu_1\dots\nu_s}dx_\alpha^{\nu_1}(V_1^{\rho_1}{{\partial}\over{\partial x_\alpha^{\rho_1}}})\dots dx_\alpha^{\nu_s}(V_s^{\rho_s}{{\partial}\over{\partial x_\alpha^{\rho_s}}})$$ 

$\times{{\partial}\over{\partial x_\alpha^{\mu_1}}}(A_{1\sigma_1}dx_\alpha^{\sigma_1})\dots {{\partial}\over{\partial x_\alpha^{\mu_r}}}(A_{r\sigma_r}dx_\alpha^{\sigma_r})=T^{\mu_1\dots\mu_r}_{\nu_1\dots\nu_s}V_1^{\rho_1}\delta^{\nu_1}_{\rho_1}\dots V_s^{\rho_s}\delta^{\nu_s}_{\rho_s}A_{1\sigma_1}\delta^{\sigma_1}_{\mu_1}\dots A_{r\sigma_r}\delta^{\sigma_r}_{\mu_r}$

\

$=T^{\mu_1\dots\mu_r}_{\nu_1\dots\nu_s}V_1^{\nu_1}\dots V_s^{\nu_s}A_{1\mu_1}\dots A_{r\mu_r}$.

\

We'll call $\tau^r_s(M)$ to the $C^\infty(M,\R)$-module of $r$-contravariant and $s$-covariant tensors on $M^n$. For example, $\tau^1_0(M)=\Gamma(TM)$: {\it vector fields} on $M$; $\tau^0_1(M)=\Gamma(T^*M)$: {\it differential 1-forms} on $M$. In general, $\tau^r_s(M)=\Gamma(T^r_sM)$: {\it sections of the bundle of $r(s)$ contravariant (covariant) tensors on $M$}, with $T^1_0M=TM$ and $T^0_1M=T^*M$.

\

Locally, given a tensor $T^{\mu_1\dots\mu_r}_{\nu_1\dots\nu_s}$ and a connection $\Gamma^\rho_{\mu\nu}$ in the manifold $M^n$, the {\it covariant derivative  of $T^{\mu_1\dots\mu_r}_{\nu_1\dots\nu_s}$ in the direction of ${{\partial}\over{\partial x_\alpha^\mu}}$} is given by $$D_\mu T^{\mu_1\dots\mu_r}_{\nu_1\dots\nu_s}\equiv T^{\mu_1\dots\mu_r}_{\nu_1\dots\nu_s ;\mu}={{\partial}\over{\partial x_\alpha^\mu}}T^{\mu_1\dots\mu_r}_{\nu_1\dots\nu_s}+\Gamma^{\mu_1}_{\mu\alpha_1}T^{\alpha_1\mu_2\dots\mu_r}_{\nu_1\dots\nu_s}+\dots +\Gamma^{\mu_r}_{\mu\alpha_r}T^{\mu_1\dots\mu_{r-1}\alpha_r}_{\nu_1\dots\nu_s}-\Gamma^{\alpha_1}_{\mu\nu_1}T^{\mu_1\dots\mu_r}_{\alpha_1\nu_2\dots\nu_s}-\dots$$ 

\

$-\Gamma^{\alpha_s}_{\mu\nu_s}T^{\mu_1\dots\mu_r}_{\nu_1\dots\nu_{s-1}\alpha_s}$.

\

It can be verified that $T^{\mu_1\dots\mu_r}_{\nu_1\dots\nu_s;\mu}$ is an $r$-contravariant and $s+1$-covariant tensor.

\

{\it Remark}: Notice that while the operators $\nabla_X$ send tensors (or sections in general) of a given order to tensors of the same order, for both covariant and contravariant indices, the operators $D_\mu$ map $(r,s)$-tensors into $(r,s+1)$-tensors.

\

{\bf Tensors in arbitrary vector bundles}

\

In $U_\alpha\cap U_\beta\subset M^n$ consider the change of local coordinates and sections: ${{\partial}\over{\partial x^\mu}}={{\partial}\over{\partial x^{\prime\nu}}}{{\partial x^{\prime\nu}}\over{\partial x^\mu}}$ and $\sigma_k={f^{-1}}^j_k\sigma^\prime_j$ with $x^\mu=x^\mu_\alpha$, $x^{\prime\nu}=x^\nu_\beta$, $\sigma_k={\sigma_k}_\alpha$, and $\sigma^\prime_j={\sigma_j}_\beta$; $\mu,\nu=1,\dots,n$; $j,k=1,\dots,m$; at each $x\in U_\alpha\cap U_\beta$, $f$ and $f^{-1}$ take values in $GL_m(\R)$ with $\sigma^\prime_j={f^k}_j \sigma_k$. We study the transformation of the Christoffel symbols $\Gamma^i_{\mu j}$ of a connection $\nabla$ in $\xi:\R^m-E\buildrel{\pi}\over\longrightarrow M^n$: $$\nabla_\mu\sigma_i=\Gamma^j_{\mu i}\sigma_j=\nabla_{{{\partial}\over{\partial x^\mu}}}\sigma_i=\nabla_{{{\partial x^{\prime\nu}}\over{\partial x^\mu}}{{\partial}\over{\partial x^{\prime \nu}}}}{f^{-1}}^j_i\sigma^\prime_j={{\partial x^{\prime \nu}}\over{\partial x^\mu}}\nabla_{{{\partial}\over{\partial x^{\prime\nu}}}}{f^{-1}}^j_i\sigma^\prime_j={{\partial x^{\prime\nu}}\over{\partial x^\mu}}(({{\partial}\over{\partial x^{\prime\nu}}}{f^{-1}}^j_i)\sigma^\prime_j+{f^{-1}}^j_i\nabla_{{{\partial}\over{\partial x^{\prime\nu}}}}\sigma^\prime_j)$$ $={{\partial x^{\prime\nu}}\over{\partial x^\mu}}({{\partial}\over{\partial x^{\prime\nu}}}{f^{-1}}^l_i+{f^{-1}}^j_i{\Gamma^\prime}^l_{\nu j})\sigma^\prime_l)=\Gamma^j_{\mu i}{f^{-1}}^l_j\sigma^\prime_l$ 

\

i.e. $$\Gamma^j_{\mu i}{f^{-1}}^l_j={{\partial x^{\prime\nu}}\over{\partial x^\mu}}({{\partial}\over{\partial x^{\prime\nu}}}{f^{-1}}^l_i +{f^{-1}}^j_i {\Gamma^\prime}^l_{\nu j});$$ multiplying by ${f^r}_l$ and using ${f^{-1}}^l_j{f^r}_l={\delta^r}_j$ we obtain $$\Gamma^r_{\mu i}={{\partial x^{\prime\nu}}\over{\partial x^\mu}}{f^{-1}}^j_i{f^r}_l{\Gamma^\prime}^l_{\nu j}+{{\partial x^{\prime\nu}}\over{\partial x^\mu}}{f^r}_l{{\partial}\over{\partial x^{\prime\nu}}}({f^{-1}}^l_i).$$ The homogeneous part in the connection coefficients gives the general law for the tensorial transformation of an object with $r(v)$ contravariant or upper (covariant or lower) internal indices, $r,v=1,\dots,m$, and $s(t)$ contravariant (covariant) space-time (external) indices, $s,t=1,\dots,n$: $$T^{\mu_1\dots\mu_s a_1\dots a_r}_{\nu_1\dots\nu_t b_1\dots b_v}={{\partial x^{\mu_1}}\over {\partial {x^{\prime}}^{\rho_1}}}\cdots{{\partial x^{\mu_s}}\over{\partial {x^{\prime}}^{\rho_s}}}{{\partial {x^\prime}^{\sigma_1}}\over{\partial x^{\nu_1}}}\cdots {{\partial{x^\prime}^{\sigma_t}}\over{\partial x^{\nu_t}}}{f^{a_1}}_{c_1}\cdots {f^{a_r}}_{c_r}{{f^{-1}}^{d_1}}_{b_1}\cdots {{f^{-1}}^{d_v}}_{b_v}{T^\prime}^{\rho_1\dots\rho_s c_1\dots c_r}_{\sigma_1\dots\sigma_t d_1\dots d_v}.$$  For example, $$T^{\mu\nu a}={{\partial x^\mu}\over{\partial {x^\prime}^\rho}}{{\partial x^\nu}\over{\partial{x^\prime}^\sigma}}{f^a}_b{T^\prime}^{\rho\sigma b}, \  \  {T^{\mu\nu}}_{ab}={{\partial x^\mu}\over{\partial {x^\prime}^\rho}}{{\partial x^\nu}\over{\partial {x^\prime}^\sigma}}{{f^{-1}}^c}_a{{f^{-1}}^d}_b {{T^\prime}^{\rho\sigma}}_{cd}.$$

\

{\bf 8. Directional covariant derivative and parallel transport of tensors; geodesics}

\

If $c:(a,b)\to M^n$, $\lambda\mapsto c(\lambda)$, with $(a,b)$ an open interval in $\R$, is a smooth path in $M^n$ locally given by $c(\lambda)=(x^1(\lambda),\dots,x^n(\lambda))$, then {\it the covariant derivative of $T^{\mu_1\dots\mu_r}_{\nu_1\dots\nu_s}$ along $c$} is the tensor defined by $$({{DT}\over {d\lambda}})^{\mu_1\dots\mu_r}_{\nu_1\dots\nu_s}:={{dx^\mu}\over{d\lambda}}D_\mu T^{\mu_1\dots\mu_r}_{\nu_1\dots\nu_s}.$$ We have then defined the {\it covariant derivative operator along the path $c$} through $${{D}\over{d\lambda}}={{dx^\mu}\over{d\lambda}}D_\mu.$$ In detail, $$({{DT}\over{d\lambda}})^{\mu_1\dots\mu_r}_{\nu_1\dots\nu_s}={{dx^\mu}\over{d\lambda}}({{\partial}\over{\partial x^\mu}}T^{\mu_1\dots\mu_r}_{\nu_1\dots\nu_s}+\Gamma^{\mu_1}_{\mu\alpha_1}T^{\alpha_1\mu_2\dots\mu_r}_{\nu_1\dots\nu_s}+\dots -\Gamma^{\alpha_s}_{\mu\nu_s}T^{\mu_1\dots\mu_r}_{\nu_1\dots\nu_{s-1}\alpha_s})$$ $$={{d}\over{d\lambda}}T^{\mu_1\dots\mu_r}_{\nu_1\dots\nu_s}+{{dx^\mu}\over{d\lambda}}\Gamma^{\mu_1}_{\mu\alpha_1}T^{\alpha_1\mu_2\dots\mu_r}_{\nu_1\dots\nu_s}+\dots -{{dx^\mu}\over{d\lambda}}\Gamma^{\alpha_s}_{\mu\nu_s}T^{\mu_1\dots\mu_r}_{\nu_1\dots\nu_{s-1}\alpha_s}.$$ Symbolically, $${{DT}\over{d\lambda}}={{d}\over{d\lambda}}T +\dot{c}\Gamma_* T^*-\dot{c}\Gamma^* T_*$$ where $*$ denotes the contractions. 

\

For a vector field $V^\mu$, $$({{DV}\over{d\lambda}})^\mu={{d}\over{d\lambda}}V^\mu+{{dx^\nu}\over {d\lambda}}\Gamma^\mu_{\nu\rho}V^\rho$$ i.e. $$({{DV}\over{d\lambda}})|_\lambda=({{dV^\mu(\lambda)}\over{d\lambda}}+{{dx^\nu(\lambda)}\over{d\lambda}}\Gamma^\mu_{\nu\rho}(\lambda)V^\rho (\lambda)){{\partial}\over{\partial x^\mu}}|_\lambda$$ where the dependence of $\lambda$ is through $c$, and for a differential 1-form $A_\mu$, $$({{DA}\over{d\lambda}})_\mu= {{dA_\mu}\over {d\lambda}}-{{dx^\nu}\over {d\lambda}}\Gamma^\rho_{\mu\nu}A_\rho$$ i.e. $$({{DA}\over{d\lambda}})|_\lambda=({{dA_\mu(\lambda)}\over{d\lambda}}-{{dx^\nu(\lambda)}\over{d\lambda}}\Gamma^\rho_{\mu\nu}(\lambda)A_\rho(\lambda))dx^\mu|_\lambda.$$ If, in particular, $V^\mu={{dx^\mu}\over{d\lambda}}$: the {\it tangent vector to the curve $c$ at $\lambda$}, i.e. $V=\dot{c}$, then $$({{D\dot{c}}\over{d\lambda}})^\mu={{d^2x^\mu}\over {d\lambda^2}}+{{dx^\nu}\over {d\lambda}}\Gamma^\mu_{\nu\rho} {{dx^\rho}\over {d\lambda}}.$$ Symbolically ${{DV}\over{d\lambda}}=({{d}\over {d\lambda}}+\dot{c}\Gamma)V$ i.e. ${{D}\over{d\lambda}}={{d}\over {d\lambda}}+\dot{c}\Gamma$ for vectors, and ${{DA}\over{d\lambda}}=({{d}\over {d\lambda}}-\dot{c}\Gamma)A$ i.e. ${{D}\over {d\lambda}}={{d}\over {d\lambda}}-\dot{c}\Gamma$ for 1-forms.

\

A tensor $T$ is said to be {\it parallel transported  by the connection $\nabla$ from $c(\lambda_0)$ to $c(\lambda_1)$ along the smooth curve $c$ in $M^n$} ($a<\lambda_0<\lambda_1<b$), if $${{DT}\over {d\lambda}}=0 \ \ \ for \ \ \ all \ \ \ \lambda \ \ \ \in [\lambda_0,\lambda_1]$$ i.e. if $${{dT(\lambda)}\over{d\lambda}}=-\dot{c}\Gamma_*(\lambda)T^*(\lambda)+\dot{c}\Gamma^*(\lambda)T_*(\lambda).$$ This is a system of $n^{r+s}$ ordinary differential equations of first order (ODE-1). By general theorems on ODE-1, if $T_{\lambda_0}\in \tau^r_s(c(\lambda_0))$ then there {\it exists} and is {\it unique} a parallel transported tensor $T(\lambda)$ along $c$, in particular at $c(\lambda_1)$, such that $T(\lambda_0)=T_{\lambda_0}$. 

\

The parallel transport of $T$ depends on $\Gamma$ i.e. on $\nabla$ and on the path $c$. There exists a {\it vector space isomorphism} $$P_c^\nabla:\tau^r_s(c(\lambda_0))\to \tau^r_s(c(\lambda_1)), \ \ \ T_{\lambda_0}\mapsto P_c^\nabla (T_{\lambda_0})=T_{\lambda_1}$$ with $(P_c ^\nabla)^{-1}=P_{c^{-1}}^\nabla$ where $c^{-1}(\lambda)=c(\lambda_1+\lambda_0-\lambda)$.

\

The equations of parallel transport for vector fields and differential 1-forms are $$({{DV}\over{d\lambda}})^\mu=0 \Leftrightarrow {{dV^\mu}\over {d\lambda}}=-{{dx^\nu}\over{d\lambda}}\Gamma^\mu_{\nu\rho}V^\rho$$ and $$({{DA}\over {d\lambda}})_\mu=0 \Leftrightarrow {{dA_\mu}\over {d\lambda}}={{dx^\nu}\over {d\lambda}}\Gamma^\rho_{\mu\nu}A_\rho$$ respectively. 

\

In particular, a curve $c$ is a {\it geodesic} in $M^n$ with respect to the connection $\nabla$, if its tangent vector $\dot{c}$ is parallel transported along $c$: $$({{D\dot{c}}\over{d\lambda}})^\mu =0 \Leftrightarrow {{d^2 x^\mu}\over {d\lambda^2}}=-\Gamma^\mu_{\nu\rho}{{dx^\nu}\over {d\lambda}}{{dx^\rho}\over {d\lambda}}.$$

\

Symbolically, $${{D\dot{c}}\over {d\lambda}}=0 \Leftrightarrow \ddot{c}=-\Gamma \dot{c}^2.$$ In more detail, $$\ddot{x}^\mu(\lambda)+\Gamma^\mu_{\nu\rho}(\lambda)\dot{x}^\nu(\lambda)\dot{x}^\rho(\lambda)=0,$$ which is a system of $n$ ordinary differential equations of second order (ODE-2) for $c(\lambda)$ ($x^\mu(\lambda)$). Given $(x_0,v_{x_0})\in TM$, there always exists a unique solution to this system of equations in an interval $(\lambda_0-\epsilon,\lambda_0+\epsilon)$, $\epsilon >0$ with the initial conditions $c(\lambda_0)=x_0$ and $\dot{c}(\lambda_0)=v_{x_0}$.  

\

The geodesic equation is invariant under the change $\lambda \mapsto a\lambda+b$, with $a,b\in \R$, $a\neq 0$. (See also section 13.) 

\

Notice the arbitrariness of $v_{x_0}$ at $x_0$, and the fact that {\it the whole geodesic} is determined in the interval $(\lambda_0-\epsilon,\lambda_0 +\epsilon)$ (``globally'') from the initial data.

\

For the trivial connection in $\R^n$ (section 6), $\Gamma^{0\mu}_{\nu\rho}=0$ and then the solutions of the geodesic equation are straight lines: $$\ddot{x}^\mu(\lambda)=0 \ \Rightarrow \ x^\mu(\lambda)=a^\mu \lambda + b^\mu.$$

\

{\bf 9. Curvature and torsion of a connection}

\

Let $\nabla$ be a connection on $\xi:\R^m-E \buildrel {\pi}\over\longrightarrow M^n$. The {\it curvature of $\nabla$} is defined as follows: $${\cal R}:\Gamma(TM)\times\Gamma(TM)\times\Gamma(E)\to \Gamma(E), \ \ (X,Y,s)\mapsto {\cal R}(X,Y,s):=([\nabla_X,\nabla_Y]-\nabla_{[X,Y]})(s)$$ i.e. $${\cal R}(X,Y,s)=\nabla_X(\nabla_Y(s))-\nabla_Y(\nabla_X(s))-\nabla_{[X,Y]}(s).$$ Clearly, ${\cal R}(X,Y,s)=-{\cal R}(Y,X,s)$.

\

We'll show that ${\cal R}$ is $C^\infty(M,\R)$-linear in its three entries. This will have as a consequence that the set of local components of ${\cal R}$ behaves as a tensor.

\

i) ${\cal R}(fX,Y,s)=\nabla_{fX}\nabla_Ys-\nabla_Y\nabla_{fX}s-\nabla_{[fX,Y]}s=f\nabla_X\nabla_Ys-\nabla_Y(f\nabla_Xs)-\nabla_{f[X,Y]-Y(f)X}s=f\nabla_X\nabla_Ys-Y(f)\nabla_Xs-f\nabla_Y\nabla_Xs-\nabla_{f[X,Y]}s+\nabla_{Y(f)X}s=f\nabla_X\nabla_Ys-Y(f)\nabla_Xs-f\nabla_Y\nabla_Xs-f\nabla_{[X,Y]}s+Y(f)\nabla_Xs=f(\nabla_X\nabla_Y-\nabla_Y\nabla_X-\nabla_{[X,Y]})(s)=f{\cal R}(X,Y,s)$;

\

ii) ${\cal R}(X,fY,s)=-{\cal R}(fY,X,s)=-f{\cal R}(Y,X,s)=f{\cal R}(X,Y,s)$;

\

iii) ${\cal R}(X,Y,fs)=\nabla_X\nabla_Y(fs)-\nabla_Y\nabla_X(fs)-\nabla_{[X,Y]}(fs)=\nabla_X(Y(f)s+f\nabla_Ys)-\nabla_Y(X(f)s+f\nabla_Xs)-[X,Y](f)s-f\nabla_{[X,Y]}s=X(Y(f))s+Y(f)\nabla_Xs+X(f)\nabla_Ys+f\nabla_X\nabla_Ys-Y(X(f))s-X(f)\nabla_Ys-Y(f)\nabla_Xs-f\nabla_Y\nabla_Xs-X(Y(f))s+Y(X(f))s-f\nabla_{[X,Y]}s=f(\nabla_X\nabla_Y-\nabla_Y\nabla_X-\nabla_{[X,Y]})(s)=f{\cal R}(X,Y,s)$.

\

{\it Locally} (in a common chart for $\xi$ and $M$), 

\

${\cal R}(\partial_\mu,\partial_\nu,\sigma_j)=[\nabla_{\partial_\mu},\nabla_{\partial_\nu}](\sigma_j)-\nabla_{[\partial_\mu,\partial_\nu]}(\sigma_j)$, but $[\partial_\mu,\partial_\nu]=0$ and $\nabla_0 s=\nabla_{0X}s=0\nabla_Xs=0$, then 

\

${\cal R}(\partial_\mu,\partial_\nu,\sigma_j)=[\nabla_{\partial_\mu},\nabla_{\partial_\nu}](\sigma_j)=\nabla_{\partial_\mu}(\nabla_{\partial_\nu}(\sigma_j))-\nabla_{\partial_\nu}(\nabla_{\partial_\mu}(\sigma_j))=\nabla_{\partial_\mu}(\Gamma^i_{\nu j}\sigma_i)-\nabla_{\partial_\nu}(\Gamma^i_{\mu j}\sigma_i)$

$=\partial_\mu (\Gamma^i_{\nu j})\sigma_i+\Gamma^i_{\nu j}\Gamma^l_{\mu i}\sigma_l-\partial_\nu(\Gamma^i_{\mu j})\sigma_i-\Gamma^i_{\mu j}\Gamma^l_{\nu i}\sigma_l={\cal R}^k_{\mu\nu j}\sigma_k$ with $${\cal R}^k_{\mu\nu j}=\partial_\mu\Gamma^k_{\nu j}-\partial_\nu\Gamma^k_{\mu j}+\Gamma^k_{\mu i}\Gamma^i_{\nu j}-\Gamma^k_{\nu i}\Gamma^i_{\mu j}.$$ Then $${\cal R}(X,Y,s)=X^\mu Y^\nu s^j {\cal R}(\partial_\mu,\partial_\nu,\sigma_j)=X^\mu Y^\nu s^j {\cal R}^i_{\mu\nu j}\sigma_i.$$ For a linear connection in a manifold, $${\cal R}^\rho_{\mu\nu\sigma}=\partial_\mu\Gamma^\rho_{\nu\sigma}-\partial_\nu\Gamma^\rho_{\mu\sigma}+\Gamma^\rho_{\mu\lambda}\Gamma^\lambda_{\nu\sigma}-\Gamma^\rho_{\nu\lambda}\Gamma^\lambda_{\mu\sigma},$$ with ${\cal R}(X,Y,Z)=X^\mu Y^\nu Z^\sigma {\cal R}(\partial_\mu,\partial_\nu,\partial_\sigma)=X^\mu Y^\nu Z^\sigma {\cal R}^\rho_{\mu\nu\sigma}\partial_\rho$. 

\

In particular, ${\cal R}(\partial_\mu,\partial_\nu,Z)=[\nabla_{\partial_\mu},\nabla_{\partial_\nu}](Z)=Z^\sigma{\cal R}^\rho_{\mu\nu\sigma}\partial_\rho$ or ${\cal R}(\partial_\mu,\partial_\nu,\partial_\sigma)=[\nabla_{\partial_\mu},\nabla_{\partial_\nu}](\partial_\sigma)={\cal R}^\rho_{\mu\nu\sigma}\partial_\rho$. 

\

Defining $$\tilde{{\cal R}}:\Gamma(TM)\times\Gamma(TM)\to End_{\C ^\infty M}(\Gamma(TM)), \ \ \tilde{{\cal R}}(X,Y):\Gamma(TM)\to\Gamma(TM), \ \ \tilde{{\cal R}}(X,Y)(Z):={\cal R}(X,Y,Z)$$ one obtains, in particular, $$\tilde{{\cal R}}(\partial_\mu,\partial_\nu)=[\nabla_{\partial_\mu},\nabla_{\partial_\nu}],$$ which is the usual expression of curvature in terms of a commutator of local covariant derivatives.

\ 

If in ${\cal R}^\rho_{\mu\nu\sigma}$ we contract $\rho$ with $\sigma$ we obtain the {\it antisymmetric} tensor $$S_{\mu\nu}=\partial_\mu\Gamma^\rho_{\nu\rho}-\partial_\nu\Gamma^\rho_{\mu\rho}=-S_{\nu\mu},$$ and if we contract $\rho$ with $\nu$ we obtain the tensor $${\cal R}_{\mu\sigma}=\partial_\mu\Gamma^\rho_{\rho\sigma}-\partial_\rho\Gamma^\rho_{\mu\sigma}+\Gamma^\rho_{\mu\lambda}\Gamma^\lambda_{\rho\sigma}-\Gamma^\rho_{\rho\lambda}\Gamma^\lambda_{\mu\sigma}.$$ In general, ${\cal R}_{\mu\sigma}$ is non symmetric, not even when $\Gamma^\mu_{\nu\rho}=\Gamma^\mu_{\rho\nu}$. In this case, however, its antisymmetric part is the half of $S_{\mu\nu}$: $${\cal R}_{\{\mu\sigma\}}={{1}\over{2}}({\cal R}_{\mu\sigma}-{\cal R}_{\sigma\mu})={{1}\over{2}}S_{\mu\sigma}.$$ But the definition of $S_{\mu\nu}$ does not require a symmetric connection.

\

In GR, where $\nabla$ is the Levi-Civita connection (section 13) uniquely determined by the metric in a pseudo-riemannian (lorentzian) manifold, it is usual to denote $${\cal R}^\rho_{\mu\nu\sigma}=R^\rho_{\sigma\mu\nu}$$ with ${\cal R}\equiv R:\Gamma(TM)\times \Gamma(TM)\times\Gamma(TM)\to\Gamma(TM)$, $(X,Y,Z)\mapsto R(X,Y,Z)=([\nabla_X,\nabla_Y]-\nabla_{[X,Y]})(Z)=X^\mu Y^\nu Z^\sigma R^\rho_{\sigma\mu\nu}\partial_\rho$. This definition holds for any connection, like the Weyl connection (non-metric symmetric), or that corresponding to the Einstein-Cartan theory (metric non-symmetric).  

\

Clearly, $R^\rho_{\sigma\mu\nu}=-R^\rho_{\sigma\nu\mu}$. Since ${\cal R}(\partial_\mu,\partial_\nu,\partial_\sigma)={\cal R}^\rho_{\mu\nu\sigma}\partial_\rho=R^\rho_{\sigma\mu\nu}\partial_\rho$, then $$<dx^\lambda ,{\cal R}(\partial_\mu,\partial_\nu,\partial_\sigma)>=<dx^\lambda ,R^\rho_{\sigma\mu\nu}\partial_\rho>=<dx^\lambda ,\partial_\rho>R^\rho_{\sigma\mu\nu}=\delta^\lambda_\rho R^\rho_{\sigma\mu\nu}=R^\lambda_{\sigma\mu\nu}$$ where $< \ , \ >$ denotes the 1-form-vector contraction, which is independent of the metric.

\

(${\cal R}:\Gamma(TU)\times\Gamma(TU)\times\Gamma(TU)\to \Gamma(TU), \ \ (\partial_\mu,\partial_\nu,\partial_\sigma)\mapsto {\cal R}(\partial_\mu,\partial_\nu,\partial_\sigma)$.)

\

For a {\it symmetric connection}, $\Gamma^\rho_{\mu\nu}=\Gamma^\rho_{\nu\mu}$ (see section 13), $$R^\rho_{\sigma\mu\nu}=\partial_\mu\Gamma^\rho_{\sigma\nu}-\partial_\nu\Gamma^\rho_{\sigma\mu}+\Gamma^\lambda_{\sigma\nu}\Gamma^\rho_{\lambda\mu}-\Gamma^\lambda_{\sigma\mu}\Gamma^\rho_{\lambda\nu}. \ \ (*)$$

\

The {\it torsion} $T$ of a linear connection $\nabla$ on a manifold $M^n$ is defined as follows: $$T:\Gamma(TM)\times\Gamma(TM)\to\Gamma(TM), (X,Y)\mapsto T(X,Y):\nabla_XY-\nabla_YX-[X,Y].$$ It holds: 

\

i) $T(X,Y)=-T(Y,X)$

\

ii) $T^0=0$ for the trivial connection $\nabla^0$ in $\R^n$.

\

iii) $T$ is $C^\infty(M,\R)$-linear in $X$ and $Y$ i.e. $T(fX,Y)=T(X,fY)=fT(X,Y)$. 

\

iv) Locally, in a chart $U_\alpha (x^\mu_\alpha), \alpha\in J$, $$T(X,Y)=(T(X,Y))^\mu{{\partial}\over {\partial x^\mu}}=((\nabla_XY)^\mu-(\nabla_YX)^\mu-[X,Y]^\mu){{\partial}\over{\partial x^\mu}}=X^\nu 2T^\mu_{\nu\rho}Y^\rho{{\partial}\over{\partial x^\mu}}$$ with $$T^\mu_{\nu\rho}={{1}\over {2}}(\Gamma^\mu_{\nu\rho}-\Gamma^\mu_{\rho\nu})=-T^\mu_{\rho\nu}=\Gamma^\mu _{[\nu\rho]}.$$ If $X^\nu=\delta^\nu_\lambda$ i.e. $X=\delta^\nu_\lambda{{\partial}\over{\partial x^\nu}}={{\partial}\over{\partial x^\lambda}}$, and $Y^\rho=\delta^\rho_\sigma$ i.e. $Y=\delta^\rho_\sigma{{\partial}\over{\partial x^\rho}}={{\partial}\over{\partial x^\sigma}}$, then $$T({{\partial}\over{\partial x^\nu}},{{\partial}\over{\partial x^\rho}})=(\Gamma^\mu_{\nu\rho}-\Gamma^\mu_{\rho\nu}){{\partial}\over{\partial x^\mu}}=2T^\mu _{\nu\rho}{{\partial}\over {\partial x^\mu}}$$ i.e. $T^\mu_{\nu\rho}={{1}\over {2}}(T({{\partial}\over{\partial x^\nu}},{{\partial}\over{\partial x^\rho}}))^\mu$. 

\

A straightforward calculation leads to: $$[D_\rho,D_\sigma]V^\mu=D_\rho(D_\sigma V^\mu)-D_\sigma(D_\rho V^\mu)=({V^\mu}_{;\sigma})_{;\rho}-({V^\mu}_{;\rho})_{;\sigma}=R^\mu_{\lambda\rho\sigma}V^\lambda-2T^\lambda_{\rho\sigma}{V^\mu}_{;\lambda}.$$ If $\varphi$ is a scalar, then $[D_\mu,D_\nu](\varphi)=D_\mu(\partial_\nu\varphi)-D_\nu(\partial_\mu\varphi)=D_\mu(\varphi_{;\nu})-D_\nu(\varphi_{;\mu})=\partial_\mu\varphi_{;\nu}-\Gamma^\rho_{\mu\nu}\varphi_{;\rho}-\partial_\nu\varphi_{;\mu}+\Gamma^\rho_{\nu\mu}\varphi_{;\rho}=-(\Gamma^\rho_{\mu\nu}-\Gamma^\rho_{\nu\mu})\partial_\rho\varphi$ i.e. $$[D_\mu,D_\nu](\varphi)=-2T^\rho_{\mu\nu}\partial_\rho\varphi.$$ So, $[D_\mu,D_\nu](\varphi)=0$ if $T^\rho_{\mu\nu}=0$.

\

Then, with $\varphi=V^\alpha W_\alpha$ and using the Leibnitz rule, for a covariant vector (1-form) one obtains $$[D_\mu,D_\nu]W_\alpha=-R^\sigma_{\alpha\mu\nu}W_\sigma -2T^\rho_{\mu\nu}W_{\alpha;\rho}.$$ The generalization for a tensor $T^{\mu_1\dots\mu_r}_{\nu_1\dots\nu_s}$ is $$[D_\mu,D_\nu](T^{\mu_1\dots\mu_r}_{\nu_1\dots\nu_s})=R^{\mu_1}_{\lambda\mu\nu}T^{\lambda\mu_2\dots\mu_r}_{\nu_1\dots\nu_s}+\cdots+R^{\mu_r}_{\lambda\mu\nu}T^{\mu_1\dots\mu_{r-1}\lambda}_{\nu_1\dots\nu_s}-R^\lambda_{\nu_1\mu\nu}T^{\mu_1\dots\mu_r}_{\lambda\nu_2\dots\nu_s}-\cdots -R^\lambda_{\nu_s\mu\nu}T^{\mu_1\dots\mu_r}_{\nu_1\dots\nu_{s-1}\lambda}$$ $$-2T^\lambda_{\mu\nu}T^{\mu_1\dots\mu_r}_{\nu_1\dots\nu_s ;\lambda}.$$

\

For a {\it symmetric} connection, $\Gamma^\mu_{\nu\rho}=\Gamma^\mu_{\rho\nu}$ and therefore $$T^\mu_{\nu\rho}=0$$ in all charts. I.e. a symmetric connection (like the Levi-Civita connection (section 13)) is torsion free. From the transformation of the $\Gamma^\mu _{\nu\rho}$'s, it is clear that $T^\mu _{\nu\rho}$ is a tensor. In particular one has $$[D_\rho,D_\sigma]V^\mu=R^\mu_{\lambda\rho\sigma}V^\lambda.$$ The {\it modified torsion tensor} is defined as $$\tilde{T}^\mu_{\nu\rho}=T^\mu_{\nu\rho}-{{1}\over {n-1}}(\delta_\rho ^\mu T_\nu -\delta_\nu ^\mu T_\rho )$$ where $T_\nu=T_{\nu\sigma}^\sigma$ is the {\it torsion ``vector"} (in fact, it is a 1-form). $\tilde{T}_{\nu\rho}^\mu$ is traceless i.e. $\tilde{T}_{\nu\mu}^\mu =0$. 

\

{\bf 10. Geometric interpretation of curvature and torsion}

\

{\bf Curvature}

\

Consider the infinitesimal parallelogram $pqrs$ in $M^n$ with coordinates $x^\mu,x^\mu+\epsilon^\mu,x^\mu+\epsilon^\mu+\delta^\mu,x^\mu+\delta^\mu$ respectively, with $|\epsilon^\mu|,|\delta^\mu|\ll$ 1. Let $c$ and $c^\prime$, not necessarily part of geodesics, be curves which join $p$ with $r$ through $q$ and $s$ respectively, and $\nabla$ (locally $\Gamma$) be an arbitrary connection in $M^n$. Let $V_p\in T_pM$; its variation from $p$ to $q$ through $c$ is obtained from the formula of the covariant derivative of a vector along a curve (section 8), which implies $DV^\mu=d\lambda({{DV}\over{d\lambda}})^\mu=dV^\mu+dx^\nu\Gamma^\mu_{\nu\rho}V^\rho$; if the transport is parallel then $DV^\mu=0$ i.e. $dV^\mu=-dx^\nu\Gamma^\mu_{\nu\rho}V^\rho$. So through $c$, $dV^\mu|_c=V_q^\mu|_c -V_p^\mu$ and with $dx^\mu=\epsilon^\mu$ one obtains $$V^\mu_q|_c=V^\mu_p-\epsilon^\nu\Gamma^\mu_{\nu\rho}(p)V_p^\rho.$$ Then, $$V_r^\mu\vert_c=V_q^\mu\vert_c-\delta^\alpha\Gamma^\mu_{\alpha\rho}(q)V_q^\rho|_c=V_p^\mu-\epsilon^\nu\Gamma^\mu_{\nu\rho}(p)V_p^\rho-\delta^\alpha\Gamma^\mu_{\alpha\lambda}(q)(V_p^\lambda-\epsilon^\nu\Gamma^\lambda_{\nu\rho}(p)V_p^\rho)$$ $$=V_p^\mu-\epsilon^\nu\Gamma^\mu_{\nu\rho}(p)V_p^\rho-\delta^\alpha(\Gamma^\mu_{\alpha\sigma}(p)+\partial_\lambda\Gamma^\mu_{\alpha\sigma}(p)\epsilon^\lambda)(V_p^\sigma-\epsilon^\nu\Gamma^\sigma_{\nu\rho}(p)V_p^\rho)$$ $$=V_p^\mu-V_p^\rho\Gamma^\mu_{\nu\rho}(p)\epsilon^\nu-V_p^\rho\Gamma^\mu_{\nu\rho}(p)\delta^\nu-V_p^\rho\partial_\nu\Gamma^\mu_{\alpha\rho}(p)\delta^\alpha\epsilon^\nu+V_p^\rho\Gamma^\sigma_{\nu\rho}(p)\Gamma^\mu_{\alpha\sigma}\delta^\alpha\epsilon^\nu$$ $$=V_p^\mu-V_p^\rho\Gamma^\mu_{\nu\rho}(p)\epsilon^\nu-V_p^\rho\Gamma^\mu_{\nu\rho}(p)\delta^\nu-V_p^\rho(\partial_\nu\Gamma^\mu_{\alpha\rho}(p)-\Gamma^\sigma_{\nu\rho}(p)\Gamma^\mu_{\alpha\sigma}(p))\delta^\alpha\epsilon^\nu,$$ where we have neglected terms of order higher than two in $\epsilon's$ and $\delta's$; to obtain $V_r^\mu|_{c^\prime}$ we simply change $\delta\leftrightarrow\epsilon$ which is equivalent to the change $\alpha\leftrightarrow\nu$: $$V_r^\mu|_{c^\prime}=V_p^\mu-V_p^\rho\Gamma^\mu_{\nu\rho}(p)\delta^\nu-V_p^\rho\Gamma^\mu_{\nu\rho}(p)\epsilon^\nu-V_p^\rho(\partial_\alpha\Gamma^\mu_{\nu\rho}(p)-\Gamma^\sigma_{\alpha\rho}(p)\Gamma^\mu_{\nu\sigma}(p))\delta^\alpha\epsilon^\nu.$$ Then the difference of the two parallel transports is $$V_r^\mu|_{c^\prime}-V_r^\mu|_c=(\partial_\nu\Gamma^\mu_{\alpha\rho}(p)-\partial_\alpha\Gamma^\mu_{\nu\rho}(p)+\Gamma^\mu_{\nu\sigma}(p)\Gamma^\sigma_{\alpha\rho}(p)-\Gamma^\sigma_{\nu\rho}(p)\Gamma^\mu_{\alpha\sigma}(p))V_p^\rho\delta^\alpha\epsilon^\nu=R^\mu_{\rho\nu\alpha}(p)V_p^\rho\epsilon^\nu\delta^\alpha$$ $\equiv R^\mu_{\rho\nu\alpha}(p)V_p^\rho A^{\nu\alpha}$ where $A^{\nu\alpha}$ is the infinitesimal area $\epsilon^\nu\delta^\alpha$ enclosed by the curves $c$ and $c^\prime$. Clearly, $$V_r^\mu|_{c^\prime}=V_r^\mu|_c \ \ if \ \ and \ \ only \ \ if \ \ R^\mu_{\rho\nu\alpha}(p)=0.$$ Then, {\it the curvature tensor measures the difference between the parallel transport of a vector through the paths $c$ and $c^\prime$, where $c\cup (-c^\prime)$ is a loop}. 

\

$V_r^\mu\vert_{c^\prime}-V_r^\mu\vert_c$ amounts to a rotation, since norms of vectors do not change by parallel transport induced by metric connections (appendix B); then one says that {\it curvature is the rotational part of the connection}.

\

When parallel transport is independent of the path, that is, for a vanishing curvature, the connection is said to be {\it integrable} (or {\it flat}).  

\

{\bf Torsion}

\

As before, consider the points $p$, $q$ and $s$ with coordinates $x^\mu$, $x^\mu+\epsilon^\mu$ and $x^\mu+\delta^\mu$, respectively. Consider the infinitesimal vectors at $p$, $\epsilon_p^\mu{{\partial}\over{\partial x^\mu}}|_p$ and $\delta_p^\nu{{\partial}\over{\partial x^\nu}}|_p$ ($\epsilon_p^\mu=\epsilon^\mu$, $\delta_p^\nu=\delta^\nu$); regarded as infinitesimal displacements (translations) in $M^n$, they respectively define the points $q$ and $s$. Make the parallel transport of $\epsilon_p^\mu{{\partial}\over{\partial x^\mu}}|_p$ along $\delta_p^\nu$: we obtain the vector at $s$, $V_s^\mu=\epsilon_p^\mu-\delta^\nu\Gamma^\mu_{\nu\rho}(p)\epsilon_p^\rho$; so the total displacement vector from $p$ to $r$ is $$\delta_p^\mu+\epsilon_p^\mu-\delta_p^\nu\Gamma^\mu_{\nu\rho}(p)\epsilon_p^\rho;$$ similarly, making the parallel displacement of $\delta^\nu{{\partial}\over{\partial x^\nu}}|_p$ along $\epsilon^\mu$ one obtains the vector at $q$, $V_q^\nu=\delta_p^\nu-\epsilon^\alpha\Gamma^\nu_{\alpha\beta}(p)\delta_p^\beta$; and the total displacement vector from $p$ to $r$ is $$\epsilon_p^\mu+\delta_p^\mu-\epsilon_p^\alpha\Gamma^\mu_{\alpha\beta}(p)\delta_p^\beta.$$ The difference between the two vectors is $$-\epsilon_p^\alpha\delta_p^\beta\Gamma^\mu_{\alpha\beta}(p)+\epsilon_p^\alpha\delta_p^\beta\Gamma^\mu_{\beta\alpha}(p)=\epsilon_p^\alpha\delta_p^\beta(\Gamma^\mu_{\beta\alpha}(p)-\Gamma^\mu_{\alpha\beta}(p))=2T^\mu_{\beta\alpha}(p)\epsilon_p^\alpha\delta_p^\beta.$$ So, {\it the torsion measures the failure of the closure of the parallelogram made of the infinitesimal displacement vectors and their parallel transports.}

\

Since the last expression is a translation, one says that {\it torsion is the translational part of the connection}.

\

{\bf 11. Exterior covariant derivative and curvature 2-form}

\

Up to here we have considered the vector bundle $\xi:\R^m-E\buildrel{\pi}\over\longrightarrow M^n$. Now we shall consider the vector bundle $\xi_k$ whose sections are the $k$-differential forms $\alpha\otimes s$ on $M^n$ with values in $E$: $$\R^{{{n!}\over{k!(n-k)!}}\times m}-\Lambda^kT^*M\otimes E\buildrel{\pi_k}\over\longrightarrow M$$ with $\Lambda^kT^*M\otimes E=\coprod_{x\in M}\Lambda^kT^*_xM\otimes E_x$; clearly, $\xi_0=\xi$. $\alpha\otimes s\in\Gamma(\Lambda^kT^*M\otimes E)$ with $(\alpha\otimes s)(x)=(x,(\alpha\otimes s)_x)$, $(\alpha\otimes s)_x\in\Lambda^kT^*_xM\otimes E_x$; if, as before, $\{\sigma_i\}^m_{i=1}$ is a basis of sections of $E$ in $U_\alpha\subset M$, $x\in U_\alpha$, and $\{x_\alpha^\mu\}^n_{\mu=1}$ are local coordinates on $U_\alpha$, then $$\{dx_\alpha^{\mu_1}|_x\wedge\dots\wedge dx_\alpha^{\mu_k}|_x \otimes\sigma_{ix}, \ i=1,\dots ,m, \ n\geq \mu_k >\dots >\mu_1\geq 1\}$$ is a basis of $\Lambda^kT^*_xM\otimes E_x$. So, $$(\alpha\otimes s)_x=\Sigma^m_{i=1}\Sigma_{n\geq \mu_k>\dots >\mu_1\geq 1} \ t^i_{\mu_1\dots\mu_k}dx^{\mu_1}_\alpha|_x\wedge\dots\wedge dx_\alpha^{\mu_k}\otimes\sigma_{ix},$$ $t^i_{\mu_1\dots\mu_k}\in\R$. 

\

We define the set of {\it total exterior covariant derivative} operators $$\{d_0^\nabla,d_1^\nabla,\dots ,d_{n-1}^\nabla\}, \ \ d_k^\nabla:\Gamma(\Lambda^kT^*M\otimes E)\to\Gamma(\Lambda^{k+1}T^*M\otimes E), \ k=0,\dots,n-1,$$ as the $\R$-linear extension of $$d_k^\nabla(\alpha\otimes s)=(d_k\alpha)\otimes s+\alpha\wedge\nabla s$$ with $$(d_k\alpha)\otimes s+\alpha\wedge\nabla s:M\to\Lambda^{k+1}T^*M\otimes E, \ x\mapsto (x,(d_k\alpha)_x\otimes s_x+\alpha_x\wedge (\nabla s)_x).$$ For $k=0$, $d_0^\nabla=\nabla :\Gamma (E)\to \Gamma(T^*M\otimes E)$.

\

Let us study the composition $d_1^\nabla\circ d_0^\nabla$. If $s\in\Gamma (E)$, then $$d_1^\nabla\circ d_0^\nabla (s)=d_1^\nabla\circ\nabla (s)\equiv {\cal R}(s)\in\Gamma(\Lambda^2T^*M\otimes E),$$ i.e. $${\cal R}(s):\Gamma (TM)\times\Gamma (TM)\to\Gamma (E), \ (X,Y)\mapsto {\cal R}(s)(X,Y):M\to E, \ x\mapsto {\cal R}(s)(X,Y)(x)=(x,{\cal R}(s)_x(X_x,Y_x)).$$ ${\cal R}(s)$, also denoted by $\nabla^2(s)$, is called the {\it second total covariant derivative of the section $s$}. (In general, $d_{k+1}^\nabla\circ d_k^\nabla\neq 0$; compare with $d_{k+1}\circ d_k=0$ in De Rham theory.)

\

Locally, $${\cal R}(s)=\nabla^2(s)=d_1^\nabla (\nabla (s))=d_1^\nabla (dx^\mu\otimes\nabla_{\partial_\mu}(s))=d_1^\nabla (dx^\mu\otimes ((\partial_\mu s^j)\sigma_j+\Gamma^j_{\mu i}s^i\sigma_j))$$ $$=d_1(dx^\mu)\otimes((\partial_\mu s^j)\sigma_j+\Gamma^j_{\mu i}s^i\sigma_j)+dx^\mu\wedge\nabla ((\partial\mu s^j)\sigma_j+\Gamma ^j_{\mu i}s^i\sigma_j)$$ $$=dx^\mu\wedge(\nabla({{\partial s^j}\over{\partial x^\mu}}\sigma_j)+\nabla(\Gamma^j_{\mu i}s^i\sigma_j))=dx^\mu\wedge(d({{\partial s^j}\over{\partial x^\mu}})\otimes \sigma_j+\Gamma ^j_i\otimes {{\partial s^i}\over{\partial x^\mu}}\sigma_j+d(\Gamma ^j_{\mu i}s^i)\otimes\sigma_j+\Gamma ^j_k\otimes\Gamma ^k_{\mu i}s^i\sigma_j)$$  $$=dx^\mu\wedge(dx^\nu{{\partial}\over{\partial x^\nu}}({{\partial s^j}\over{\partial x^\mu}})\otimes\sigma_j+dx^\nu{{\partial}\over{\partial x^\nu}}(\Gamma^j_{\mu i}s^i)\otimes\sigma_j+dx^\nu\Gamma^j_{\nu i}\otimes{{\partial s^i}\over{\partial x^\mu}}\sigma_j+dx^\nu\Gamma^j_{\nu k}\otimes\Gamma^k_{\mu i}s^i\sigma_j)$$ $$=ds^i\wedge\Gamma^j_i\otimes\sigma_j+dx^\mu\wedge dx^\nu{{\partial}\over{\partial x^\nu}}(\Gamma^j_{\mu i}s^i)\otimes\sigma_j+\Gamma^k_i\wedge\Gamma^j_ks^i\otimes\sigma_j$$ $$=ds^i\wedge\Gamma^j_i\otimes\sigma_j+dx^\mu\wedge dx^\nu{{\partial}\over{\partial x^\nu}}(\Gamma^j_{\mu i})s^i\otimes\sigma_j+dx^\mu\wedge dx^\nu\Gamma^j_{\mu i}{{\partial s^i}\over{\partial x^\nu}}\otimes \sigma_j+\Gamma^k_i\wedge\Gamma^j_ks^i\times\sigma_j$$ $$=(dx^\mu\wedge d(\Gamma^j_{\mu i})s^i+\Gamma^k_i\wedge\gamma^j_ks^i)\otimes\sigma_j=(-d(\Gamma^j_{\mu i})\wedge dx^\mu s^i+s^i\Gamma^k_i\wedge\Gamma^j_k)\otimes\sigma_j=-(d\Gamma^j_i+\Gamma^j_k\wedge\Gamma^k_i)s^i\otimes\sigma_j$$ i.e. $${\cal R}(s)=\nabla^2(s)=-{\cal R}^i_js^j\otimes\sigma_i$$ where $({\cal R}^i_j)$ is the $m\times m$ matrix with entries in the differential 2-forms on $U\subset M$ i.e. $({\cal R}^i_j)\in \Gamma(\Lambda^2T^*U)(m)$ (or ${\cal R}^i_j\in\Omega^2_{DR}(U)$) given by $${\cal R}^i_j=d\Gamma^i_j+\Gamma^i_k\wedge\Gamma^k_j.$$ (Symbolically, ${\cal R}=d\Gamma+\Gamma\wedge\Gamma=(d+\Gamma\wedge)\Gamma$.)

\

From the last expression for ${\cal R}^i_j$ it is clear that ${\cal R}(s)$ is closely related to the curvature tensor ${\cal R}$. Let us see this relation: $$d\Gamma^i_j=d(\Gamma^i_{\nu j}dx^\nu)=\partial_\mu\Gamma^i_{\nu j}dx^\mu\wedge dx^\nu={{1}\over{2}}(\partial_\mu\Gamma^i_{\nu j}-\partial_\nu\Gamma^i_{\mu j})dx^\mu\wedge dx^\nu,$$ $$\Gamma^i_k\wedge\Gamma^k_j=\Gamma^i_{\mu k}dx^\mu\wedge\Gamma^k_{\nu j}dx^\nu=\Gamma^i_{\mu k}\Gamma^k_{\nu j}dx^\mu\wedge dx^\nu={{1}\over{2}}(\Gamma^i_{\mu k}\Gamma^k_{\nu j}-\Gamma^i_{\nu k}\Gamma^k_{\mu j})dx^\mu\wedge dx^\nu,$$ then $${\cal R}^i_j={{1}\over{2}}{\cal R}^i_{\mu\nu j}dx^\mu\wedge dx^\nu.$$ Also, $${\cal R}(s)(X,Y)=-{\cal R}^i_js^j\sigma_i(X,Y)=-{{1}\over{2}}{\cal R}^i_{\mu\nu j}dx^\mu\wedge dx^\nu (X,Y)s^j\sigma_i=-{{1}\over{2}}{\cal R}^i_{\mu\nu j}{{1}\over{2}}(dx^\mu\otimes dx^\nu-dx^\nu\otimes dx^\mu)(X,Y)s^j\sigma_i$$ $$=-{{1}\over{2}}{\cal R}^i_{\mu\nu j}{{1}\over{2}}(X^\mu Y^\nu-X^\nu Y^\mu)s^j\sigma_i=-{{1}\over{2}}{\cal R}^i_{\mu\nu j}X^\mu Y^\nu s^j\sigma_i$$ i.e. $${\cal R}(X,Y,s)=-2{\cal R}(s)(X,Y).$$ The matrix $({\cal R}^i_j)$ is called {\it the curvature 2-form matrix}.

\

{\bf 12. Bianchi identities}

\
 
For each pair $i,j$ in $\{1,\dots,m\}$, ${\cal R}^i_j$ is a local 2-form. Its exterior derivative is a local 3-form. Since $d^2=0$, we have $d{\cal R}^i_j=d(\Gamma^i_k\wedge d\Gamma^k_j)=d\Gamma^i_k\wedge\Gamma^k_j-\Gamma^i_k\wedge d\Gamma^k_j$, then $d{\cal R}^i_j+\Gamma^i_k\wedge d\Gamma^k_j-d\Gamma^i_k\wedge\Gamma^k_j=0$ or $d{\cal R}^i_j+\Gamma^i_k\wedge({\cal R}^i_j-\Gamma^k_i\wedge\Gamma^i_j)-({\cal R}^i_k-\Gamma^i_l\wedge\Gamma^l_k)\wedge\Gamma^k_j=0$. I.e.  $$d{\cal R}^i_j+\Gamma^i_k\wedge{\cal R}^k_j-{\cal R}^i_k\wedge\Gamma^k_j=0.$$ These are {\it the Bianchi equations.} Simbolically, $d{\cal R}+\Gamma\wedge{\cal R}-{\cal R}\wedge\Gamma=0$.

\

The l.h.s. is an $m\times m$ matrix of 3-forms. So, in terms of the curvature form and the connection coefficients one has $m^2$ equations. However, when the ${\cal R}^i_j$'s are written in terms of the $\Gamma^i_j$'s one obtains $m^2$ identities: $d(d\Gamma^i_j+\Gamma^i_k\wedge\Gamma^k_j)+(\Gamma^i_k\wedge(d\Gamma^k_j+\Gamma^k_l\wedge\Gamma^l_j)-(d\Gamma^i_k+\Gamma^i_l\wedge\Gamma^l_k)\wedge\Gamma^k_j=(d\Gamma^i_k)\wedge\Gamma^k_j-\Gamma^i_k\wedge d\Gamma^k_j+\Gamma^i_k\wedge d\Gamma^k_j+\Gamma^i_k\wedge\Gamma^k_l\wedge\Gamma^l_j-(d\Gamma^i_k)\wedge\Gamma^k_j-\Gamma^i_l\wedge\Gamma^l_k\wedge\Gamma^k_j=0$ {\it identically}.

\

From the explicit expression of ${\cal R}^i_j$ and $\Gamma^k_l$ in terms of local coordinates, in the Bianchi equations one has: $d{\cal R}^i_j={{1}\over{2}}d({\cal R}^i_{\nu\rho j}dx^\nu\wedge dx^\rho)={{1}\over{2}}{\cal R}^i_{\nu\rho j,\mu}dx^\mu\wedge dx^\nu\wedge dx^\rho$, $\Gamma^i_k\wedge{\cal R}^k_j=\Gamma^i_{\mu k}dx^\mu\wedge{{1}\over{2}}{\cal R}^k_{\nu\rho j}dx^\nu\wedge dx^\rho={{1}\over{2}}\Gamma^i_{\mu k}{\cal R}^k_{\nu\rho j}dx^\mu\wedge dx^\nu\wedge dx^\rho$, ${\cal R}^i_k\wedge\Gamma^k_j={{1}\over{2}}{\cal R}^i_{\mu\nu k}dx^\mu\wedge dx^\nu\wedge\Gamma^k_{\rho j}dx^\rho={{1}\over{2}}{\cal R}^i_{\mu\nu k}\Gamma^k_{\rho j}dx^\mu\wedge dx^\nu \wedge dx^\rho$; then $${{1}\over{2}}({\cal R}^i_{\nu\rho j,\mu}+\Gamma^i_{\mu k}{\cal R}^k_{\nu\rho j}-{\cal R}^i_{\mu\nu k}\Gamma^k_{\rho j})dx^\mu\wedge dx^\nu\wedge dx^\rho=0$$ i.e. $${\cal R}^i_{\nu\rho j,\mu}+\Gamma^i_{\mu k}{\cal R}^k_{\nu\rho j}-{\cal R}^i_{\mu\nu k}\Gamma^k_{\rho j}=0.$$

\

For a linear connection in $M^n$, with $i,j,k=\sigma,\lambda,\alpha$, $${\cal R}^\sigma_{\nu\rho\lambda,\mu}+\Gamma^\sigma_{\mu\alpha}{\cal R}^\alpha_{\nu\rho\lambda}-{\cal R}^\sigma_{\mu\nu\alpha}\Gamma^\alpha_{\rho\lambda}=0.$$ In GR, ${\cal R}^\sigma_{\nu\rho\lambda}=R^\sigma_{\lambda\nu\rho}$, then the Bianchi equations are: $$R^\sigma_{\lambda\nu\rho,\mu}+\Gamma^\sigma_{\mu\alpha}R^\alpha_{\lambda\nu\rho}-R^\sigma_{\alpha\mu\nu}\Gamma^\alpha_{\rho\lambda}=0.$$

\

{\it Note 1}. In section 22, we'll see that in the case $E=TM$ and the connection is that of Levi-Civita (section 13), when the Bianchi equations are written in terms of the Ricci tensor (see section 16) and the scalar curvature (section 18), we obtain the vanishing of the covariant divergence of the Einstein's tensor (see section 19): $G^{\mu\nu};_\nu=0$.

\

{\it Note 2}. In electromagnetism, $dF=0$ in terms of the curvature tensor $F$ (field strength), amounts to the homogeneous Maxwell equations. Instead, if $F=dA$ is used, we obtain an identity.

\

{\it Remark}: Up to here, {\it all} the results have been independent of the existence of a {\it metric} $g_{\mu\nu}$ in the manifold $M^n$ i.e. of a non degenerate symmetric scalar product at each tangent space $T_xM^n$. This metric is introduced in the next section.

\

{\bf 13. The Levi-Civita connection}

\

In Appendix A we shall prove the {\it Fundamental Theorem of Riemannian or Pseudo-Riemannian Geometry}, which states that in a riemannian or pseudo-riemannian manifold $(M^n,g_{\mu\nu})$ there exists a unique symmetric and metric linear connection, the Levi-Civita connection, given by $$\Gamma^\mu_{\nu\rho}={{1}\over{2}}g^{\mu\sigma}(\partial_\nu g_{\rho\sigma}+\partial_\rho g_{\nu\sigma}-\partial_\sigma g_{\nu\rho})=g^{\mu\sigma}\Gamma_{\sigma\nu\rho}$$ with $$\Gamma_{\sigma\nu\rho}={{1}\over{2}}(\partial_\nu g_{\rho\sigma}+\partial_\rho g_{\nu\sigma}-\partial_\sigma g_{\nu\rho}).$$ Then $g_{\lambda\mu}\Gamma^\mu_{\nu\rho}=g_{\lambda\mu}g^{\mu\sigma}\Gamma_{\sigma\nu\rho}=\delta^\sigma_\lambda\Gamma_{\sigma\nu\rho}=\Gamma_{\lambda\nu\rho}$.

\

It holds: 

\

i) $D_\mu g_{\nu\rho}=g_{\nu\rho;\mu}=0$ (and also $D_\mu g^{\nu\rho}={g^{\nu\rho}};\mu=0$). A consequence of this is that for any smooth path $c:(a,b)\to M^n$ the metric tensor $g_{\mu\nu}$ is parallel transported along $c$: $$({{Dg}\over{d\lambda}})_{\mu\nu}={{dx^\rho}\over{d\lambda}}D_\rho g_{\mu\nu}=0,$$ where ${{dx^\mu}\over {d\lambda}}=\dot{c}^\mu$. Then it follows that the scalar product of two parallel transported vectors along $c$ by the Levi-Civita connection is also parallel transported i.e. covariantly constant: $${{D}\over{d\lambda}}(g_{\mu\nu}V^\mu W^\nu)=({{Dg}\over{d\lambda}})_{\mu\nu}V^\mu W^\nu +g_{\mu\nu}({{DV}\over{d\lambda}})^\mu W^\nu +g_{\mu\nu} V^\mu ({{DW}\over{d\lambda}})^\nu=0.$$

\

In particular, if $V=W=\dot{c}$, then $g_{\mu\nu}\dot{c}^\mu \dot{c}^\nu\equiv (\dot{c},\dot{c})\equiv ||\dot{c}||^2$ remains constant by parallel transport; if $||\dot{c}||^2>0$,=0,$<$0 the geodesic is respectively called {\it timelike}, {\it null} or {\it lightlike}, and {\it spacelike}. Since $\delta^\mu_\nu=g^{\mu\alpha}g_{\alpha\nu}$, then $\delta^\mu_{\nu;\rho}=0$. 

\

ii) $$D_\mu V_\rho=D_\mu (g_{\rho\nu}V^\nu)=g_{\rho\nu}D_\mu V^\nu$$ i.e. the covariant derivative commutes with the raising or lowering of the indices. 

\

It can be shown that if $c:(a,b)\to M^n$ is a smooth path that extremizes the proper time (or path length) $\tau=\int_c d\lambda f^{1/2}$, with $f=g_{\mu\nu}{{dx^\mu}\over{d\lambda}} {{dx^\nu}\over{d\lambda}}$, then $c$ is a geodesic of the Levi-Civita connection. Also, a change of parameter $\tau\to \lambda=\lambda(\tau)$ preserves the form of the geodesic equation if and only if $\tau\to\lambda$ is an {\it affine transformation}, i.e. $$\lambda=a\tau+b,$$ where $a,b\in \R$ and $a\neq 0$. For an arbitrary transformation one obtains $${{d^2x^\mu}\over{d\lambda^2}}+\Gamma^\mu_{\nu\rho}{{dx^\nu}\over{d\lambda}}{{dx^\rho}\over{d\lambda}}=-{{d^2\lambda}\over{d\tau^2}}({{d\lambda}\over{d\tau}})^{-2} {{dx^\mu}\over{d\lambda}}.$$ This means that the derivation of the geodesic equation in section 8, forces the parameter $\lambda$ to be an {\it affine parameter} i.e. a parameter linearly related, up to an additive constant, to the proper time $\tau$. 

\

It is important to mention that the fact that the connection coefficients $\Gamma_{\nu\rho}^\mu$ depend on the metric function $g_{\mu\nu}$, is the usual argument in the literature for denying to G.R. the character of a gauge theory of gravity. More on this below. 

\

{\bf 14. Physics 1: Equivalence principle in GR}

\

{\it Massive free point particles move along timelike geodesics. Massless free point particles move along lightlike geodesics; in this case} $\lambda$ {\it can not be the proper time since} $(d\tau)^2=0$, (Dirac, 1975).

\

{\bf 15. Covariant components of the curvature tensor}

\

Starting from the expression $(*)$ in section 9, a long but straightforward calculation leads to the result, for the Levi-Civita connection, $$R_{\mu\nu\rho\sigma}=g_{\mu\lambda}R^\lambda_{\nu\rho\sigma}={{1}\over{2}}(\partial_\mu\partial_\sigma g_{\nu\rho}+\partial_\nu\partial_\rho g_{\mu\sigma}-\partial_\mu\partial_\rho g_{\nu\sigma}-\partial_\nu\partial_\sigma g_{\mu\rho})+g_{\alpha\beta}(\Gamma^\alpha_{\mu\sigma}\Gamma^\beta_{\nu\rho}-\Gamma^\alpha_{\mu\rho}\Gamma^\beta_{\nu\sigma}).$$ Clearly, $R_{\mu\nu\rho\sigma}$ is a covariant 4-rank tensor. It has $n^4$ components (e.g. if $n=4$, $4^4=256)$.

\

{\it Algebraic properties of $R_{\mu\nu\rho\sigma}$}

\

i) $R_{\rho\sigma\mu\nu}={{1}\over{2}}(\partial_\rho\partial_\nu g_{\sigma\mu}+\partial_\sigma\partial_\mu g_{\rho\nu}-\partial_\rho\partial_\mu g_{\sigma\nu}-\partial_\sigma\partial_\nu g_{\rho\mu})+g_{\alpha\beta}(\Gamma^\alpha_{\rho\nu}\Gamma^\beta_{\sigma\mu}-\Gamma^\alpha_{\rho\mu}\gamma^\beta_{\sigma\nu})=R_{\mu\nu\rho\sigma}$ 

(symmetry under the interchange between the first pair of indices with the second pair of indices i.e. $R_{AB}=R_{BA}$ with $A=\rho\sigma$ and $B=\mu\nu$).

\

ii) $R_{\nu\mu\rho\sigma}=-R_{\mu\nu\rho\sigma}$, $R_{\mu\nu\sigma\rho}=-R_{\mu\nu\rho\sigma}$, then $R_{\nu\mu\sigma\rho}=R_{\mu\nu\rho\sigma}$ which can be summarized as $$R_{\mu\nu\rho\sigma}=-R_{\nu\mu\rho\sigma}=-R_{\mu\nu\sigma\rho}.$$

iii) $R_{\mu\nu\rho\sigma}+R_{\mu\rho\sigma\nu}+R_{\mu\sigma\nu\rho}=0$ (cyclicity).

\

If one defines $$A_{\mu\nu\rho\sigma}:=R_{\mu\nu\rho\sigma}+R_{\mu\rho\sigma\nu}+R_{\mu\sigma\nu\rho},$$ it can be proved that it is totally antisymmetric in its four indices. Since $A_{\mu\nu\rho\sigma}=0$, this imposes $\pmatrix{m \cr 4 \cr}={{m!}\over{4!(m-4)!}}$ conditions on $R_{\mu\nu\rho\sigma}$. (Number  of ways one can take four distinct elements among $m$; obviously it must be $m\geq 4$.) 

\

Let us determine the number of algebraically {\it independent components} of $R_{\mu\nu\rho\sigma}$. Let $S_{ab}$ and $A_{ab}$ be respectively a symmetric and antisymmetric tensor in $m$ dimensions. The corresponding number of independent components are $N(S_{ab};m)={{m(m+1)}\over{2}}$ and $N(A_{ab};m)={{m(m-1)}\over{2}}$. So, we have $$\matrix{m & 1 & 2 & 3 & 4 & 5 & 6 & \dots & 10 & \dots \cr N(S_{ab};m) & 1 & 3 & 6 & 10 & 15 & 21 & \dots & 55 & \dots \cr N(A_{ab};m) & 0 & 1 & 3 & 6 & 10 & 15 & \dots & 45 & \dots \cr}$$ Notice that ${{N(A_{ab};m)}\over {N(S_{ab};m)}} \to 1_-$ as $m \to \infty$. If we write $R_{\mu\nu\rho\sigma}\equiv R_{AB}$, since under $\mu\leftrightarrow\nu$ or $\rho\leftrightarrow\sigma$ $R_{\mu\nu\rho\sigma}$ is antisymmetric, each index $A$ or $B$ contributes with ${{m(m-1)}\over{2}}$ independent components; but now one has a ``two-index'' symmetric matrix $R_{AB}$ with $A,B\in \{1,\dots,{{m(m-1)}\over{2}}\}$, which gives ${{1}\over{2}}{{m(m-1)}\over{2}}({{m(m-1)}\over{2}}+1)={{1}\over{8}}m(m-1)(m(m-1)+2)={{1}\over{8}}m(m-1)(m^2-m+2)$ independent components for $R_{\mu\nu\rho\sigma}$. But iii) and then the antisymmetry of $A_{\mu\nu\rho\sigma}$ imposes ${{m!}\over{4!(m-4)!}}={{m(m-1)(m-2)(m-3)}\over{4!}}$ conditions. Then, $$N(R_{\mu\nu\rho\sigma};m)={{1}\over{8}}m(m-1)(m^2-m+2)-{{1}\over{4!}}m(m-1)(m-2)(m-3)={{m^2(m^2-1)}\over{12}}.$$ So, we have $$\matrix{m & 1 & 2 & 3 & 4 & 5 & 6 & \dots & 10 & 11 & \dots & 26 & \dots \cr N(R_{\mu\nu\rho\sigma};m) & 0 & 1 & 6 & 20 & 50 & 105 & \dots & 825 & 1210 & \dots & 38025 & \dots \cr}$$

\

{\bf 16. Ricci tensor for the Levi-Civita connection}

\

Define the covariant 2-tensor $$R_{\nu\sigma}:=g^{\mu\rho}R_{\mu\nu\rho\sigma}\equiv R^\rho_{\nu\rho\sigma}.$$ We contracted indices 1 and 3; contracting 1-2 and 3-4 gives zero; contracting 1-4, 2-3 and 2-4 gives $\pm R_{\nu\sigma}$: $g^{\mu\sigma}R_{\mu\nu\rho\sigma}=-g^{\mu\sigma}R_{\mu\nu\sigma\rho}=-R_{\nu\rho}$, $g^{\nu\rho}R_{\mu\nu\rho\sigma}=-g^{\nu\rho}R_{\nu\mu\rho\sigma}=-R_{\mu\sigma}$, $g^{\nu\sigma}R_{\mu\nu\rho\sigma}=g^{\nu\sigma}R_{\nu\mu\sigma\rho}=R_{\mu\rho}$. So, up to a sign, the Ricci tensor is {\it uniquely defined} from $R_{\mu\nu\rho\sigma}$ and $g^{\mu\nu}$. 

\

$R_{\nu\sigma}$ is {\it symmetric}: $R_{\sigma\nu}=g^{\mu\rho}R_{\mu\sigma\rho\nu}=g^{\mu\rho}R_{\rho\nu\mu\sigma}=g^{\rho\mu}R_{\rho\nu\mu\sigma}=R_{\nu\sigma}$. Then, $$N(R_{\mu\nu};m)={{m(m+1)}\over{2}}.$$ We have $$\matrix{m & 1 & 2 & 3 & 4 & 5 & 6 & \dots & 10 & 11 & \dots \cr N(R_{\mu\nu};m) & 1 & 3 & 6 & 10 & 15 & 21 & \dots & 55 & 66 & \dots \cr}$$ We can write $$R_{\nu\sigma}=g^{\mu\rho}g_{\mu\lambda}R^\lambda_{\nu\rho\sigma}=\delta^\rho_\lambda R^\lambda_{\nu\rho\sigma}=R^\rho_{\nu\rho\sigma}=<dx^\rho,{\cal R}(\partial_\rho,\partial_\sigma,\partial_\nu)>=<dx^\rho,{\cal R}(\partial_\rho,\partial_\nu,\partial_\sigma)>.$$

\

{\bf 17. Physics 2: Einstein's hypotesis: (Local) equations for empty (``vacuum'', without matter, vanishing cosmological constant $\Lambda$, only $g_{\mu\nu}$) space-time}

\

In all charts of the manifold i.e. {\it in all reference frames}: $$R_{\mu\nu}=0,$$ $\mu,\nu=1,\dots,m$ or $0,\dots,m-1$. $R_{\mu\nu}\in C^\infty (U_\alpha;\R)$,   (Dirac, 1975). These are equations at each chart or reference system $(U_\alpha, x_\alpha^\mu)$ of $M^m$. The number of algebraically independent components of $R_{\mu\nu}$ equals the number of independent components of $g_{\mu\nu}$. $R_{\mu\nu}=0$ {\it does not} imply $R_{\mu\nu\rho\sigma}=0$. In other words, in GR empty space-time can be curved.

\

{\bf 18. Ricci (or curvature) scalar (with Levi-Civita connection)}

\

$$R:=g^{\mu\nu}R_{\mu\nu}\equiv R^\mu_\mu.$$ $R\in C^\infty(U_\alpha;\R)$. 

\

Then, in empty space-time, $$R=0.$$

\

{\bf 19.} The {\it Einstein tensor} is defined by $$G_{\mu\nu}:=R_{\mu\nu}-{{1}\over{2}}g_{\mu\nu}R.$$ $G_{\mu\nu}\in C^\infty(U_\alpha;\R)$. Also, it is symmetric (10 algebraically independent components in 4 dimensions).

\

{\it Proposition} (Mathematics) $$G_{\mu\nu}=0 \Leftrightarrow R_{\mu\nu}=0$$ for $m=D\neq 2$, $D=d+1$ (space-time dimension). 

\

Proof. 

\

$\Rightarrow$) $G_{\mu\nu}=0\Rightarrow R_{\mu\nu}={{1}\over{2}}g_{\mu\nu}R \Rightarrow R^\mu_\mu=R={{1}\over{2}}g_{\mu}^\mu R={{D}\over{2}}R \Rightarrow ({{D}\over{2}}-1)R=0$. If ${{D}\over{2}}\neq 1$ i.e. $D\neq 2$, then $R=0$ and then $R_{\mu\nu}=G_{\mu\nu}=0$. In particular this holds for $D=3+1=4$. 

\

$\Leftarrow$) $R_{\mu\nu}=0 \Rightarrow R^\mu_{\mu}=R=0 \Rightarrow G_{\mu\nu}=0$. This holds for all $D$'s=$m$'s.

\

{\bf 20. Physics 2': (Local) Einstein equations in empty space-time}

\

In all charts of the manifold i.e. {\it in all reference frames}: $$G_{\mu\nu}=0$$ for {\it all} space-time dimensions $D=m=d+1$. (For $D=4$ these are ten equations.)

\

{\bf 21. Examples in $m=D=1,2,3.$ Generalization to $m\geq 4$ and Weyl tensor}

\

$D=1$. $N(R_{\mu\nu\rho\sigma};1)=0$; then $R_{1111}$ (or $R_{0000}$)=0. This reflects the fact that the curvature tensor represents {\it intrinsic} properties of the space in question, and not how the space (in this case a line, straight or curved) is embedded in a higher dimensional space. Let $g_{11}(x)\equiv g(x)$ be the metric tensor in $D=1$. If $x\mapsto x^\prime=x^\prime (x)$ then $g^\prime (x^\prime)=({{dx}\over{dx^\prime}})^2 g(x)$. Choose $x^\prime (x)=\int ^x dy \sqrt{\vert g(y)\vert }$; then $({{dx}\over{dx^\prime}})^2={{1}\over{\vert g(x)\vert}}$ and so $g^\prime(x^\prime)={{g(x)}\over{\vert g(x)\vert}}$ which equals +1 (-1) if $g(x)>0$ ($<0$). From the constancy of $g(x)$, $\Gamma(x)={{1}\over{2g(x)}}{{d}\over{dx}}g(x)=0$.

\

$D=2$. $N(R_{\mu\nu\rho\sigma};2)=1$. By antisymmetry in $\mu\nu$ and $\rho\sigma$, the only possibilities for a non-vanishing $R_{\mu\nu\rho\sigma}$ are $R_{1010}$, $R_{1001}$, $R_{0110}$ and  $R_{0101}$. We choose $R_{0101}$ and it is easily verified that the unique $R_{\mu\nu\rho\sigma}$ which satisfies the algebraic properties of section {\bf 15} and gives $R_{0110}=-R_{1010}=R_{1001}=-R_{0101}$ is $$R_{\mu\nu\rho\sigma}=(g_{\mu\rho}g_{\nu\sigma}-g_{\mu\sigma}g_{\nu\rho}){{R_{0101}}\over{g}},$$ with $g=det \pmatrix{g_{00} & g_{01} \cr g_{10} & g_{11} \cr}=g_{00}g_{11}-g_{01}g_{10}=g_{00}g_{11}-g_{01}^2$. In the presence of matter, Einstein's equations are $$G_{\mu\nu}=const. \times T_{\mu\nu}$$ where $T_{\mu\nu}$ is the energy-momentum tensor of matter. Then, for $D=2=1+1$, $$T_{\mu\nu}=0.$$ This means that in $D=2$, the {\it unique} solutions to Einstein's equations are those corresponding to the ``vacua'' i.e. $T_{\mu\nu}=0$. 
 
\

$D=3$. Since $N(R_{\mu\nu\rho\sigma};3)=N(R_{\mu\nu};3)=3$, the curvature tensor can be expressed in terms of $g_{\mu\nu}$ and $R_{\mu\nu}$; the most general form satisfying the symmetry properties of section {\bf 15} is $$R_{\mu\nu\rho\sigma}=A(g_{\mu\rho}g_{\nu\sigma}-g_{\nu\rho}g_{\mu\sigma})+B(g_{\mu\rho}R_{\nu\sigma}-g_{\nu\rho}R_{\mu\sigma}-g_{\mu\sigma}R_{\nu\rho}+g_{\nu\sigma}R_{\mu\rho})$$ with $A$ and $B$ numerical constants. From the definition of the Ricci tensor, $R_{\nu\sigma}=g^{\mu\rho}R_{\mu\nu\rho\sigma}=(2A+BR)g_{\nu\sigma}+BR_{\nu\sigma}$, where $R$ is the Ricci scalar; then $B=1$ and $A=-{{1}\over{2}}R$. Therefore, $$R_{\mu\nu\rho\sigma}=-{{R}\over{2}}(g_{\mu\rho}g_{\nu\sigma}-g_{\nu\rho}g_{\mu\sigma})+g_{\mu\rho}R_{\nu\sigma}-g_{\nu\rho}R_{\mu\sigma}-g_{\mu\sigma}R_{\nu\rho}+g_{\nu\sigma}R_{\mu\rho}.$$

\

$D=m\geq 4$. In all these cases $$N(R_{\mu\nu\rho\sigma};m)-N(R_{\mu\nu};m):=N(C_{\mu\nu\rho\sigma};m)={{m(m+1)(m+2)(m-3)}\over{12}}>0$$ where $C_{\mu\nu\rho\sigma}$, the {\it Weyl tensor}, has the same algebraic properties as $R_{\mu\nu\rho\sigma}$ but can't be obtained from $g_{\mu\nu}$ and $R_{\mu\nu}$. One writes $$R_{\mu\nu\rho\sigma}=C(g_{\mu\rho}g_{\nu\sigma}-g_{\nu\rho}g_{\mu\sigma})+D(g_{\mu\rho}R_{\nu\sigma}-g_{\nu\rho}R_{\mu\sigma}-g_{\mu\sigma}R_{\nu\rho}+g_{\nu\sigma}R_{\mu\rho})+C_{\mu\nu\rho\sigma}$$ with all traces of $C_{\mu\nu\rho\sigma}$ vanishing i.e. $$C^\mu_{\nu\mu\sigma}=0.$$ Then $$R_{\nu\sigma}=g^{\mu\rho}R_{\mu\nu\rho\sigma}=C(mg_{\nu\sigma}-g_{\nu\sigma})+D(mR_{\nu\sigma}-2R_{\nu\sigma}+g_{\nu\sigma}R)=g_{\nu\sigma}((m-1)C+DR)+(m-2)DR_{\nu\sigma}.$$ Then $D={{1}\over{m-2}}$ and $C=-{{R}\over{(m-1)(m-2)}}$. Therefore, $$R_{\mu\nu\rho\sigma}=-{{R}\over{(m-1)(m-2)}}(g_{\mu\rho}g_{\nu\sigma}-g_{\nu\rho}g_{\mu\sigma})+{{1}\over{m-2}}(g_{\mu\rho}R_{\nu\sigma}-g_{\nu\rho}R_{\mu\sigma}-g_{\mu\sigma}R_{\nu\rho}+g_{\nu\sigma}R_{\mu\rho})+C_{\mu\nu\rho\sigma}.$$ The Weyl tensor in terms of curvature and metric is given by $$C_{\mu\nu\rho\sigma}=R_{\mu\nu\rho\sigma}+{{2R}\over{(m-1)(m-2)}}g_{\rho [\mu}g_{\nu]\sigma}-{{2}\over{m-2}}(g_{\rho [\mu}R_{\nu ]\sigma}-g_{\sigma [\nu}R_{\mu ]\rho}).$$ One has the table $$\matrix{m  & 4 & 5 & 6 & \cdots \cr N(C_{\mu\nu\rho\sigma};m) & 10 & 35 & 84 & \cdots \cr}$$ Clearly $C_{\mu\nu\rho\sigma}=0$ for $m=2,3$. 

\

{\bf 22. For the Levi-Civita connection, $G^{\mu\nu};_\nu=0$}

\

From the general expression for the covariant derivative of a tensor $T^{\mu_1\dots\mu_r}_{\nu_1\dots\nu_r}$ in section 7, the relation between the covariant and the ordinary derivatives of $R^\sigma_{\lambda\nu\rho}$ is given by $$R^\sigma_{\lambda\nu\rho,\mu}=R^\sigma_{\lambda\nu\rho;\mu}-\Gamma^\sigma_{\mu\alpha}R^\alpha_{\lambda\nu\rho}+\Gamma^\beta_{\mu\lambda}R^\sigma_{\beta\nu\rho}+\Gamma^\beta_{\mu\nu}R^\sigma_{\lambda\beta\rho}+\Gamma^\beta_{\mu\rho}R^\sigma_{\lambda\nu\beta}.$$ Then, the last form of the Bianchi equations in section 12 becomes $$R^\sigma_{\lambda\nu\rho;\mu}+R^\sigma_{\lambda\rho\mu;\nu}+R^\sigma_{\lambda\mu\nu;\rho}=0.$$ Since the covariant derivative commutes with the lowering of indices, contracting with $g_{\alpha\sigma}$ we obtain $$R_{\alpha\lambda\nu\rho;\mu}+R_{\alpha\lambda\rho\mu;\nu}+R_{\alpha\lambda\mu\nu;\rho}=0.$$ Contracting $\alpha$ and $\rho$ with $g^{\alpha\rho}$, $$R^\alpha_{\lambda\nu\alpha;\mu}+R^\alpha_{\lambda\alpha\mu;\nu}+R^\alpha_{\lambda\mu\nu;\alpha}=0,$$ and we obtain, in terms of the Ricci and curvature tensors, $$-R_{\lambda\nu;\mu}+R_{\lambda\mu;\nu}+R^\alpha_{\lambda\mu\nu;\alpha}=0.$$ Finally, contracting $\lambda$ and $\nu$ with $g^{\lambda\nu}$, we obtain, in terms of the Ricci tensor and the scalar curvature, $$(R^\nu_\mu-{{1}\over{2}}\delta^\nu_\mu R);_\nu=0$$ i.e. $$G^{\mu\nu};_\nu=0.$$ So, {\it the Einstein tensor}, apparently arbitrarily defined in section 19, appears {\it naturally} when the Bianchi equations for the Levi-Civita connection are expressed in terms of the Ricci tensor and the scalar curvature, and some contractions are done. $G_{\mu\nu}$ is, then, a {\it purely geometrical object}.

\

{\bf 23. Physics 3: (Local) Einstein equations in the presence of matter}

\

In all charts of the manifold i.e. {\it in all reference frames}: $$G_{\mu\nu}={{8\pi G}\over{c^4}}T_{\mu\nu},$$ (Einstein, 1956), where $G$ is the Newton gravitational constant, $c$ is the velocity of light in the vacuum, and $T_{\mu\nu}$ is the energy momentum tensor of matter: all other fields than $g_{\mu\nu}$. Clearly, $T_{\mu\nu}$ is symmetric ($T_{\mu\nu}=T_{\nu\mu}$) and covariantly conserved: $$ T^{\mu\nu};\nu=0.$$ 

{\it Units}: $[G_{\mu\nu}]=[R_{\mu\nu}]=[R_{\mu\nu\rho\sigma}]=[L]^{-2}$, ${{[G]}\over{[c^4]}}={{[t]^2}\over{[M][L]}}={{1}\over{[force]}}$; then $[T_{\mu\nu}]={{[M]}\over{[L][t]}}=[energy \ density]$, where $[L]$, $[M]$, and $[t]$ denote the units of length, mass, and time, respectively.

\

{\bf 24. Tensor bundles as associated bundles to the bundle of frames of $M^n$}

\

$T^r_sM^n$ is the total space of the ($n+n^{r+s}$)-dimensional real vector bundle of $r$-contravariant and $s$-covariant tensors on $M^n$, with fibre $\R^{n^{r+s}}\cong \{\lambda^{i_1\dots i_r}_{j_1\dots j_s}\in \R, i_k,j_l\in \{1,\dots,n\},k=1,\dots,r,l=1,\dots,s\}\equiv\{\vec{\lambda}\}$. The {\it bundle of frames of $M^n$}, ${\cal F}_{M^n}$, is the principal bundle with structure group $GL_n(\R)$ (the fibre of the bundle) on $M^n$ (the base space), and with total space $FM^n$ consisting of the set of all ordered basis of the tangent space at each point of $M^n$, namely $$FM^n=\coprod_{x\in M^n}\{r_x\equiv(v_{1x},\dots,v_{nx}),\{v_{kx}\}_{k=1}^n:\ basis \ of \ T_xM^n\}=\cup_{x\in M^n}\{x\}\times\{(v_{1x},\dots,v_{nx})\}$$ $$\equiv \coprod_{x\in M^n}(FM^n)_x,$$ where $(FM^n)_x$ is the fibre over $x$, with $dim_\R (FM^n)_x=n^2$. The bundles of orthogonal frames, Lorentz frames, restricted Lorentz frames, etc. of $M^n$, are obtained by reducing the group $GL_n(\R)$ respectively to $O(n)$, $O(n-1,1)$, $SO^0(n-1,1)$, etc. If $x\in U_\alpha\equiv U$, then $v_{kx}=\sum_{\mu=1}^n v_k^\mu (x){{\partial}\over{\partial x^\mu}}|_x$; also, $dim_{\R} FM^n=n+n^2$. The $n+n^2$ local coordinates on ${\cal F}_{U_\alpha}$ is the set $(x^\rho,X^\mu_\nu)$ with $x^\rho(x,r_x)=x^\rho(x)$ and $X^\mu_\nu(x,r_x)=v^\mu_{\nu x}$, $\rho,\mu, \nu \in\{1,\dots,n\}$.

\

One has: $$\matrix{n & 1 & 2 & 3 & 4 & 5 & \dots & 10 & \dots \cr dim_{\R} FM^n & 2 & 6 & 12 & 20 & 30 & \dots & 110 & \dots \cr}$$ The bundle structure of ${\cal F}_{M^n}$ is represented by $${\cal F}_{M^n}:GL_n(\R)\to FM^n\buildrel {\pi_F}\over\longrightarrow M^n$$ where $\pi_F$ is the projection $\pi_F(x,(v_{1x},\dots,v_{nx}))=x$ and $GL_n(\R)\to FM^n$ represents the right action of $GL_n(\R)$ on $FM^n$ given by $$FM^n\times GL_n(\R)\buildrel{\psi}\over\longrightarrow FM^n, \ ((v_{1x},\dots,v_{nx}),a)\mapsto (v_{1x}{a^1}_1+\cdots+v_{nx}{a^n}_1,\dots,v_{1x}{a^1}_n+\cdots+v_{nx}{a^n}_n)$$ $\equiv (v_{1x},\dots,v_{nx})a.$

\

The left action of $GL_n(\R)$ on $\R^{n^{r+s}}$, given by $$GL_n(\R)\times \R^{n^{r+s}}\buildrel{\mu}\over\longrightarrow\R^{n^{r+s}}, \ (a,\vec{\lambda})\mapsto (a\vec{\lambda})^{i_1\dots i_r}_{j_1\dots j_s}={a^{i_1}}_{k_1}\dots{a^{i_r}}_{k_r}{{a^{-1}}^{l_1}}_{j_1}\dots{{a^{-1}}^{l_s}}_{j_s}\lambda^{k_1\dots k_r}_{l_1\dots l_s},$$ induces the associated bundle $FM^n\times_{GL_n(\R)}\R^{n^{r+s}}$ which turns out to be isomorphic (through $\varphi$, see below) to $T^r_sM^n$. One has the following commutative diagram: $$\matrix{FM^n\times_{GL_n(\R)}\R^{n^{r+s}} & \buildrel{\varphi}\over\longrightarrow & T^r_sM^n \cr \pi^\prime_F\downarrow & &  \downarrow\pi^r_s \cr M^n & \buildrel{Id_{M^n}}\over\longrightarrow & M^n \cr}$$ where:

\

$\varphi([((v_{1x},\dots,v_{nx}),\vec{\lambda})])=\sum^n_{i_k,j_l=1}\lambda^{i_1\dots i_r}_{j_1\dots j_s}v_{i_1x}\otimes\dots v_{i_rx}\otimes w_x^{j_1}
\otimes\dots\otimes w_x^{j_s}$,

\

with $[((v_{1x},\dots,v_{nx}),\vec{\lambda})]=\{((v_{1x},\dots,v_{nx})a,a^{-1}\vec{\lambda})\}_{a\in GL_n(\R)}$;
$\{w_x^1,\dots,w_x^n\}$ is the dual basis of 

\

$\{v_{1x},\dots,v_{nx}\}$ i.e. $w_x^i(v_{jx})=\delta^i_j$; and $\pi^\prime_F([((v_{1x},\dots,v_{nx}),\vec{\lambda})])=\pi_F(x,(v_{1x},\dots,v_{nx}))=x.$ Using the fact that the dual basis vectors $w^j_x$ transform with $a^{-1}$, it is easily verified that $\varphi$ is well defined i.e. it is independent of the representative element of the class $[((v_{1x},\dots,v_{nx}),\vec{\lambda})]$.

\

A {\it section} of the bundle of frames of $M^n$ i.e. a smooth function $s:M^n\to FM^n$ with $\pi_F\circ s=Id_{M^n}$, trivializes ${\cal F}_{M^n}$ and therefore all the tensor bundles associated with it (in particular the tangent bundle of $M^n$). The same occurs for any of the reductions of ${\cal F}_{M^n}$ (bundle of Lorentz frames, restricted Lorentz frames, etc.).

\

{\bf 25. Vertical bundle of a principal fibre bundle}

\

Let $\eta$ be a principal fibre bundle (p.f.b.), $\eta=(P^{r+s},B^s,\pi,G^r,\psi,{\cal U}): \ G^r\to P^{r+s}\buildrel{\pi}\over\longrightarrow B^s$, where $B^s\equiv B$ (base space) and $P^{r+s}\equiv P$ (total space) are differentiable manifolds of dimensions $s$ and $r+s$ respectively, $G^r\equiv G$ is an $r$-dimensional Lie group with right action on $P$, $P\times G\to P$, $(p,g)\mapsto pg$, and ${\cal U}$ is a system of local trivializations $\pi^{-1}(U_\alpha)\buildrel{\varphi_\alpha}\over\longrightarrow U_\alpha\times G$ with $\pi_1\circ \varphi_\alpha=\pi$. 

\

For each $p\in P$ there exists a {\it canonical vector space isomorphism} $\varphi_p$ between ${\cal G}=Lie(G)$ : the Lie algebra of $G$, and $V_p=T_pP_{\pi(p)}$: the tangent space to the fibre over $\pi(p)$ at $p$, {\it the vertical space} at $p$: $$\varphi_p:{\cal G}\to V_p, \ A\mapsto \varphi_p(A)\equiv A^*_p,$$ with $$A^*_p:C^\infty (P,\R)\to\R, \ f\mapsto{{d}\over{dt}}f(pe^{tA})|_{t=0}.$$ We used the fact that $T_pP_{\pi(p)}\subset T_pP$; if $A_i$, $i=1,\dots,r$ is a basis of ${\cal G}$, then $\varphi_p(A_i)$ is a basis of $V_p$; in general, neither $A_i$ nor $\varphi_p(A_i)$ are canonical basis.

\

Given $p,p^\prime\in P$, since $\varphi_p:{\cal G}\to V_p$ and $\varphi_{p^\prime}:{\cal G}\to V_{p^\prime}$ are isomorphisms, there is a {\it canonical} vector space isomorphism ({\it absolute teleparallelism}) between $V_p$ and $V_{p^\prime}$, for {\it all} $p,p^\prime\in P$: $$V_p\buildrel{\varphi_{p^\prime}\circ\varphi^{-1}_p}\over\longrightarrow V_{p^\prime}.$$ {\it Remark}: This result, namely, the existence of $\varphi_p$ at each $p\in P$, is independent of any connection. 

\

This implies the {\it triviality} of the {\it vertical bundle} $V_\eta$ of the p.f.b. $\eta$: $$V_\eta: \ \R^r-V^{2r+s}\buildrel{\pi_{V_\eta}}\over\longrightarrow P,$$ with $V^{2r+s}=\coprod_{p\in P}V_p=\cup_{p\in P}\{p\}\times V_p$ and $\pi_{V_\eta }(p,v_p)=p.$

\

In fact, $V_\eta$ admits $r$ independent global sections $\sigma_i:P\to V^{2r+s}$, $\sigma_i(p)=(p,\varphi_p(A_i))$; then there is the following vector bundle isomorphism: $$\matrix{\R^r & & \R^r \cr \vert & & \vert \cr V^{2r+s} & \buildrel{\phi}\over\longrightarrow & P\times\R^r \cr \pi_{V_\eta}\downarrow & &  \downarrow\pi_1 \cr P & \buildrel{Id}\over\longrightarrow & P \cr}$$ with $\phi(p,v_p)=\phi (p,\sum_{i=1}^r\lambda^i\varphi_p(A_i))=(p,(\lambda^1,\dots,\lambda^r))$. $\phi$ is not canonical since it depends on the basis $A_i$ of ${\cal G}$. 

\

Let $\omega$ be a connection on $\eta$, i.e. $\omega\in\Gamma(T^*P\otimes {\cal G})$ with $\omega:P\to T^*P\otimes{\cal G}$, $p\mapsto\omega(p)=(p,\omega_p)$, $\omega_p:T_pP\to{\cal G}$, $v_p\mapsto\omega_p(v_p)=\varphi^{-1}_p(ver(v_p))$. Since $ker(\omega_p)=H^s_p\equiv H_p$: the horizontal vector space at $p$, then $\omega_p$ is $|H_p|=\infty\to 1$. However, $\omega_p|_{V_p}:V_p\to{\cal G}$, $\omega_p|_{V_p}(v_p)=\varphi^{-1}_p(v_p)$ is a vector space isomorphism i.e. $$\omega_p|_{V_p}=\varphi^{-1}_p.$$ In other words, if $\omega$ is a connection on $\eta$, then at each $p\in P$, $\omega$ gives an inverse of $\varphi_p$. Therefore, for the isomorphism between $V_p$ and $V_{p^\prime}$, one has $$V_p\buildrel{{\omega_{p^\prime}\vert_{V_{p^\prime}}}^{-1}\circ\omega_p}\vert_{V_{p}}\over\longrightarrow V_{p^\prime}.$$ In particular, we are interested in the case $P={\cal F}_{M^n}$, the frame bundle of a differentiable manifold $M^n$, where $(P,B,\pi,G,\psi,{\cal U})=(FM^n,M^n,\pi_F,GL_n(\R),\psi,{\cal U})$; $p=(x,r_x)\in FM^n$, and $r_x=(v_{1x},\dots,v_{nx})$. Its vertical bundle is isomorphic to the product bundle $FM^n\times\R^{n^2}$: $$\matrix{\R^{n^2} & & \R^{n^2} \cr \vert & & \vert \cr V_{{\cal F}_{M^n}} & \buildrel{\phi_c}\over\longrightarrow & FM^n\times\R^{n^2} \cr \pi_V\downarrow & & \downarrow\pi_1\cr FM^n & \buildrel{Id}\over\longrightarrow & FM^n \cr}$$ where $\phi_c$ is the canonical isomorphism determined by the canonical basis of $gl_n(\R)=\R(n)$ given by the $n^2$ matrices $(A_{ij})_{kl}=\delta_{ik}\delta_{jl}$. The similar result holds for the reductions of $GL_n(\R)$ to $O(n)$, $SO^0(n-1,1)$, etc. mentioned in section 24. 

\

In particular, for the case $n=4$ and $G=SO^0(3,1)$, with $dim_\R (SO^0(3,1))=dim_\R (so(3,1))=dim_\R (o(3,1))=6$, case relevant in GR, $FM^4$ is the bundle of Lorentz frames ${\cal F}^L_{M^4}$, and we have the vector bundle isomorphism $$\matrix{\R^{6} & & \R^{6} \cr \vert & & \vert \cr V_{{\cal F}^L_{M^4}} & \buildrel{\phi^L_c}\over\longrightarrow & F_LM^4\times \R^{6} \cr \pi_V\vert & & \vert\pi_1 \cr F_LM^4 & \buildrel{Id}\over\longrightarrow & F_LM^4 \cr}$$ with $dim_\R(F_LM^4)=4+6=10$ and $dim_\R(V_{{\cal F}^L_{M^4}})=16$. In this case, the canonical basis of $so(3,1)$ (or of $o(3,1)$) is the set of matrices $$\{\pmatrix{0 & 0 & 0 & 0 \cr 0 & 0 & 0 & 0 \cr 0 & 0 & 0 & -1 \cr 0 & 0 & 1 & 0}, \pmatrix{0 & 0 & 0 & 0 \cr 0 & 0 & 0 & 1 \cr 0 & 0 & 0 & 0 \cr 0 & -1 & 0 & 0 \cr}, \pmatrix{0 & 0 & 0 & 0 \cr 0 & 0 & -1 & 0 \cr 0 & 1 & 0 & 0 \cr 0 & 0 & 0 & 0 \cr},\pmatrix{0 & 1 & 0 & 0 \cr 1 & 0 & 0 & 0 \cr 0 & 0 & 0 & 0 \cr 0 & 0 & 0 & 0 \cr}, \pmatrix{0 & 0 & 1 & 0 \cr 0 & 0 & 0 & 0 \cr 1 & 0 & 0 & 0 \cr 0 & 0 & 0 & 0 \cr},$$ $\pmatrix{0 & 0 & 0 & 1 \cr 0 & 0 & 0 & 0 \cr 0 & 0 & 0 & 0 \cr 1 & 0 & 0 & 0 \cr}\}$, respectively $\{l_{23}\equiv a_1,l_{31}\equiv a_2, l_{12}\equiv a_3,l_{01}\equiv b_1,l_{02}\equiv b_2,l_{03}\equiv b_3\}$, where the first three matrices generate rotations around the axis $x$, $y$, and $z$, and the second three matrices generate boosts along the same axis, respectively. The derivation of the canonical basis is as follows: one starts from the definition of the Lorentz transformations $\Lambda$: $\eta_L:=\Lambda^T\eta_L\Lambda$, with $\eta_L\equiv\eta=(\eta_{00},\eta_{11},\eta_{22},\eta_{33})=(+1,-1,-1,-1)$ (or $(-1,+1,+1,+1)$) and $\eta_{ab}=0$ if $a\neq b$; if $\Lambda(\lambda)$ is a smooth path through the identity $\Lambda(0)=I$, the corresponding tangent vector at $I$, $\dot{\Lambda}(0)=L$, obeys the equation $L^T\eta=-\eta L$. The generators $a_i$ and $b_i$ obey $[a_i,a_j]=\epsilon_{ijk}a_k$, $[b_i,b_j]=-\epsilon_{ijk}a_k$, $[a_i,b_j]=\epsilon_{ijk}b_k$. If $l=\sum_{i=1}^3(\beta_ib_i+\alpha_ia_i)$ and $l^\prime=\sum_{i=1}^3(\beta_i^\prime b_i+\alpha_i^\prime a_i)$ are in $o(3,1)$ ($\alpha's,\beta's\in\R$), then is Lie bracket $[l,l^\prime]=l^{\prime\prime}=\sum_{i=1}^3(\beta^{\prime\prime}_ib_i+\alpha^{\prime\prime}_ia_i)\in o(3,1)$ with $\beta^{\prime\prime}_1=\beta_2\alpha^\prime_3+\alpha_2\beta^\prime_3-\beta_3\alpha^\prime_2-\alpha_3\beta^\prime_2$, $\beta^{\prime\prime}_2=\beta_3\alpha^\prime_1+\alpha_3\beta^\prime_1-\beta_1\alpha^\prime_3-\alpha_1\beta^\prime_3$, $\beta^{\prime\prime}_3=\beta_1\alpha^\prime_2+\alpha_1\beta^\prime_2-\beta_2\alpha^\prime_1-\alpha_2\beta^\prime_1$, $\alpha^{\prime\prime}_1=-\beta_2\beta^\prime_3+\alpha_2\alpha^\prime_3+\beta_3\beta^\prime_2-\alpha_3\alpha^\prime_2$, $\alpha^{\prime\prime}_2=-\beta_3\beta^\prime_1+\alpha_3\alpha^\prime_1+\beta_1\beta^\prime_3-\alpha_1\alpha^\prime_3$, $\alpha^{\prime\prime}_3=-\beta_1\beta^\prime_2+\alpha_1\alpha^\prime_2+\beta_2\beta^\prime_1-\alpha_2\alpha^\prime_1$. 

\

{\bf 26. Soldering form on ${\cal F}_{M^n}$}

\

Given the differentiable manifold $M^n$, the {\it soldering} or {\it canonical form} $\theta$ on ${\cal F}_{M^n}$ is the $\R^n$-valued differential 1-form on $FM^n$ i.e. $\theta\in\Gamma(T^*FM^n\otimes\R^n)$ defined as follows:

\

$\theta:FM^n\to T^*FM^n\otimes\R^n$, $(x,r_x)\mapsto\theta((x,r_x))=((x,r_x),\theta_{(x,r_x)})$,

\

$\theta_{(x,r_x)}: T_{(x,r_x)}FM^n\to\R^n$, $v_{(x,r_x)}\mapsto \theta_{(x,r_x)}(v_{(x,r_x)})=\tilde{r}_x^{-1}\circ\pi_{F_{*(x,r_x)}}(v_{(x,r_x)})$ 

\

i.e. $$\theta_{(x,r_x)}=\tilde{r}_x^{-1}\circ d\pi_F|_{(x,r_x)},$$ where $\pi_F$ is the projection in the bundle ${\cal F}_{M^n}$ (section 24) and $\tilde{r}_x$ is the vector space isomorphism 

\

$\tilde{r}_x:\R^n\to T_xM$, $(\lambda^1,\dots,\lambda^n)\mapsto\tilde{r}_x(\lambda^1,\dots,\lambda^n)=\sum_{i=1}^n\lambda^iv_{ix}$

\

with inverse

\

$\tilde{r}_x^{-1}(\sum_{i=1}^n\lambda^iv_{ix})=(\lambda^1,\dots,\lambda^n)$. We have the commutative diagram $$\matrix{\R^n & \buildrel{Id}\over\longrightarrow & \R^n \cr \theta_{(x,r_x)}\uparrow & & \uparrow\tilde{r}_x^{-1} \cr T_{(x,r_x)}FM^n & \buildrel{d\pi_F|_{(x,r_x)}}\over\longrightarrow & T_xM^n \cr}$$ Notice that $dim_\R (T_{(x,r_x)}FM^n)=n+n^2$. Since $d\pi_F|_{(x,r_x)}$ is onto, $\theta_{(x,r_x)}$ is a vector space epimorphism, with $ker(\theta_{(x,r_x)})=V_{(x,r_x)}$, the vertical space of the bundle $FM^n$ at $(x,r_x)$, with $dim_\R ker(\theta_{(x,r_x)})=n^2$. The existence of $\theta$ is independent of any connection. Also, it is clearly a {\it global} section of the bundle $T^*FM^n\otimes\R^n$. 

\

$(\vec{e}_\mu)_j=\delta_{\mu j}$, $\mu,j=1,\dots,n$ is the canonical basis of $\R^n$, then $$\theta_{(x,r_x)}=\sum_{\mu=1}^n\theta^\mu_{(x,r_x)}\otimes\vec{e}_\mu$$ where $\theta^\mu_{(x,r_x)}\in T^*_{(x,r_x)}FM^n$. 

\

In local coordinates on ${\cal F}_{U_\alpha}$, $$\theta^\mu=\sum_{\nu=1}^n(X^{-1})^\mu_\nu dx^\nu$$ where $(X^{-1})^\mu_\nu(x,r_x)=(X^\mu_\nu(x,r_x))^{-1}$. [In fact, if $v_{(x,r_x)}\in T_{(x,r_x)}FU_\alpha$, then $v_{(x,r_x)}=\sum_{\mu=1}^n\lambda^\mu{{\partial}\over{\partial x^\mu}}|_{(x,r_x)}+\sum_{\mu,\nu=1}^n\lambda^\mu_\nu{{\partial}\over{\partial X^\mu_\nu}}|_{(x,r_x)}$ with $d\pi_F|_{(x,r_x)}(v_{(x,r_x)})=\sum_{\mu=1}^n\lambda^\mu{{\partial}\over{\partial x^\mu}}|_x\in T_xU_\alpha$; then $\theta_{(x,r_x)}(v_{(x,r_x)})$

$=\tilde{r}^{-1}_x\circ d\pi_F|_{(x,r_x)}(v_{(x,r_x)})=\tilde{r}^{-1}_x(\sum_{\mu=1}^n\lambda^\mu{{\partial}\over{\partial x^\mu}}|_x)=\sum_{\mu=1}^n\lambda^\mu\tilde{r}^{-1}_x({{\partial}\over{\partial x^\mu}}|_x)=\sum_{\mu,\nu=1}^n\lambda^\mu(X^\nu_\mu(x,r_x))^{-1}\vec{e}_\nu$; 

\

on the other hand, $\theta_{(x,r_x)}(v_{(x,r_x)})=(\sum_{\mu=1}^n\theta^\mu_{(x,r_x)}\otimes\vec{e}_\mu )(\sum_{\nu=1}^n\lambda^\mu{{\partial}\over{\partial x^\nu}}|_{(x,r_x)}+\sum_{\nu,\rho=1}^n\lambda^\nu_\rho{{\partial}\over{\partial X^\nu_\rho}}|_{(x,r_x)})$

\

$=(\sum_{\mu,\alpha=1}^n (X^{-1})^\mu_\alpha (x,r_x)dx^\alpha|_{(x,r_x)}\otimes \vec{e}_\mu )(\sum_\nu^n\lambda^\nu{{\partial}\over{\partial x^\nu}}|_{(x,r_x)}+\sum_{\nu,\rho=1}^n\lambda^\nu_\rho{{\partial}\over{\partial X^\nu_\rho}}|_{(x,r_x)})$

\

$=\sum_{\mu,\nu,\alpha=1}^n(X^\mu_\alpha (x,r_x))^{-1}\delta^\alpha_\nu\lambda^\nu\vec{e}_\mu=\sum_{\mu,\nu=1}^n(X^\mu_\nu (x,r_x))^{-1}\lambda^\nu\vec{e}_\mu$.]

\

Thus, a local section on ${\cal F}_{M^n}$, $s_\alpha:U_\alpha\to FU_\alpha$, $x\to s_\alpha(x)=(x,r_x)$, gives rise to a set of $n$ local differential 1-forms $\theta^\mu_\alpha\equiv\theta^\mu$ on $FU_\alpha$.

\

If $\omega$ is a connection on ${\cal F}_{M^n}$, and $H_{(x,r_x)}$ is the horizontal space at $(x,r_x)$, then $$\theta_{(x,r_x)}|_{H_{(x,r_x)}}=\tilde{r}^{-1}_x\circ d\pi_F|_{(x,r_x)}|_{H(x,r_x)}$$ is a vector space isomorphism, since $d\pi_F|_{{(x,r_x)}}$ is a canonical isomorphism between $H_{(x,r_x)}$ and $T_xM^n$: $$\matrix{\R^n & \buildrel{Id}\over\longrightarrow & \R^n \cr \theta_{(x,r_x)}|_{H_{(x,r_x)}}\uparrow & & \uparrow\tilde{r}_x^{-1} \cr H_{(x,r_x)} & \buildrel{d\pi_F|_{(x,r_x)}}\over\longrightarrow & T_xM^n \cr}$$ We emphasize that $\theta_{(x,r_x)}|_{H_{(x,r_x)}}$ depends on both the frame at $x$ ($r_x$) and the connection $\omega$. 

\

Any connection $\omega$ on the frame bundle ${\cal F}_{M^n}$, together with the canonical soldering form $\theta$, trivializes the tangent bundle of ${\cal F}_{M^n}$. This fact is known as {\it absolute parallelism}. The canonical bundle isomorphism (only depending on $\omega$) is given through the following diagram: $$\matrix{\R^{n+n^2} & & \R^{n+n^2}\cr \vert & & \vert \cr (TFM^n)^{2(n+n^2)} & \buildrel{\phi_c}\over\longrightarrow & (FM^n)^{n+n^2}\times\R^{n+n^2}\cr \pi_{F_*}\downarrow & & \downarrow \pi_1 \cr (FM^n)^{n+n^2} & \buildrel{Id}\over \longrightarrow & (FM^n)^{n+n^2} \cr}$$ with $$\phi_c(((x,r_x),v_{(x,r_x)}))=((x,r_x),(\theta_{(x,r_x)}\vert_{H_{(x,r_x)}}\times\omega_{(x,r_x)}\vert_{V_{(x,r_x)}})(hor(v_{(x,r_x)}),ver(v_{(x,r_x)}))),$$ where $v_{(x,r_x)}\in T_{(x,r_x)}FM^n$ and $gl_n(\R)=\R(n)\cong\R^{n^2}$. 

\

Absolute parallelism in the bundle of Lorentz frames ${\cal F}^L_{M^n}$ is given by the diagram $$\matrix{\R^{{{n(n+1)}\over{2}}} & & \R^{{{n(n+1)}\over{2}}}\cr \vert & & \vert \cr (TF^LM^n)^{n(n+1)} & \buildrel{\phi^L_c}\over\longrightarrow & (F^LM^n)^{{{n(n+1)}\over{2}}}\times\R^{{{n(n+1)}\over{2}}}\cr \pi_{L_*}\downarrow & & \downarrow \pi_1 \cr (F^LM^n)^{{{n(n+1)}\over{2}}} & \buildrel{Id}\over \longrightarrow & (F^LM^n)^{{{n(n+1)}\over{2}}} \cr}$$ with $$\phi^L_c(((x,e_x),v_{(x,e_x)}))=((x,e_x),(\theta_{(x,e_x)}\vert_{H_{(x,e_x)}}\times\omega^L_{(x,e_x)}\vert_{V_{(x,e_x)}})(hor(v_{(x,e_x)}),ver(v_{(x,e_x)}))),$$ $e_x=(e_{1x},\dots,e_{nx})$, and $H_{(x,e_x)}=ker(\omega^L_{(x,e_x)})$. 

\

In particular, for the $n=4$ case: $$\matrix{\R^{10} & & \R^{10}\cr \vert & & \vert \cr (TF^LM^4)^{20} & \buildrel{\phi^L_c}\over\longrightarrow & (F^LM^4)^{10}\times\R^{10}\cr \pi_{L_*}\downarrow & & \downarrow \pi_1 \cr (F^LM^4)^{10} & \buildrel{Id}\over \longrightarrow & (F^LM^4)^{10} \cr}.$$

\

{\bf 27. Linear connection in a manifold $M^n$ on ${\cal F}_{M^n}$}

\

A $gl_n(\R)$-valued linear connection $\nabla$ on $M^n$ is locally given by $$\omega_U=\Gamma^\mu_{\nu\rho}dx^\nu\otimes{E^\rho}_\mu$$ where $({E^\rho}_\mu)_{\alpha\beta}={\delta^\rho}_\alpha\delta_{\mu\beta}$ with $\rho,\mu,\alpha,\beta=1,\dots,n$ is the canonical basis of $\R(n)=gl_n(\R)=Lie(GL_n(\R))$, and $\Gamma^\mu_{\nu\rho}$ are the Christoffel symbols (section 4). 

\

On $FU$, the connection $\omega_{FU}$ such that $\omega_U=\sigma^*(\omega_{FU})$ with $\sigma:U\to FU$ the local section given by $x^\mu\mapsto\sigma(x^\mu)=(x^\mu,\delta^\nu_\rho))$ is given by $$\omega_{FU}=(X^{-1})^\mu_\sigma(dX^\sigma_\rho +\Gamma^\sigma_{\nu\lambda}X^\lambda_\rho dx^\nu)\otimes{E^\rho}_\mu.$$ (Kobayashi and Nomizu, 1963; pp 140-143)) Clearly, $\omega_U\in \Gamma(T^*U\otimes gl_n(\R))$ and $\omega_{FU}\in\Gamma(T^*FU\otimes gl_n(\R))$. 

\

Real-valued connection 1-forms ${{\omega_U}^\mu}_\rho$ and ${{\omega_{FU}}^\mu}_\rho$ are defined by $$\omega_U={{\omega_U}^\mu}_\rho\otimes{E^\rho}_\mu \ and \ \omega_{FU}={{\omega_{FU}}^\mu}_\rho\otimes{E^\rho}_\mu.$$  The horizontal lift of a local vector field ${{\partial}\over{\partial x^\mu}}$ by the connection $\omega$ in ${\cal F}_{M^n}$ is then given by $$({{\partial}\over{\partial x^\mu}})^\uparrow={{\partial}\over{\partial x^\mu}}-\Gamma^\rho_{\mu\nu}X^\nu_\sigma{{\partial}\over{\partial X^\rho_\sigma}}.$$ In fact, ${{\omega_{FU}}^\alpha}_\beta (({{\partial}\over{\partial x^\mu}})^\uparrow)=(X^{-1})^\alpha_\lambda (dX^\lambda_\beta +\Gamma^\lambda_{\gamma\xi}X^\xi_\beta dx^\gamma)({{\partial}\over{\partial x^\mu}}-\Gamma^\rho_{\mu\nu}X^\nu_\sigma{{\partial}\over{\partial X^\rho_\sigma}})$

\

$=(X^{-1})^\alpha_\lambda(-\Gamma^\rho_{\mu\nu}X^\nu_\sigma dX^\lambda_\beta ({{\partial}\over{\partial X^\rho_\sigma}})+\Gamma^\lambda_{\gamma\xi}X^\xi_\beta dx^\gamma ({{\partial}\over{\partial x^\mu}}))$

\

$=(X^{-1})^\alpha_\lambda (-\Gamma^\rho_{\mu\nu}X^\nu_\sigma\delta^\lambda_\rho\delta^\sigma_\beta +\Gamma^\lambda_{\gamma\xi}X^\xi_\beta\delta^\gamma_\mu)=(X^{-1})^\alpha_\lambda(-\Gamma^\lambda_{\mu\nu}X^\nu_\beta+\Gamma^\lambda_{\mu\nu}X^\nu_\beta)=0.$

\

{\bf 28. Tetrads and spin connection}

\

1. At each chart $U\subset M^n$ we can take as a basis of $\Gamma(TU)$ the local vector fields (Vielbeine) $$e_a={e_a}^\mu\partial_\mu, \ a=1,\dots,n$$ with $r_x=(e_{1x},\dots,e_{nx})\in (FU)_x$. Since the $n\times n$ matrices $({e_a}^\mu(x))\in GL_n(\R)$, there exist the inverse vector fields $e_a^{-1}\equiv e^a=e_\mu ^a dx^\mu$: 1-forms with $e_\mu ^a=(e_a^\mu)^{-1}\in GL_n(\R)$ and 

$${e_a}^\mu {e_\mu}^b=\delta^b_a \ and \ {e_a}^\mu {e_\nu}^a=\delta^\mu_\nu.$$ Then ${e_\nu}^ae_a={e_\nu}^a{e_a}^\mu \partial_\mu=\delta^\mu_\nu\partial_\mu$ i.e. $$\partial_\nu={e_\nu}^ae_a.$$ In general, the ${e_a}'s$ are called {\it non-coordinate basis} and the ${e^a}'s$ {\it anholonomic coordinates}. For $n=4$, the Vierbeine $e_a$ are called {\it tetrads}.

\

2. While $[\partial_\mu,\partial_\nu]=0$, the Vielbeine have non-vanishing Lie brackets. In fact, applying the commutator $[e_a,e_b]$ to a function $f\in C^\infty(U,\R)$, one easily obtains $$[e_a,e_b]=\lambda^c_{ab}e_c$$ with $\lambda^c_{ab}={e_\mu}^c({e_a}^\nu(\partial_\nu{e_b}^\mu)-{e_b}^\nu(\partial_\nu{e_a}^\mu))=-\lambda^c_{ba}$. 

\

3. For a local vector field, $V=V^\mu\partial_\mu=V^\mu{e_\mu}^ae_a=V^ae_a$, with $$V^a={e_\mu}^aV^\mu, \ V^\mu={e_a}^\mu V^a.$$

\

4. At each $x\in M^n$, the fibre of the {\it co-frame bundle} of $M^n$, $${\cal F}^*_{M^n}:GL_n(\R)\to (FM^n)^*\buildrel{\pi_*}\over\longrightarrow M^n,$$ is the set $$(FM^n)^*_x=\{ordered \ basis \ of \ T^*_xM\}=\{(f^1_x,\dots,f^n_x), \ f^a_x={f_\nu}^a(x)dx^\nu|_x\}.$$ Again, $({f_\nu}^a(x))\in GL_n(\R)$ and, locally, $({{f_\nu}^a})^{-1}={f_a}^\nu$ with $${f_a}^\nu{f_\nu}^b=\delta_a^b \ and \ {f_a}^\nu{f_\mu}^a=\delta^\nu_\mu.$$ Also, $$f^a={f_\nu}^adx^\nu \ and \ dx^\nu={f_a}^\nu f^a.$$ From the duality relation $dx^\mu(\partial_\nu)=\delta^\mu_\nu$ we obtain $\delta^\mu_\nu={f_a}^\mu f^a({e_\nu}^be_b)={f_a}^\mu{e_\nu}^bf^a(e_b)$; imposing the duality relation between $f's$ and $e's$, $$f^a(e_b)=\delta^a_b$$ one obtains $${f_\mu}^d={e_\mu}^d \ and \ {f_a}^\nu={e_a}^\nu.$$ Then, $$f^a={e_\nu}^adx^\nu \ and \ dx^\nu={e_a}^\nu f^a.$$ (Another usual notation for $f^a$ is $\theta^a$.)

\

5. Given an $(r,s)$-tensor in $M^n$, $$T=T^{\mu_1\dots\mu_r}_{\nu_1\dots\nu_s}\partial_{\mu_1}\otimes\dots\otimes\partial_{\mu_r}\otimes dx^{\nu_1}\otimes\dots\otimes dx^{\nu_s},$$ we obtain $$T=T^{\mu_1\dots\mu_r}_{\nu_1\dots\nu_s}{e_{\mu_1}}^{a_1}e_{a_1}\otimes\dots\otimes{e_{\mu_r}}^{a_r}e_{a_r}\otimes{f_{b_1}}^{\nu_1}f^{b_1}\otimes\dots\otimes{f_{b_s}}^{\nu_s}f^{b_s}=T^{a_1\dots a_r}_{b_1\dots b_s}e_{a_1}\otimes\dots\otimes e_{a_r}\otimes f^{b_1}\otimes\dots\otimes f^{b_s}$$ with $$T^{a_1\dots a_r}_{b_1\dots b_s}={e_{\mu_1}}^{a_1}\dots{e_{\mu_r}}^{a_r}{f_{b_1}}^{\nu_1}\dots{f_{b_s}}^{a_s}T^{\mu_1\dots\mu_r}_{\nu_1\dots\nu_s}.$$ 

\

For example, $$T={T^\mu}_\nu\partial_\mu\otimes dx^\nu={T^\mu}_\nu{e_\mu}^ae_a\otimes dx^\nu={T^a}_\nu e_a\otimes dx^\nu={T^\mu}_\nu\partial_\mu\otimes{f_b}^\nu f^b={T^\mu}_b\partial_\mu\otimes f^b={T^\mu}_\nu{e_\mu}^ae_a\otimes{f_b}^\nu f^b$$ $={T^a}_be_a\otimes f^b$.

\

6. Let $g=g_{\mu\nu}dx^\mu\otimes dx^\nu$ be a metric in $M^n$. $g$ has a signature, given by the diagonal metric $\eta_{ab}$, equal to $\delta_{ab}$ in the euclidean case ($\eta=\eta_E$), or with -1's and +1's in the general riemannian case; the lorentzian case, relevant for GR, has $\eta=\eta_L$ (section {\bf 25.}). $g_{\mu\nu}$ being a symmetric matrix, at any point $x\in M^n$ it can be diagonalized to $\eta_{ab}$. The metric and its signature distinguish the subset of Vielbeine which obey the following {\it orthonormality} condition: $$g(e_a,e_b)=\eta_{ab}.$$ In detail, $$g_{\mu\nu}dx^\mu\otimes dx^\nu({e_a}^\rho\partial_\rho,{e_b}^\sigma\partial_\sigma)=g_{\mu\nu}{e_a}^\rho{e_b}^\sigma dx^\mu(\partial_\rho)dx^\nu(\partial_\sigma)=g_{\mu\nu}{e_a}^\rho{e_b}^\sigma\delta^\mu_\rho\delta^\nu_\sigma=g_{\mu\nu}{e_a}^\mu{e_b}^\nu$$ i.e. $$g_{\mu\nu}{e_a}^\mu{e_b}^\nu=\eta_{ab} \eqno{(a)}.$$ This relation can be easily inverted, namely, $g_{\mu\nu}{e_a}^\mu{e_b}^\nu{e_\rho}^a{e_\sigma}^b=g_{\mu\nu}\delta^\mu_\rho\delta^\nu_\sigma=g_{\rho\sigma}$ i.e. $$\eta_{ab}{e_\mu}^a{e_\nu}^b=g_{\mu\nu} \ \ or \ \ g=\eta_{ab}e^a\otimes e^b. \eqno{(b)}$$ The unique solution of $g_{\mu\nu}(\bar{x})=\eta_{ab}{e_\mu}^a (\bar{x}){e_\nu}^b (\bar{x})=\eta_{\mu\nu}$ is ${e_\mu}^a (\bar{x})=\delta_\mu ^a$. Such a frame is called {\it inertial} at $x$. 

\

Notice that formula $(b)$ is fundamental: since it holds in all charts in $M^n$, we have, up to topological obstructions, ``trivialized'' the metric everywhere i.e. globally, at the expense of the $x$-dependence of the coframes $e^a$.

\

It is usual the rough statement that the duals of the Vielbeine are the ``square roots" of the metric. In particular, for the lorentzian $n=4$ case, $det(g_{\mu\nu})=-(det({e_\nu}^a))^2$. Also, equation $(a)$ allows to interprete the $n\times n$ matrices ${e_c}^\rho$ as the matrices which {\it diagonalize} the metric $g_{\mu\nu}$ to the Lorentz metric $\eta_{ab}$. $(b)$ says that the ${e^a}'s$ are more fundamental than the metric.

\

7. Equation $(b)$ in the last subsection appears naturally when describing spinor fields in curved space-times. If $\gamma_\mu(x)$ are the Dirac matrices in $M^n$, then $$\{\gamma_\mu(x),\gamma_\nu(x)\}=2g_{\mu\nu}(x)I.$$ The solution $$\gamma_\mu(x)={E_\mu}^a(x)\gamma_a \eqno{(*)}$$ with $\gamma_a$ the ``flat" Dirac matrices obeying $\{\gamma_a,\gamma_b\}=2\eta_{ab}I$, leads to $$\eta_{ab}{E_\mu}^a(x){E_\nu}^b(x)=g_{\mu\nu}$$ which says that the ${{E_\mu}^a} \ 's$ are the duals of the Vielbeine ${e_a}^\mu$. It can be proved that the solution $(*)$ is unique (O'Raifeartaigh, 1997).

\

8. It is clear that through $(b)$ in subsection 6., the $n^2$ quantities involved by a Vielbein determine uniquely a metric $g_{\mu\nu}$; however, the set of ${{n(n+1)}\over{2}}$ components of a metric determines a Vielbein only up to an equivalence relation i.e. it determines a class of Vielbeine, whose elements are related by a group $G$ of ${{n(n-1)}\over{2}}$ elements. 

\

Let ${e_a}^\mu={h_a}^c{e^\prime_c}^\mu$; then 

\

$\eta_{ab}=g_{\mu\nu}{h_a}^c{e^\prime_c}^\mu {h_b}^d{e^\prime_d}^\nu=g_{\mu\nu}{e^\prime_c}^\mu{e^\prime_d}^\nu {h_a}^c {h_b}^d=\eta_{cd}{h_a}^c {h_b}^d={h_a}^c\eta_{cd}{h_b}^d$ i.e. $$\eta=h\eta h^T.$$ So, if $\eta=\eta_L$ then $h\in{\cal L}_n=O(1,n-1)$; if $\eta=\eta_E$ then $h\in O_n=O(n)$; etc.

\

In the following we shall restrict to the case of orthonormal frames in the sense defined in 6., so one has a principal $G$-bundle over $M^n$: $$G^{{{n(n-1)}\over{2}}}\to F_G\to M^n,$$ with $G={\cal L}_n$ or $O_n$, and $F_G \subset FM^n$. The interest in GR is for $G={\cal L}_4$ and one has the lorentzian bundle $${\cal L}_4\to F_{{\cal L}_4}\to M^4$$ and the bundle reduction $$\matrix{{\cal L}_4 & \buildrel {\iota}\over \longrightarrow & GL_4(\R) \cr \downarrow & & \downarrow \cr F_{{\cal L}_4} & \buildrel {\iota} \over \longrightarrow & FM^4 \cr \pi_{{\cal L}_4}\downarrow & & \downarrow \pi_F \cr M^4 & \buildrel {Id} \over \longrightarrow & M^4 \cr}$$ where $\iota$ is the inclusion.

\

We want to emphasize that the bundle is trivial, that is, $F_{{\cal L}_4}\cong M^4\times{\cal L}_4$, if $M^4$ is contractible i.e. if it is of the same homotopy type as a point.

\

At each $U\subset M$ one has a {\it local Lorentz group} ${\cal L}_U$ which, at $x\in U$, ``takes the value" ${\cal L}(x)$. Since the Vielbeine are natural basis of $\Gamma(TU)$, and the metric necessarily has a signature $\eta_{ab}$, the principal ${\cal L}_n$-bundles over $M^n$ are also natural. (We emphasize that here, we do not use the word `` natural" in its technical sense, but in a coloquial sense.) We also notice the natural appearance of a new group at each $x\in M^n$, besides the group of general coordinate transformations in the intersections of open sets: the Lorentz group.

\

9. In section {\bf 4.}, the covariant derivative of a local section $\sigma_i$ of an arbitrary vector bundle $E$ was defined by $\nabla_\mu\sigma_i=\Gamma^j_{\mu i}\sigma_j$. Let $\sigma_i=e_a$; then $$\nabla_\mu e_a=\nabla_\mu({e_a}^\nu\partial_\nu)=(\partial_\mu{e_a}^\nu)\partial_\nu+{e_a}^\nu\nabla_\mu\partial_\nu=(\partial_\mu{e_a}^\rho+{e_a}^\nu\Gamma^\rho_{\mu\nu})\partial_\rho.$$ We now define the {\it spin connection} coefficients $\omega_{\mu a}^b$ through $$\nabla_\mu e_a:=\omega^b_{\mu a}e_b  \ \ and  \ \ \nabla_\mu e^a:=\omega_{\mu b}^a e^b.$$ (The $\omega^b_{\mu a}$'s are, in the present case, nothing but the $\Gamma^j_{\mu i}$'s.) Then, $\omega_{\mu a}^b{e_b}^\rho\partial_\rho=(\partial_\mu {e_a}^\rho+{e_a}^\nu\Gamma_{\mu\nu}^\rho)\partial_\rho$; from the linear independence of the coordinate basis, and multiplying by ${e_\rho}^c$ we obtain $$\omega^c_{\mu a}={e_\rho}^c\partial_\mu{e_a}^\rho +{e_\rho}^c{e_a}^\nu\Gamma^\rho_{\mu\nu} \eqno{(c)}$$ or $$\omega^c _{ab}=-{e_b}^\rho {e_{\rho ,a}}^c+{e_ a}^\mu {e_ b}^\nu {e_\rho}^c \Gamma^\rho_{\mu\nu} \eqno{(c^\prime)}$$ with $$\omega^c_{ab}={e_\mu}^c \omega^\mu_{ab}, \ \ \omega^\mu_{ab}=g^{\mu\nu}\omega_{\nu ab}, \ \ {e_{\rho ,a}}^c=\partial_a {e_\rho}^c, \ \ \partial_a={e_a}^\nu\partial_\nu=e_a.$$ Multiplying by ${e_c}^\sigma{e_\lambda}^a$ we obtain the inverse relation: $$\Gamma_{\mu\lambda}^\sigma={e_a}^\sigma\partial_\mu{e_\lambda}^a+{e_c}^\sigma{e_\lambda}^a\omega_{\mu a}^c \eqno{(d)}$$ or $$\Gamma^\sigma _{\mu\lambda}=\omega^c_{da} {e_c}^\sigma {e_\lambda}^a {e_\mu}^d -{e_{a,d}}^\sigma {e_a}^\lambda {e_\mu}^d \eqno{(d^\prime)}$$ with ${e_{a,d}}^\sigma=\partial_d {e_a}^\sigma$.

\

Multiplying $(c)$ by ${e_c}^\sigma$ and $(d)$ by ${e_\sigma}^b$ we obtain, respectively, $$\partial_\mu{e_a}^\sigma +\Gamma^\sigma_{\mu\nu}{e_a}^\nu-\omega^c_{\mu a}{e_c}^\sigma=0 \eqno{(e)}$$ and $$\partial_\mu{e_\lambda}^b-\Gamma^\sigma_{\mu\lambda}{e_\sigma}^b+\omega^b_{\mu a}{e_\lambda}^a=0.\eqno{(f)}$$ The covariant derivative of tensors with upper and lower ``internal" (Lorentz) and ``external" (space-time) indices is {\it defined} by: 
$${\cal D}_\mu T^{\mu_1\dots\mu_r a_1\dots a_t}_{\nu_1\dots\nu_s b_1\dots b_u}=\partial_\mu T^{\mu_1\dots a_t}_{\nu_1\dots b_u}+\Gamma^{\mu_1}_{\mu\lambda_1}T^{\lambda_1\mu_2\dots a_t}_{\nu_1\dots b_u}+\dots +\omega^{a_t}_{\mu c_t}T^{\mu_1\dots a_{t-1}c_t}_{\nu_1\dots b_u}-\Gamma^{\lambda_1}_{\mu\nu_1}T^{\mu_1\dots a_t}_{\lambda_1\nu_2\dots b_u}-\dots -\omega^{c_u}_{\mu b_u}T^{\mu_1\dots a_t}_{\nu_1\dots b_{u-1}c_u}.$$ With this definition, the equations (e) and (f) read
$${\cal D}_\mu{e_a}^\sigma=0 \ \ and \ \ {\cal D}_\mu{e_\lambda}^b=0$$ respectively. (We also denote ${\cal D}_\mu T^{\mu_1\dots\mu_r a_1\dots a_t}_{\nu_1\dots\nu_s b_1\dots b_u}=T^{\mu_1\dots\mu_r a_1\dots a_t}_{\nu_1\dots\nu_s b_1\dots b_u \ ;\mu}$.)

\

From the Lorentz transformations of the tetrads ${e_a}^\mu$ and their inverses, ${e_\mu}^a={e_\mu^\prime}^c {h^{-1}_c}^a$ we obtain: $${e_\sigma}^c\partial_\mu{e_a}^\sigma={e^\prime_\sigma}^r {h^{-1}_r}^c\partial_\mu ({h_a}^d {e^\prime_d}^\sigma)=
{e^\prime_\sigma}^r {h^{-1}_r}^c \partial_\mu ({h_a}^d){e^\prime_d}^\sigma+{e^\prime_\sigma}^r {h^{-1}_r}^c {h_a}^d \partial_\mu {e^\prime_d}^\sigma$$ $=\partial_\mu ({h_a}^d){h^{-1}_d}^c+{h_a}^d ({e^\prime_\sigma}^r \partial_\mu {e^\prime_d}^\sigma ){h^{-1}_r}^c,$

\

and $${e_\sigma}^c {e_a}^\nu\Gamma^\sigma_{\mu\nu}={e^\prime_\sigma}^r {h^{-1}_r}^c {h_a}^d {e^\prime_d}^\nu\Gamma^\sigma_{\mu\nu}={h_a}^d({e^\prime_\sigma}^r {e^\prime_d}^\nu \Gamma^\sigma_{\mu\nu}){h^{-1}_r}^c.$$                              

Then $$\omega^c_{\mu a}={h_a}^d ({e^\prime_\sigma}^r\partial_\mu{e^\prime_d}^\sigma +{e^\prime_\sigma}^r{e^\prime_d}^\nu\Gamma^\sigma_{\mu\nu}){h^{-1}_r}^c+\partial_\mu({h_a}^d){h^{-1}_d}^c={h_a}^d{\omega^\prime}_{\mu d}^r{h^{-1}_r}^c +\partial_\mu({h_a}^d){h^{-1}_d}^c.$$ I.e. $$\omega=h\omega^\prime h^{-1}+(dh)h^{-1}$$ or, equivalently, $$\omega^\prime=h^{-1}\omega h-h^{-1}dh.$$ So, the 1-form ${\omega^a}_b:=\omega^a_{\mu b}dx^\mu$ is not a 1-1 ${\cal L}$-tensor, since its transformation has an inhomogeneous term. 

\

Notice that from (c), $\omega$ has the structure $$\omega=e^{-1}(\partial+\Gamma)e,$$ and from (d), the structure of $\Gamma$ is $$\Gamma=e(\partial +\omega)e^{-1}.$$

\

It can be easily shown that for a {\it metric connection} $\nabla$, for which $$D_\rho g_{\mu\nu}=0,$$ the spin connection with lower Lorentz indices $$\omega_{\mu bc}=\omega^a_{\mu c}\eta_{ab}$$ is antisymmetric in these indices. In fact, using (e) and (f), $$0=g_{\mu\nu;\rho}=(\eta_{ab}{e_\mu}^a{e_\nu}^b)_{;\rho}=\eta_{ab;\rho}{e_\mu}^a{e_\nu}^b+\eta_{ab}{{e_\mu}^a}_{;\rho}{e_\nu}^b+\eta_{ab}{e_\mu}^a{{e_\nu}^b}_{;\rho}$$ $$=\eta_{ab;\rho}{e_\mu}^a{e_\nu}^b=(-\omega^c_{\rho a}\eta_{cb}-\omega^c_{\rho b}\eta_{ac}){e_\mu}^a{e_\nu}^b=-(\omega_{\rho ba}+\omega_{\rho ab}){e_\mu}^a{e_\nu}^b,$$ then $$0=-(\omega_{\rho ba}+\omega_{\rho ab}){e_\mu}^a{e_\nu}^b{e_c}^\mu{e_d}^\nu =-(\omega_{\rho ba}+\omega_{\rho ab})\delta^a_c \delta^b_d =-(\omega_{\rho dc}+\omega_{\rho cd})$$ i.e. $$\omega_{\rho dc}=-\omega_{\rho cd}.$$ Thus, we see that it is the condition of metric compatibility which {\it reduces} the Lie algebra of the gauge group from $gl_n(\R)$ to $o(1,n-1)$ (or $o(n)$), where the 1-form $\omega_{bc}=\omega_{\mu bc}dx^\mu$ takes values. The reduced gauge group can be $O(1,n-1)$ (or $O(n)$) or, for the case $n=4$, $SL(2,\C)$ if ${h_a}^c \in {\cal L}_+^\uparrow =SO^0(1,3)$ at each $x\in M^4$ (Randono, 2010).

\

Up to here the content of this section does not depend on the symmetry properties of $\Gamma^\mu_{\nu\rho}$ in its lower indices. If in particular the Levi-Civita connection of section {\bf 13.} is inserted in $(b)$, using $(c)$ 
(valid for any connection $\nabla$ with local coefficients $\Gamma^\rho_{\mu\nu}$), we obtain the spin connection coefficients $\omega^c_{\mu a}$ in terms of the Vielbeine, their derivatives, and the Lorentz metric $\eta_{ab}$: $$\omega^c_{\mu a}={e_\rho}^c\partial_\mu{e_a}^\rho+{{1}\over{2}}{e_\rho}^c{e_a}^\nu \eta^{fd}{e_f}^\rho{e_d}^\sigma \eta_{hk}(\partial_\mu({e_\sigma}^h{e_\nu}^k)+\partial_\nu({e_\sigma}^h{e_\mu}^k)-\partial_\sigma({e_\mu}^h{e_\nu}^k)).$$ So, we have the result analogous to the dependence of the Levi-Civita connection on the metric: the dependence of the spin connection on the tetrads.

\

10. Explicitly, on each chart $(U,x^\mu)$ in $M$, the (metric) spin connection with values in $so(3,1)$, is constructed as follows: 

\

$\omega={{1}\over{2}}\omega_{\mu ab}dx^\mu\otimes l_{ab}\in \Gamma(T^*U\otimes so(3,1))$, $x\mapsto\omega(x)=(x,\omega_x)$ with $\omega_x={{1}\over{2}}\omega_{\mu ab}(x)dx^\mu|_x\otimes l_{ab}\in T^*_x U\otimes so(3,1)$ 

\

i.e. $$\omega_x:T_xU\to so(3,1), \ v_x\mapsto\omega_x(v_x)={{1}\over{2}}\omega_{\mu ab}(x)dx^\mu|_x(v_x)l_{ab}={{1}\over{2}}\omega_{\mu ab}(x)v_x^\mu l_{ab}={{1}\over{2}}\omega_{ab}^\prime(x)l_{ab}$$ with $\omega_{ab}^\prime(x)=-\omega_{ba}^\prime(x):=\omega_{\mu ab}(x)v_x^\mu$ and $\omega_x(v_x)=\omega_{01}^\prime l_{01}+\dots \omega_{31}^\prime l_{31}$.

\

For later use, consider the connected component of the {\it Poincar\'e group} ${\cal P}_4$, the semidirect sum of the of the translation group ${\cal T}_4$ and the connected component of the Lorentz group ${\cal L}_4$: $${\cal P}_4={\cal T}_4 \odot SO^0(3,1), \ (a^\prime,\Lambda ^\prime)(a,\Lambda)=(a^\prime+\Lambda ^\prime a,\Lambda ^\prime\Lambda)$$ with Lie algebra $$p_4=\R ^4 \odot so(3,1), \ (\vec{\lambda}^\prime,l^\prime)(\vec{\lambda},l)=(l^\prime\vec{\lambda}-l\vec{\lambda}^\prime,[l^\prime,l])=-(\vec{\lambda},l)(\vec{\lambda}^\prime,l^\prime).$$ (${\cal P}_4$ is a subgroup of the {\it affine group} $A_4$; for arbitrary $n$, $A_n=\R ^n \odot GL_n(\R)$ with Lie algebra $a_n=\R ^n \odot \R (n)$.)

\ 

Also, on each chart $(U,x^n)$ in $M$, the tetrad (1-form) $\tilde{e}^a$ with values in $Lie ({\cal T}_4)=\R ^4$ i.e. $\tilde{e}^a \in \Gamma(T^* U \otimes \R ^4)$ is constructed as follows: $$\tilde{e}^a=e^a _\mu dx^\mu \otimes \vec{\lambda}:U \to T^*U \otimes \R ^4, \ \tilde{e}^a(x)=(x,\tilde{e}^a_x), \ \tilde{e}^a_x=e^a_\mu (x)dx^\mu \vert_x \otimes\vec{\lambda}\in T^*_xU\otimes \R ^4,$$ $\ \tilde{e}^a_x:T_xU \to \R ^4, \ e^a_\mu(x)dx^\mu\vert_x(v_x)\vec{\lambda}=\epsilon_x\vec{\lambda}$ with $\epsilon_x=e^a_\mu(x)v_x^\mu\in \R$. 

\

11. The Lorentz bundle ${\cal L}_4\to F_{{\cal L}_4}\to M^4$ in subsection 8. extends the symmetry group of GR, the group of general coordinate transformations of $M^4$, ${\cal D}$, to the {\it semidirect sum} $$G_{GR}={\cal L}_4\odot{\cal D},$$ with composition law given by $$(h^\prime,g^\prime)(h,g)=(h^\prime(g^\prime h {g^\prime}^{-1}), g^\prime g).$$ In fact, it is easy to verify that ${\cal D}$ has a left action on ${\cal L}_4$ at each fibre $F_{{{\cal L}_4}_x}$ of the bundle, given by the commutative diagram $$\matrix{F_{{{\cal L}_4}_x} & \buildrel{h}\over\longrightarrow & F_{{{\cal L}_4}_x} \cr g\downarrow & & \downarrow g \cr F_{{{\cal L}_4}_x} & \buildrel{h^\prime}\over\longrightarrow & F_{{{\cal L}_4}_x} \cr}$$ which defines $$h^\prime=g h g^{-1}\equiv L_g(h)$$ (conjugation action), with $g(x,(e_1(x),\dots,e_4(x)))=g(e_a(x))=g({e_a}^\mu(x){{\partial}\over{\partial x^\mu}}|_x)={e_a^\prime}^\mu(x){{\partial}\over{\partial x^{\prime\mu}}}|_x$, ${e_a^\prime}^\mu(x)={{\partial x^{\prime\mu}}\over{\partial x^\nu}}{e_a}^\nu(x)$. The action is left since $h^\prime\mapsto h^{\prime\prime}=g^\prime h^\prime {g^\prime}^{-1}=g^\prime(g h g^{-1}) {g^\prime}^{-1}=(g^\prime g) h (g^\prime g)^{-1}$ i.e. $L_{g^\prime g}(h)=L_{g^\prime} L_g(h)$. (See the extension to $G_{GR}={\cal P}_4\odot{\cal D}$ in section {\bf 34}.)   

\

{\bf 29. Curvature and torsion in terms of spin connection and tetrads. Cartan structure equations; Bianchi identities}

\

In what follows we shall designate by $\Omega^k(L^r_s)$ the real vector space of $k$ differential forms on $M$ with values in the $(r,s)$-Lorentz tensors. 

\

Given the Vielbeine ${e_\mu}^a$ and the spin connection $\omega^a_{\mu b}$ on the chart $(U,x^\mu)$ on $M$, we have the differential forms $$e^a={e_\mu}^adx^\mu\in \Omega^1(L^1) \ and \ {\omega^a}_b=\omega^a_{\mu b}dx^\mu.$$ (${\omega^a}_b\notin\Omega^1(L^1_1)$ since ${\omega^a}_b$ is not an $L^1_1$-tensor, but a connection on the Lorentz bundle ${\cal L}_4\to F_{{\cal L}_4}\to M^4$.) Then we have the 2-forms $$T^a=de^a+{\omega^a}_b\wedge e^b={{1}\over{2}}{T^a}_{\mu\nu}dx^\mu\wedge dx^\nu \in \Omega^2(L^1), \eqno{(*)}$$ with $${T^a}_{\mu\nu}=\partial_\mu {e_\nu}^a -\partial_\nu {e_\mu}^a +\omega^a_{\mu b}{e_\nu}^b-\omega^a_{\nu b}{e_\mu}^b=-{T^a}_{\nu\mu},$$
and $${R^a}_b=d{\omega^a}_b+{\omega^a}_c\wedge{\omega^c}_b={{1}\over{2}}{R^a}_{b\mu\nu} dx^\mu\wedge dx^\nu\in\Omega^2(L^1_1), \eqno{(**)}$$ with $$R^a_{b\mu\nu}=\partial_\mu\omega^a_{\nu b}-\partial_\nu\omega^a_{\mu b}+\omega^a_{\mu c}\omega^c_{\nu b}-\omega^a_{\nu c}\omega^c_{\mu b}.$$
$(*)$ and $(**)$ are known as the {\it Cartan structure equations}. As we shall show below, $T^a$ and ${R^a}_b$ are, respectively, the torsion and curvature 2-forms of section {\bf 9.}

\

For $de^a$ one has $$de^a=d({e_\nu}^a dx^\nu )=\partial_\mu {e_\nu}^a dx^\mu\wedge dx^\nu =\Omega^a_{\mu\nu}dx^\mu\wedge dx^\nu \equiv \Omega ^a$$ with $$\Omega^a_{\mu\nu}=(de^a)_{\mu\nu}={{1}\over{2}}({e_{\nu ,\mu}}^a -{e_{\mu ,\nu}}^a)=-\Omega^a_{\nu \mu}.$$ Also, $$\Omega^a_{bc}=(de^a)_{bc}={e_b}^\mu {e_c}^\nu \Omega^a_{\mu\nu}={{1}\over{2}}{e_\nu}^a ({e_ c}^\mu \partial_\mu {e_ b}^\nu -{e_ b}^\mu \partial_\mu {e_ c}^\nu)={{1}\over {2}}{e_\mu}^a ({e_{b,c}}^\mu -{e_{c,b}}^\mu)=-\Omega^a_{cb}.$$ Comparing with $\lambda^a_{bc}$ of {\bf 28}.2, we have $$\Omega^a_{bc}=-{{1}\over {2}}\lambda^a_{bc}$$ and therefore $$[e_b,e_c]=-2\Omega^a_{bc}e_a =-2(de^a)_{bc}e_a.$$ So $\Omega^a_{bc}$ also measures the non-commutativity of the Vielbeine. By the Jacobi identity, $$\Omega^f_{ad}\Omega^d_{bc}+\Omega^f_{bd}\Omega^d_{ca}+\Omega^f_{cd}\Omega^d_{ab}=0.$$

\

It is easy to show that for a metric connection, the curvature tensor with lower Lorentz indices $$R_{ab}=\eta_{ad}{R^d}_b$$ is antisymmetric i.e. $$R_{ab}=-R_{ba}.$$ In fact, $R_{ab}=\eta_{ad}({d\omega^d}_b+{\omega^d}_c \wedge {\omega^c}_b)=\eta_{ad}d{\omega^d}_b+\eta_{ad}{\omega^d}_c\wedge{\omega^c}_b=d\omega_{ab}+\omega_{ac}\wedge{\omega^c}_b=-(d\omega_{ba}+\omega_{ca}\wedge{\omega^c}_b)=-(d\omega_{ba}-{\omega^c}_b\wedge\omega_{ca})=-(d\omega_{ba}-\omega_{cb}\wedge{\omega^c}_a)$, while $R_{ba}=\eta_{bc}{R^c}_a=\eta_{bc}(d{\omega^c}_a+{\omega^c}_d\wedge{\omega^d}_a)=d\omega_{ba}+\omega_{bd}\wedge{\omega^d}_a=d\omega_{ba}-\omega_{db}\wedge{\omega^d}_a$.

\

Symbolically we write $${\bf T}=de+\omega\wedge e, \ \ {\bf R}=d\omega+\omega\wedge \omega.$$ We notice that {\it torsion is related to the tetrads} as {\it curvature is related to the spin connection}. 

\

On the other hand, while curvature involves only $\omega$, torsion involves both $e$ and $\omega$ (not only $e$). This is related to the fact that the Poincar\'e group is the semidirect (not direct) sum of $\R ^4$ (translations) and $SO^0(3,1)$ (spacetime rotations).

\ 

A manifold equipped with a metric $g_{\mu\nu}$ and a connection $\Gamma ^\mu _{\nu\rho}$ compatible with the metric but with non-vanishing torsion, is called an {\it Einstein-Cartan manifold}. The metric induces the Levi-Civita connection, ${(\Gamma _{LC})}^\mu _{\nu\rho}$ (section {\bf 13}) with $\Gamma ^\mu _{\nu\rho}={(\Gamma _{LC})}^\mu_{\nu\rho}+$ contortion tensor.

\

In the Einstein-Cartan (E-C) theory of gravity, the 1-forms $$\{e^a,\omega_{ab}\}$$ are called {\it gauge or gravitational potentials}, respectively {\it translational} and {\it rotational}, while the 2-forms $$\{T^a,{R^a}_b\}$$ are called {\it gauge or gravitational field strengths}, respectively {\it translational} and {\it rotational}. (See, however, section {\bf 34}.) At a point, it is always possible to set $e_{pt}=1$ and $\omega_{pt}=0$, i.e. respectively ${e_\mu}^a ={\delta_\mu}^a$ (16 conditions) and $\omega_{\mu ab}=0$ (24 conditions). (Hehl, 1985; Hartley, 1995.) The total number of conditions, 40, coincides with that for making zero the Christoffel symbols in the case of the Levi-Civita connection ($|\{{(\Gamma_{LC})}^\mu _{\nu\rho}\}|=40$).

\

{\it Comment}. Together with the comments in section {\bf 10}, we have the following relations: $$curvature \longleftrightarrow spin \ connection \longleftrightarrow spacetime \ rotations,$$ $$torsion \longleftrightarrow tetrads \longleftrightarrow spacetime \ translations.$$ On the other hand, from Noether theorem, we have the relations: $$spacetime \ rotations \longleftrightarrow angular \ momentum,$$ $$spacetime \ translations \longleftrightarrow energy \ momentum.$$ Naively, one should then expect the following relations: $$curvature \longleftrightarrow angular \ momentum,$$ $$torsion \longleftrightarrow energy \ momentum.$$

However, in Einstein-Cartan theory, based on a {\it non-symmetric metric connection}, the sources of curvature and torsion are respectively energy momentum and spin angular momentum i.e. $$curvature \longleftrightarrow energy \ momentum,$$ $$torsion \longleftrightarrow spin \ angular \ momentum.$$ This ``crossing" of relations is due to holonomy theorems (Trautman, 1973). 

\

These facts can be better understood as follows: In (special) relativistic field theory (r.f.t.), fields belong to irreducible representations of the Poincar\'e group ${\cal P}_4$, which are characterized by two parameters: mass and spin. Invariance under translations (${\cal T}_4$) and rotations (${\cal L}_4$) respectively leads, by Noether theorem, to the conservation of energy-momentum ($T_{\mu\nu}$) and angular momentum: orbital + intrinsic (spin, with density $S^\mu_{\nu\rho}$). On the other hand, differential geometry (d.g.), through holonomy theorems, relates curvature ($R^\mu_{\nu\rho\sigma}$) with the Lorentz group and torsion ($T^\rho_{\mu\nu}$) with translations (section {\bf 10}). Finally, Einstein (E) equations make energy-momentum the source of curvature, while Cartan (C) equations makes spin the source of torsion. This is summarized in the following diagram: $$\matrix{  & {\cal L}_4 & & \cr d.g.\nearrow & \downarrow & \searrow r.f.t. \cr R^\mu_{\nu\rho\sigma} & \delta_\omega\searrow & S^\mu_{\nu\rho} \cr \uparrow E &  & C\downarrow \cr T^\mu_\nu & \nwarrow\delta_e & T^\rho_{\mu\nu} \cr r.f.t. \nwarrow & \uparrow & \swarrow d.g. \cr  & {\cal T}_4 & & \cr}$$ In a formulation of the Einstein-Cartan theory based on tetrads and spin connection, the Einstein equations are obtained by variation with respect to the tetrads ($\delta_e$), related to translations, and the Cartan equations by variation with respect to the spin connection ($\delta_\omega$), related to rotations. (See section {\bf 32}.)

\

Locally, as differential 2-forms with values in $so(3,1)$ and $Lie({\cal T}_4)=\R ^4$, ${\bf R}$ and $T^a$ are respectively given as follows: $${\bf R}={{1}\over{2}}R_{ab\rho\sigma}dx^\rho\wedge dx^\sigma\otimes l_{ab}\in\Gamma(\Lambda^2 U\otimes so(3,1)), \ R_{ab\rho\sigma}=\eta_{ad}{R^d}_{b\rho\sigma}, \ {\bf R}(x)=(x,{\bf R}_x),$$ $${\bf R}_x={{1}\over{2}}R_{ab\rho\sigma}(x)dx^\rho\vert_x\wedge dx^\sigma\vert_x \otimes l_{ab}\in\Lambda^2_x U\otimes so(3,1), \ {\bf R}_x:T_xU\otimes T_xU\to so(3,1), \ R_x(v_x,w_x)$$  $$={{1}\over{4}}R_{ab\rho\sigma}(x)(v^\rho_x w^\sigma_x-v^\sigma_xw^\rho_x)l_{ab};$$  $$T^a={T^a}_{\mu\nu} dx^\mu\wedge dx^\nu\otimes \vec{\lambda}\in \Gamma(\Lambda^2 U\otimes \R ^4), \ T^a(x)=(x,T^a_x), \ T^a_x=T^a_{\mu\nu}(x)dx^\mu\vert_x\wedge dx^\nu\vert_x\otimes \vec{\lambda},$$ $$\ T^a_x:T_xU\otimes T_xU\to \R ^4, \ T^a_x(v_x,w_x)={{1}\over{2}}T^a_{\mu\nu}(x)(v_x^\mu w_x^\nu-v_x^\nu w_x^\mu)\vec{\lambda}.$$ 

\

From the definition of ${\bf T}$, we have, since $d^2=0$, $d{\bf T}=d\omega\wedge e-\omega\wedge de=d\omega\wedge e-\omega\wedge({\bf T}-\omega\wedge e)=d\omega\wedge e-\omega\wedge {\bf T}+\omega\wedge\omega\wedge e$, i.e. $d{\bf T}+\omega\wedge {\bf T}=(d\omega+\omega\wedge\omega)\wedge e$, that is $$d{\bf T}+\omega\wedge {\bf T}={\bf R}\wedge e\in \Omega^3(L^1). \eqno{(\alpha)}$$ In Lorentz components, $$dT^a+{\omega^a}_b\wedge T^b={R^a}_b\wedge e^b. \eqno{(\alpha^\prime)}$$ For ${\bf R}$ one has $d{\bf R}=d\omega\wedge\omega-\omega\wedge d\omega=({\bf R}-\omega\wedge\omega)\wedge\omega-\omega\wedge({\bf R}-\omega\wedge\omega)={\bf R}\wedge\omega-\omega\wedge {\bf R}$, i.e. $$d{\bf R}+\omega\wedge {\bf R}-{\bf R}\wedge\omega=0\in\Omega^3(L^1_1). \eqno{(\beta)}$$ In components, $$d{R^a}_b+{\omega^a}_c\wedge {R^c}_b-{R^a}_c\wedge{\omega^c}_b=0. \eqno{(\beta^\prime)}$$ $(\alpha)$ and $(\beta)$ (or $(\alpha^\prime)$ and $(\beta^\prime)$) are the so called Bianchi identities. (Compare ($\beta$) with the corresponding equation in section {\bf 12.})

\

Defining the {\it covariant exterior derivative} operator acting on Lorentz tensors-valued differential forms $${\cal D}_\omega=d+\omega\wedge$$ we have the equations $${\bf T}={\cal D}_\omega e, \ \ {\cal D}_\omega {\bf T}={\bf R}\wedge e, \ \ {\cal D}_\omega {\bf R}={\bf R}\wedge\omega.$$

\

Though $\omega$ is not a Lorentz tensor, one has ${\bf R}={\cal D}_\omega \omega$.

\

It is easy to verify that $T^a$ is nothing but twice the torsion tensor of section {\bf 9.}: 

\

${e_a}^\lambda {T^a}_{\mu\nu}={e_a}^\lambda((de^a)_{\mu\nu}+({\omega^a}_b\wedge e^b)_{\mu\nu})={e_a}^\lambda(\partial_\mu{e_\nu}^a-\partial_\nu{e_\mu}^a+\omega^a_{\mu b}{e_\nu}^b-\omega^a_{\nu b}{e_\mu}^b)=({e_a}^\lambda\partial_\mu{e_\nu}^a+{e_a}^\lambda{e_\nu}^b\omega^a_{\mu b})$

\

$-({e_a}^\lambda\partial_\nu{e_\mu}^a+{e_a}^\lambda{e_\mu}^b\omega^a_{\nu b})=\Gamma^\lambda_{\mu\nu}-\Gamma^\lambda_{\nu\mu}=2T^\lambda_{\mu\nu}$. 

\

A similar calculation leads to 

\

${e_a}^\rho{e_\sigma}^b{R^a}_{b\mu\nu}={e_a}^\rho{e_\sigma}^b(d{\omega^a}_b+{\omega^a}_c\wedge{\omega^c}_b)_{\mu\nu}={e_a}^\rho{e_\sigma}^b(\partial_\mu{\omega^a}_{\nu b}-\partial_\nu{\omega^a}_{\mu b}+{\omega^a}_{\mu c}{\omega^c}_{\nu b}-{\omega^a}_{\nu c}{\omega^c}_{\mu b})={R^\rho}_{\sigma\mu\nu}$.

\

For the Ricci tensor and the Ricci scalar of sections {\bf 16} and {\bf 18} respectively (but now not restricted to the Levi-Civita connection) we have $$R_{\sigma\nu}={R^\mu}_{\sigma\mu\nu}={e_a}^\mu {e_\sigma}^b {R^a}_{b\mu\nu} \ \ and \ \ R={R^\nu}_\nu={e_a}^\mu {e_\sigma}^b {R^a}_{b\mu\nu}g^{\sigma\nu}={e_a}^\mu {e_b}^\nu {R^{ab}}_{\mu\nu}.$$

\

We summarize the above formulae in the following table: $$\matrix{  & \Omega^1(L^1) & \Omega^1(L^1_1) & \Omega^2(L^1) & \Omega^2(L^1_1) & \Omega^3(L^1) & \Omega^3(L^1_1) & Components \cr  e & \times & - & - & - & - & - & e^a \cr  \omega & - & - & - & - & - & - & \omega^a_b \cr  {\bf T}=de+\omega\wedge e & - & - & \times & - & - & - & T^a=de^a+\omega^a_b\wedge e^b \cr  {\bf R}=d\omega+\omega\wedge\omega & - & - & - & \times & - & - & R^a_b=d\omega^a_b+\omega^a_c\wedge \omega^c_b \cr  d{\bf T}+\omega\wedge {\bf T}={\bf R}\wedge e & - & - & - & - & \times & - & dT^a+\omega^a_b\wedge T^b=R^a_b\wedge e^b \cr d{\bf R}+\omega\wedge {\bf R}={\bf R}\wedge\omega & - & - & - & - & - & \times & dR^a_b+\omega^a_c\wedge R^c_b=R^a_c\wedge \omega^c_b \cr}$$

\

{\bf 30. Spin connection in non-coordinate basis}

\

The Christoffel symbols for a metric connection with torsion is given in Appendix B: $$\Gamma^\mu_{\nu\rho}=(\Gamma_{LC})^\mu_{\nu\rho}+K^\mu_{\nu\rho}$$ where the contortion tensor depends on the metric and the torsion, while $(\Gamma_{LC})^\mu_{\nu\rho}$ only depends on the metric and its derivatives. Contracting with ${e^a}'s$ and ${e_b}'s$ one has $${e_a}^\nu {e_b}^\rho {e_\mu}^c (\Gamma_{LC})^\mu_{\nu\rho}+{e_ a}^\nu {e_ b}^\rho {e_\mu}^c K^\mu_{\nu\rho}={e_ a}^\nu {e_b}^\rho {e_\mu}^c \Gamma^\mu_{\nu\rho},$$ and using $(c^\prime)$ in {\bf 28}.9 we obtain $$\omega^c_{ab}+{e_ b}^\rho {e_{\rho ,a}}^c={e_ a}^\nu {e_ b}^\rho {e_\mu}^c (\Gamma_{LC})^\mu_{\nu\rho}+K^c_{ab}.$$ Using the expressions for $g_{\mu\nu}$, $\partial_\rho g_{\mu\nu}$, etc. in $\Gamma_{LC}$ in terms of ${e}'s$ and their derivatives, a straightforward calculation leads to $$\omega_{dab}=\gamma_{abd}+K_{dab} \ \ \ \ \ (U_4-space) \eqno{(*)}$$ where $$\gamma_{abd}=-\Omega_{dab}+\Omega_{abd}-\Omega_{bda}=-\gamma_{adb}$$ are the {\it Ricci rotation coefficients}, with $\vert\{\gamma_{abd}\}\vert={{n^2(n-1)}\over{2}}$, (24 for $n=4$), and $$X_{abc}=\eta_{ad}X^d_{bc}, \ \ X=\omega,\gamma,\Omega,K; \ \ X^d_{cb}=-X^d_{bc}, \ \ X=\omega,\gamma,\Omega.$$ If $T^a=0$, then $$\omega_{dab}=\gamma_{abd}. \ \ \ \ \ (V_4-space)$$ We emphasize that the ${\gamma_{abd}}'s$ come from the metric, the Vielbeine and their inverses and derivatives. So, the parallel transport and concomitant rotations of vectors by $(*)$ has two sources: {\it metric} (g) and {\it torsion} ($\tau$): from $$\omega_{\nu ab}=\gamma_{ab\nu}+K_{\nu ab},$$ we have $$(\delta_{\vert\vert}A)_a\vert_x=-\omega_{\nu ab}(x)A^b(x)dx^\nu\vert_x=(\delta_{\vert\vert}^{(g)}A)_a\vert_x +(\delta_{\vert\vert}^{(\tau)}A)\vert_x$$ with $$(\delta_{\vert\vert}^{(g)}A)_a\vert_x=-\gamma_{ab\nu}(x)A^b(x)dx^\nu\vert_x$$ and $$(\delta_{\vert\vert}^{(\tau)}A)_a\vert_x=-K_{\nu ab}(x)A^b(x)dx^\nu\vert_x.$$ ($A_a={e_a}^\mu A_\mu$, $A^b=\eta^{bc}A_c$.)

\

{\bf 31. Locally inertial coordinates}

\ 

Let $(M^n,g,\Gamma)$ be a $U^n$-space (see Appendix B), $x\in M^n$ and $(U,\varphi=(x^\mu))$ a chart on $M^n$ with $x\in U$ and $x^\mu(x)=0$, $\mu=0,\dots,n-1$. Let $(U^\prime,\varphi^\prime=(x^{\prime\mu}))$ be an intersecting chart with $x^{\prime\mu}(x)=0$ and $$x^\mu=x^{\prime\mu}-{{1}\over{2}}\Gamma^\mu_{(\nu\rho)}(x)x^{\prime\nu}x^{\prime\rho},$$ where $(\nu\rho)$ means symmetrization. The antisymmetric part $\Gamma^\mu_{[\nu\rho]}=T^\mu_{\nu\rho}=-T^\mu_{\rho\nu}$ (torsion) does not contribute to the change of coordinates. 

\

The condition of metricity at $x$, $$0=g_{\mu\nu ;\lambda}(x)=g_{\mu\nu ,\lambda}(x)-\Gamma^\rho_{\lambda\mu}(x)g_{\nu\rho}(x)-\Gamma^\rho_{\lambda\nu}(x)g_{\mu\rho}(x),$$ which, being a tensor also holds in the chart $U^\prime$, the tensor transformation formula of $g_{\mu\nu}$ (section {\bf 7}), and the diagonalization of $g_{\mu\nu}$ to $\eta_{\mu\nu}$ at $x$ (uniquely determined up to a Lorentz transformation (section {\bf 28}.8)), lead to the equations: $$g^{\prime}_{\mu\nu}(x^{\prime\lambda})=\eta_{\mu\nu}+(\eta_{\mu\rho}T^\rho_{\lambda\nu}(x)+\eta_{\nu\rho}T^\rho_{\lambda\mu}(x))x^{\prime\lambda}+O({x^{\prime\mu}}^2)=\eta_{\mu\nu}+{{\partial}\over{\partial x^{\prime\lambda}}}g^\prime_{\mu\nu}(x)x^{\prime\lambda}+O({x^{\prime\mu}}^2), \eqno{(a)}$$ and $${(\Gamma_{LC}^\prime)}^\mu_{\nu\rho}(x)={{1}\over{2}}\eta^{\mu\sigma}(\partial^\prime_\nu (g^\prime_{\rho\sigma})(x)+\partial^\prime_\rho (g^\prime_{\sigma\nu})(x)-\partial^\prime_\sigma (g^\prime_{\nu\rho})(x)). \eqno{(b)}$$ So, $$T^\mu_{\nu\rho}(x)=0 \Rightarrow \partial^\prime_\lambda (g^\prime_{\mu\nu})(x)=0  \  \ and \ \ {(\Gamma^\prime_{LC})}^\mu_{\nu\rho}(x)=0,$$ i.e. the vanishing of the torsion at $x$ is a {\it sufficient} condition for having a local inertial system at $x$. 

\ 

However, the condition is not {\it necessary}: in fact, $$\eta_{\mu\rho}T^\rho_{\lambda\nu}(p)+\eta_{\nu\rho}T^\rho_{\lambda\mu}(p)=T_{\lambda\nu\mu}(p)+T_{\lambda\mu\nu}(p)=0 $$ implies that $T_{\mu\nu\rho}$ is also antisymmetric in its second and third indices, and then it is totally antisymmetric, since $T_{\mu\nu\lambda}=-T_{\mu\lambda\nu}=T_{\lambda\mu\nu}=-T_{\lambda\nu\mu}$. 

\

A calculation gives: 

\

$n=2$:$$T^0_{01}=T^1_{01}=0$$

\ 

$n=3$: $$T^0_{01}=T^0_{02}=T^1_{01}=T^1_{12}=T^2_{02}=T^2_{12}=0,$$ $$T^0_{12}=T^2_{10}=T^1_{02}$$

\

$n=4$: $$T^0_{01}=T^0_{02}=T^0_{03}=T^1_{01}=T^1_{12}=T^1_{13}=T^2_{02}=T^2_{12}=T^2_{32}=T^3_{03}=T^3_{13}=T^3_{23}=0,$$ $$T^0_{12}=T^2_{10}=T^1_{02},$$ $$T^0_{13}=T^3_{10}=T^1_{03},$$ $$T^0_{23}=T^3_{20}=T^2_{03},$$ $$T^1_{23}=T^2_{31}=T^3_{12}$$ In each case, the number of independent but not necessarily zero components of the torsion tensor coincides with the number of independent components of the totally antisymmetric torsion tensor with covariant indices, number which results from the condition that the definition of geodesics as ``world-lines of particles" (parallel transported velocities, section {\bf 8}) to coincide with their definition as extremals of length. This last fact can be seen as follows:

\ 

As world-lines, geodesics are defined in section {\bf 8}, the equation being $${{d^2 x^\alpha}\over{d\lambda ^2}}+\Gamma^\alpha_{(\nu\mu)}{{dx^\nu}\over{d\lambda}}{{dx^\mu}\over{d\lambda}}=0 \eqno{(*)}$$ where only the symmetric part of $\Gamma^\mu_{\nu\rho}$ contributes: $$\Gamma^\alpha_{(\nu\mu)}={(\Gamma_{LC})}^\alpha_{\nu\mu}-g^{\alpha\rho}(T^\lambda_{\mu\rho}g_{\lambda\nu}+T^\lambda_{\nu\rho}g_{\lambda\mu})={(\Gamma_{LC})}^\alpha_{\nu\mu}-(T_\mu \ ^\alpha \ _\nu +T_\nu \ ^\alpha \ _\mu )={(\Gamma_{LC})}^\alpha_{\nu\mu}-2T_{(\mu} \ ^\alpha \ _{\nu )}$$ with $g^{\delta\gamma}T^\alpha_{\beta\gamma}g_{\alpha\sigma}=g^{\delta\gamma}T_{\beta\gamma\sigma}=T_\beta \ ^\delta \ _\sigma$; notice that the covariant form of the torsion tensor, $T_{\beta\gamma\delta}$, is antisymmetric in the first two indices: $T_{\beta\gamma\delta}=-T_{\gamma\beta\delta}$. (With this definition of $T_{\alpha\beta\gamma}$, the covariant form of the contortion tensor is $$K_{\mu\nu\rho}=g_{\rho\alpha}K^\alpha_{\mu\nu}=T_{\mu\nu\rho}-T_{\nu\rho\mu}+T_{\rho\mu\nu},$$ which is antisymmetric in the last two indices i.e. $K_{\mu\nu\rho}=-K_{\mu\rho\nu}$.)

\

On the other hand, the equation of geodesics defined as extremals of arc-length: $$0=\delta\int ds=\delta\int(g_{\mu\nu}dx^\mu dx^\nu )^{{1}\over{2}},$$ turns out to be (Carroll, 2004, pp.106-109) $${{d^2 x^\alpha}\over{d\lambda ^2}}+{(\Gamma_{LC})}^\alpha_{\nu\mu}{{dx^\nu}\over{d\lambda}}{{dx^\mu}\over{d\lambda}}=0. \eqno{(**)}$$ Then, for the definitions $(*)$ and $(**)$ to coincide, $T_{(\mu} \ ^\alpha \ _{\nu )}$ must vanish i.e. $T_\mu \ ^\alpha \ _\nu =-T_\nu \ ^\alpha \ _\mu \Leftrightarrow g^{\alpha\rho}T_{\mu\rho\nu}=-g^{\alpha\rho}T_{\nu\rho\mu} \Leftrightarrow T_{\mu\sigma\nu}=-T_{\nu\sigma\mu}$ i.e. $T_{\alpha\beta\gamma}$ must be 1-3 antisymmetric; but this implies that $T_{\alpha\beta\gamma}$ is also 2-3 antisymmetric: $T_{\mu\sigma\nu}=-T_{\nu\sigma\mu}=T_{\sigma\nu\mu}=-T_{\mu\nu\sigma}$. Since, by definition, $T_{\mu\nu\rho}$ is antisymmetric in the first two indices, it then turns out to be totally antisymmetric; in $n$ dimensions, its number of independent components is $\pmatrix{n \cr 3 \cr}={{n(n-1)(n-2)}\over{6}}\equiv N$. Some values are: $$\matrix{n & 2 & 3 & 4 \cr N & 0 & 1 & 4 \cr}$$ The set of allowed non-vanishing components of the torsion tensor still leads to ``physical" (geometrical) effects in the sense of section {\bf 10}. The non-closure of a parallelogram with infinitesimal sides $\epsilon^\mu$ and $\delta^\nu$ is measured by the vector $$\Delta^\mu=2T^\mu_{\beta\alpha}\delta^\beta\epsilon^\alpha=T^\mu_{\beta\alpha}(\delta^\beta\epsilon^\alpha-\delta^\alpha\epsilon^\beta)$$. For $n=4$ its components are:

$$\Delta^0=T^0_{12}(\delta^1\epsilon^2-\delta^2\epsilon^1)+T^0_{13}(\delta^1\epsilon^3-\delta^3\epsilon^1)+T^0_{23}(\delta^2\epsilon^3-\delta^3\epsilon^2),$$
$$\Delta^1=T^1_{23}(\delta^2\epsilon^3-\delta^3\epsilon^2)+T^1_{02}(\delta^0\epsilon^2-\delta^2\epsilon^0)+T^1_{03}(\delta^0\epsilon^3-\delta^3\epsilon^0),$$
$$\Delta^2=T^2_{31}(\delta^3\epsilon^1-\delta^1\epsilon^3)+T^2_{10}(\delta^1\epsilon^0-\delta^0\epsilon^1)+T^2_{03}(\delta^0\epsilon^3-\delta^3\epsilon^0),$$
$$\Delta^3=T^3_{10}(\delta^1\epsilon^0-\delta^0\epsilon^1)+T^3_{20}(\delta^2\epsilon^0-\delta^0\epsilon^2)+T^3_{12}(\delta^1\epsilon^2-\delta^2\epsilon^1),$$ which can be distinct from zero.

\

In summary, the {\it necessary} and {\it sufficient} condition for erecting a locally inertial {\it coordinate} system at a point $x$ in a $U^4$-space, is that the symmetric part of the contortion tensor vanish up to terms of order ${(x^\mu)}^2$, where $x^\mu(x)=0$. (Socolovsky, 2010.)

\

In the above sense, the {\it weak equivalence principle}, which only refers to the free motion of point-like and therefore classical particles, still {\it holds} in a $U^4$-space ($U^4=(M^4,g,\Gamma)$ with $\Gamma$ a metric connection). (A similar result was recently found by Fabbri (Fabbri, 2011).)

\

{\bf 32. Einstein-Cartan equations}

\

(We owe this derivation to L. Fabbri, 2010.)

\

A. {\it Pure gravitational case (``vacuum'')}

\

We start from the curvature 2-form $R^a_b$ with components ${R^a}_{b\mu\nu}$ given in section {\bf 29}; we define the Ricci tensor $$R_{b\nu}:={R^a}_{b\mu\nu}{e_a}^\mu$$ and the Ricci scalar $$R=R_{b\nu}\eta^{bc}{e_c}^\nu.$$ Then $$R=\eta^{bc}{R^a}_{b\mu\nu}{e_a}^\mu{e_c}^\nu=R({\omega^a}_b,e_c).$$ The gravitational action is given by the Einstein-Hilbert lagrangian density $eR$, $$S_G=\int d^4x \ eR$$ with $e=\sqrt{-detg_{\mu\nu}}\equiv\surd$. 

\

A.1. Variation with respect to the spin connection: $\delta=\delta_\omega$ ($\to${\it Cartan equations})

\

Varying ${R^a}_{b\mu\nu}$ w.r.t. $\omega$, using $\delta(\partial_\alpha\omega^c_{\beta d})=\partial_\alpha (\delta\omega^c_{\beta d})$, and adding and subtracting $(\delta\omega^a_{\rho b})\Gamma^\rho_{\mu\nu}$ one obtains $$\delta {R^a}_{b\mu\nu}={\cal D}_\mu(\delta\omega^a_{\nu b})-{\cal D}_\nu(\delta\omega^a_{\mu b})+2\delta\omega^a_{\rho b}T^\rho_{\mu\nu}$$ where $${\cal D}_\mu(\delta\omega^a_{\nu b})=\partial_\mu (\delta\omega^a_{\nu b})-\omega^c_{\mu b}(\delta\omega^a_{\nu c})+\omega^a_{\mu c}(\delta\omega^c_{\nu b})-\Gamma^\rho_{\mu\nu}(\delta\omega^a_{\rho b})$$ and $${\cal D}_\nu(\delta\omega^a_{\mu b})=\partial_\nu (\delta\omega^a_{\mu b})-\omega^c_{\nu b}(\delta\omega^a_{\mu c})+\omega^a_{\nu c}(\delta\omega^c_{\mu b})-\Gamma^\rho_{\nu\mu}(\delta\omega^a_{\rho b})$$ are covariant derivatives since $\delta\omega^c_{\alpha d}$ is a tensor ($\omega^c_{\alpha d}$ is a connection,  not a tensor, but the difference of two connections is a tensor). Then, $$\delta_\omega S_G=\int d^4x \ e{e_a}^\mu\eta^{bc}{e_c}^\nu\delta {R^a}_{b\mu\nu}=\int d^4x \ e({\cal D}_\mu((\delta\omega^a_{\nu b}){e_a}^\mu\eta^{bc}{e_c}^\nu)-{\cal D}_\nu((\delta\omega^a_{\mu b}){e_a}^\mu\eta^{bc}{e_c}^\nu)$$ $$+2{e_a}^\mu\eta^{bc}{e_c}^\nu\delta\omega^a_{\rho b}T^\rho_{\mu\nu})=\int d^4x \ e(D_\mu V^\mu +2(\delta{\omega_\rho}^{ac})T^\rho_{ac})$$ with $V^\mu$ the 4-vector given by $$V^\mu=(\delta{\omega_\nu}^{ac})({e_a}^\mu{e_c}^\nu-{e_a}^\nu{e_c}^\mu),$$ where we have used the Leibnitz rule for ${\cal D}_\alpha$, ${\cal D}_\alpha {e_a}^\sigma =0$, ${\cal D}_\mu\eta^{bc}=0$, and raised Lorentz indices with $\eta^{ab}$. For $D_\mu V^\nu$ one has $$D_\mu V^\nu=(D_{LC})_\mu V^\nu+K^\nu_{\mu\rho}V^\rho=\partial_\mu V^\nu+(\Gamma_{LC})^\nu_{\mu\lambda}V^\lambda +K^\nu_{\mu\rho}V^\rho ,$$ then $$D_\mu V^\mu=(D_{LC})_\mu V^\mu +K^\mu_{\mu\rho}V^\rho=e^{-1}(eV^\rho)_{,\rho}-2T_\lambda V^\lambda$$ where we have used $({\Gamma_{LC}})^\mu_{\mu\lambda}=\surd^{-1}\partial_\lambda\surd$ (appendix C), appendix B, and the definition of the torsion 1-form $T_\lambda=T^\rho_{\lambda\rho}$ (section {\bf 9}). Neglecting the surface term $\int d^4x \ (eV^\rho)_{,\rho}$ one obtains $$0=\delta_\omega S_G=\int d^4x \ e(-T_\mu V^\mu +\delta{\omega_\rho}^{ac}T^\rho_{ac})=\int d^4x \ e(\delta{\omega_\nu}^{ac})(-{e_c}^\nu T_a+{e_a}^\nu T_c+T^\nu_{ac})$$ with $T_a=T_\mu{e_a}^\mu$, which, due to the arbitrariness of $\delta{\omega_\nu}^{ac}$ leads to the {\it Cartan equations for torsion} (in ``vacuum''): $$T^\nu_{ac}+{e_a}^\nu T_c-{e_c}^\nu T_a=0. \eqno{(i)}$$ Multiplying $(i)$ by ${e_\rho}^a{e_\sigma}^c$ one obtains the Cartan equations in local coordinates: $$T^\nu_{\rho\sigma}+\delta^\nu_\rho T_\sigma-\delta^\nu_\sigma T_\rho=0. \eqno{(ii)}$$ 

\

{\it Proposition 1}: Torsion vanishes.

\

Proof. Taking the $\nu-\sigma$ trace in $(ii)$ leads to $T_\rho+T_\rho-4T_\rho=-2T_\rho=0$; then $T_\rho=0$ and from $(ii)$ again, $$T^\nu_{\rho\sigma}=0. \eqno{qed}$$ {\it Note}: The above result holds in $n$ dimensions for $n\neq 2$: the $\nu-\sigma$ trace gives $(2-n)T_\rho=0$. For $n=2$, $T_\rho$ is arbitrary with independent components $T^0_{01}=-T_1$ and $T^1_{01}=T_0$. Also, notice that $(i)$ (or $(ii)$) is not a differential equation, but an algebraic one; this is the mathematical expression of the fact that in E-C theory {\it torsion does not propagate}.

\

{\it Proposition 2}: Let $$T^\nu_{\rho\sigma}+\delta^\nu_\rho T_\sigma-\delta^\nu_\sigma T_\rho=\kappa S^\nu_{\rho\sigma} \eqno{(iii)}$$ with $S^\nu_{\rho\sigma}=-S^\nu_{\sigma\rho}$, and $\kappa$ a constant. ($(iii)$ corresponds to a non-vacuum case and will be used in part B.) Then, $$T^\nu_{\rho\sigma}=\kappa (S^\nu_{\rho\sigma}-{{1}\over{2-n}}(\delta^\nu_\rho S_\sigma-\delta^\nu_\sigma S_\rho)) \eqno{(iv)}$$ where $n\neq 2$ is the dimension of the manifold and $S_\mu= S^\nu_{\mu\nu}$. In particular, for $n=4$, $$T^\nu_{\rho\sigma}=\kappa (S^\nu_{\rho\sigma}+{{1}\over{2}}(\delta^\nu_\rho S_\sigma-\delta^\nu_\sigma S_\rho)). \eqno{(v)}$$ Proof. Again taking the trace $\nu-\sigma$ in $(iii)$, $(2-n)T_\rho=\kappa S_\rho$, then $T_\rho={{1}\over{2-n}}S_\rho$ and $T^\nu_{\rho\sigma}$ is $(iv)$. \ \ \ \ \ \ \ \ \ qed 

\

For $n=2$, the unique solution of $(iii)$ is $S^\mu_{\nu\rho}=0$: in fact, $\delta^\nu_\nu=2$ and then $S_\rho=0$; so $S_0=S^1_{01}=0$ and $S_1=S^0_{10}=0$. 

\

A.2. Variation with respect to the tetrads: $\delta=\delta_e$ ($\to$ {\it Einstein equations})

\

Again from $S_G$, $$\delta_eS_G=\int d^4x \ ((\delta R)e+R\delta e)=\int d^4x \ ({R^a}_{b\mu\nu}\eta^{bc}e((\delta {e_a}^\mu){e_c}^\nu+{e_a}^\mu\delta{e_c}^\nu)-Re{e_\lambda}^d\delta{e_d}^\lambda)$$ $$=\int d^4x \ (2{R^a}_\mu-R{e_\mu}^a)e\delta{e_a}^\mu=0$$ where we used $\delta e=-e{e_\lambda}^d\delta {e_d}^\lambda$ (appendix C), and from the arbitrariness of $\delta{e_a}^\mu$, we obtain the {\it Einstein equations for curvature} (in ``vacuum''): $${R^a}_\mu-{{1}\over{2}}R{e_\mu}^a=0 \eqno{(vi)}$$ or $${G^a}_\mu=0 \eqno{(vii)}$$ with $${G^a}_\mu={R^a}_\mu-{{1}\over{2}}R{e_\mu}^a. \eqno{(viii)}$$ (${R^a}_\mu=\eta^{ac}R_{c\mu}$.) Since in vacuum $R=0$ (section {\bf 18}), $(vi)$ amounts to $${R^a}_\mu=0. \eqno{(ix)}$$ Of course, multiplying $(vi)$ by ${e_a}^\nu$ we obtain Einstein equations in local coordinates (section {\bf 20}). 

\

In summary, for the pure gravitational case, {\it Einstein theory=Einstein-Cartan theory}; this is a consequence of the form of the Einstein-Hilbert action $S_G$. From the form of $R$, gravity has been expressed as an interacting gauge theory (see section {\bf 33}) between the spin connection ${\omega^a}_b$ and the coframes field $e^a$; both ${\omega^a}_b$ and $e^a$ are pure geometric fields, which live in the frame and coframe bundles ${\cal L}_4\to F_{{\cal L}_4}\to M^4$ and ${\cal L}_4\to F^*_{{\cal L}_4}\to M^4$ respectively.

\

B. {\it Minimal coupling to Dirac fields}

\

The Dirac-Einstein action is given by $$S_{D-E}=k\int d^4x \ e L_{D-E}=k\int d^4x \ e \ ({{i}\over{2}}(\bar{\psi}\gamma^a(D_a\psi)-(\bar{D}_a\bar{\psi})\gamma^a\psi ) -m\bar{\psi}\psi)$$ where $$D_a\psi=(e_a-{{i}\over{4}}\omega_{abc}\sigma^{bc})\psi={e_a}^\mu(\partial_\mu-{{i}\over{4}}\omega_{\mu bc}\sigma^{bc})\psi={e_a}^\mu D_\mu\psi$$ and $$\bar{D}_a\bar{\psi}=e_a\bar{\psi}+{{i}\over{4}}\omega_{abc}\bar{\psi}\sigma^{bc}={e_a}^\mu(\partial_\mu \bar{\psi}+{{i}\over{4}}\omega_{\mu bc}\bar{\psi}\sigma^{bc})={e_a}^\mu \bar{D}_\mu\bar{\psi}$$ are the covariant derivatives of the Dirac field $\psi$ and its conjugate $\bar{\psi}=\psi^\dagger \gamma_0$ with respect to the spin connection, which give the minimal coupling between fermions and gravity; they are obtained through the replacement $$d_a\psi\to D_a \psi  \ \ i.e. \ \ e_a\psi={e_a}^\mu\partial_\mu\psi\to D_a\psi$$ which amounts to the ``comma goes to semicolon'' rule for tensors but here adapted to spinor fields. $\sigma ^{bc}={{i}\over{2}} [\gamma ^b,\gamma ^c]$, and the $\gamma ^a$'s are the usual numerical (constant) Dirac gamma matrices satisfying $\{\gamma ^a,\gamma ^b\}=2\eta ^{ab}I$, $\gamma ^{0\dag}=\gamma ^0$ and $\gamma ^{j\dag}=-\gamma ^j$. $k=-16\pi{{G}\over{c^4}}$ ($-16\pi$ in natural units). Then the action is $$S_{D-E}=k\int d^4x \ e \ ({{i}\over{2}}\bar{\psi}\gamma^\mu(\partial_\mu\psi +{{1}\over{8}}\omega_{\mu bc}[\gamma^b,\gamma^c]\psi )-{{i}\over{2}}(\partial_\mu\bar{\psi}-{{1}\over{8}}\omega_{\mu bc}\bar{\psi}[\gamma^b,\gamma^c])\gamma^\mu\psi -m\bar{\psi}\psi)$$ where $\gamma^\mu={e_a}^\mu\gamma^a=\gamma^\mu (x)$. 

\

B.1. Variation with respect to the spin connection: $\delta=\delta_\omega$ 

\

$$\delta_\omega S_{D-E}={{k}\over{8}}\int d^4x \ e \ \bar{\psi}\{\gamma^\mu,\sigma^{bc}\}\psi\delta\omega_{\mu bc}={{k}\over{2}}\int d^4x \ e \ S^{\mu bc}\delta\omega_{\mu bc}$$ with $S^{\mu bc}={e_a}^\mu S^{abc}$, where $$S^{abc}={{1}\over{4}}\bar{\psi}\{\gamma^a,\sigma^{bc}\}\psi$$ is the {\it spin density tensor} of the Dirac field. $S^{abc}$ is totally antisymmetric and therefore in 4 dimensions it has 4 independent components: $S^{012}$, $S^{123}$, $S^{230}$ and $S^{301}$.  

\

Combining this result with the corresponding variation for the pure gravitational field (part A), we obtain $$0=\delta_\omega (S_G+S_{D-E})=\int d^4x \ e \ \delta{\omega_\nu}^{ac}(T^\nu_{ac}+{e_a}^\nu T_c-{e_c}^\nu T_a+{{k}\over{2}}S^\nu_{ac})$$ and therefore $$T^\nu_{ac}+{e_a}^\nu T_c-{e_c}^\nu T_a=-{{k}\over{2}}S^\nu_{ac},$$ the {\it Cartan equation}. Multiplying by ${e_\rho}^a{e_\sigma}^c$ one obtains $$T^\nu_{\rho\sigma}+\delta^\nu_\rho T_\sigma-\delta^\nu_\sigma T_\rho=-{{k}\over{2}}S^\nu_{\rho\sigma}$$ with $$S^\nu_{\rho\sigma}={{1}\over{4}}\bar{\psi}\{\gamma^\mu, \sigma_{\rho\sigma}\}\psi.$$ 

\

From (iii) in part A, with $\kappa=-{{k}\over{2}}$, we obtain the {\it torsion in terms of the spin tensor}: $$T^\nu_{\rho\sigma}={{8\pi G}\over{c^4}}(S^\nu_{\rho\sigma}+{{1}\over{2}}(\delta^\nu_\rho S_\sigma-\delta^\nu_\sigma S_\rho))$$ with $S_\rho=S^\nu_{\rho\nu}$. In natural units, $G=c=\hbar=1$ and so $T^\nu_{\rho\sigma}=8\pi(S^\nu_{\rho\sigma}+{{1}\over{2}}(\delta^\nu_\rho S_\sigma-\delta^\nu_\sigma S_\rho))$. 

\

B.2. Variation with respect to the tetrads: $\delta=\delta_e$

\

From $S_{D-E}$ and using appendix C for $\delta e$, one obtains $$\delta_e S_{D-E}=k\int \ e \ ({{i}\over{2}}(\bar{\psi}\gamma^a(D_\mu\psi)-(\bar{D}_\mu\bar{\psi})\gamma^a\psi)-{e_\mu}^aL_{D-E})\delta{e_a}^\mu.$$ For the Dirac fields which obey the equations of motion $${{\delta S_{D-E}}\over{\delta\bar{\psi}}}={{\delta S_{D-E}}\over{\delta\psi}}=0$$ i.e. $$i\gamma^a(\bar{D}_a\bar{\psi})+m\bar{\psi}=i\gamma^aD_a\psi-m\psi=0$$ the Dirac-Einstein lagrangian vanishes i.e. $L_{D-E}\vert_{eq. \ mot.}=0$. Then, combining this result with the corresponding variation for the pure gravitational field (part A), $$0=\delta_e(S_G+S_{D-E})=\int d^4x \ e \ (2{R^a}_\mu-R{e_\mu}^a+k{{i}\over{2}}(\bar{\psi}\gamma^a(D_\mu\psi)-(\bar{D}_\mu\bar{\psi})\gamma^a\psi))\delta{e_a}^\mu,$$ and from the arbitrariness of $\delta{e_a}^\mu$, $${R^a}_\mu-{{1}\over{2}}R{e_\mu}^a=-{{k}\over{2}}{T^a}_\mu \eqno{(*)}$$ with $${T^a}_\mu={{i}\over{2}}(\bar{\psi}\gamma^a(D_\mu\psi)-(\bar{D}_\mu\bar{\psi})\gamma^a\psi)$$ the {\it energy-momentum tensor} of the Dirac field. Multiplying $(*)$ by ${e_a}^\nu$ one obtains $${R^\nu}_\mu-{{1}\over{2}}R\delta^\nu_\mu=-{{k}\over{2}}{T^\nu}_\mu \ \ or \ \ R_{\lambda\mu}-{{1}\over{2}}Rg_{\lambda\mu}=-{{k}\over{2}}T_{\lambda\mu},$$ the {\it Einstein equation}.

\

{\it Note}: For $L_{D-E}$ one has $$L_{D-E}={e_a}^\mu{T^a}_\mu-m\bar{\psi}\psi$$ i.e. ${T^a}_\mu$ {\it couples to the tetrad}. On the other hand, $${T^a}_\mu={\theta^a}_\mu+\omega_{\mu bc}S^{abc}$$ where $${\theta^a}_\mu={{i}\over{2}}(\bar{\psi}\gamma^a\partial_\mu\psi-(\partial_\mu\bar{\psi})\gamma^a\psi)$$ is the {\it canonical energy-momentum tensor} of the Dirac field. Then, $$L_{D-E}={e_a}^\mu{\theta^a}_\mu+{e_a}^\mu\omega_{\mu bc}S^{abc}-m\bar{\psi}\psi={e_a}^\mu{\theta^a}_\mu+\omega_{abc}S^{abc}-m\bar{\psi}\psi.$$ So, ${\theta^a}_\mu$ couples to the tetrad while spin couples to the spin connection; moreover, since $S^{abc}$ is totally antisymmetric, {\it the Dirac field only interacts with the totally antisymmetric part of the connection}.

\

{\bf 33. Lorentz gauge invariance of Einstein and Einstein-Cartan theories}

\

Under Lorentz transformations ${h_a}^b(x)$ in the tangent, cotangent and, in general, tensor spaces at each point $x$ of the manifold $M^4$ (for simplicity we restrict the discussion to four dimensions), the frames $e_a$ (or ${e_a}^\mu$), coframes $e^a$ (or ${e_\mu}^a$) and spin connection ${\omega^a}_b$ (or $\omega^a_{\mu b}$) transform as indicated in sections {\bf 28}.8 and {\bf 28}.9. {\it The volume element} $d^4x \ e$ in $S_G$ (or $S_{D-E}$) {\it is invariant}; in fact, $$g_{\mu\nu}(x)=\eta_{ab}{e_\mu}^a(x){e_\nu}^b(x)=\eta_{ab}{e^\prime_\mu}^c{h^{-1}_c}^a{e^\prime_\nu}^d{h^{-1}_d}^b={e^\prime_\mu}^c{e^\prime_\nu}^d{h^{-1}_c}^a\eta_{ab}{h^{-1}_d}^b={e^\prime_\mu}^c{e^\prime_\nu}^d\eta_{cd}=g^\prime_{\mu\nu}(x)$$ and then $det(-g_{\mu\nu}(x))=det(-g^\prime_{\mu\nu}(x))$ i.e. $e^\prime (x)=e(x)$. Then $d^4x \ e=d^4x^\prime \ e^\prime$ since $x^{\prime\mu}=x^\mu$. 

\

A. Pure gravitational case (``vacuum'')

\

Though ${\omega^a}_b=\omega^a_{\mu ab}dx^\mu$ is a connection and transforms as $${\omega^c}_a={h_c}^d{\omega^{\prime r}}_d{h^{-1}_r}^c + (d{h_a}^d){h^{-1}_d}^c,$$ its curvature ${R^a}_b$ is a Lorentz tensor: $${R^a}_b={h_b}^d{h_c^{-1}}^a{R^{\prime c}}_d \in \Omega^2({L^1}_1)$$ (see table in section {\bf 29}); then the Ricci scalar is gauge invariant: $$R={R^a}_b e_a\eta^{bc}e_c={h_b}^d{R^{\prime c}}_d{h^{-1}_c}^a{h_a}^f e^\prime_f \eta^{bg}{h_g}^l e^\prime_l={R^{\prime c}}_d\delta^f_c e^\prime_f\eta^{dl}e^\prime_l={R^{\prime c}}_d e^\prime_c\eta^{dl}e^\prime_l=R^\prime.$$ 

\

(An explicit proof of the gauge invariance of $R$ is given in Appendix D.) 

\

{\it Remark}: In this section, the tetrads $e_a$ (or their duals $e^a$) are {\it not} gauge potentials; only the spin connection ${\omega^a}_b$ is a gauge potential, related to the Lorentz group and therefore to space-time rotations (at each point of the manifold). So, if only the Lorentz group is gauged, the metric $g=\eta_{ab}e^a \otimes e^b$ (see eq. (b) in section {\bf 28}.6) does not come from the connection. 

\

B. Minimal coupling to Dirac field

\

See O'Raifeartaigh (1997), Ch. 5, $\oint$ 3, pp 115-116.

\

{\bf 34. Poincar\'e gauge invariance of Einstein and Einstein-Cartan theories}

\

We need the concepts of affine structures: spaces, bundles and connections ($\oint$3, chapter III, Kobayashi and Nomizu, 1963). 

\

The general linear group $GL_n(\R)$ in $n$ real dimensions acts from the left on the vector space $\R^n$ by simple matrix multiplication: $(g,\lambda)\mapsto g\lambda$, which is a linear operation.

\

The {\it affine space} $$\A^n=\{\pmatrix{\lambda\cr 1 \cr}, \ \lambda\in \R^n\}$$ is acted by the {\it affine group} in $n$ real dimensions $$GA_n(\R)=\{\pmatrix{g & \xi \cr 0 & 1\cr}, \ g\in GL_n(\R), \ \xi\in \R^n\}$$ as follows: $$GA_n(\R)\times\A^n\to\A^n, \ (\pmatrix{g & \xi \cr 0 & 1\cr},\pmatrix{\lambda\cr 1 \cr})\mapsto \pmatrix{g\lambda+\xi \cr 1 \cr}.$$ Then, one has the following diagram of short exact sequences (s.e.s.'s) of groups and group homomorphisms: $$\matrix{0 & \longrightarrow & \R^n & \buildrel{\mu}\over\longrightarrow & GA_n(\R) & \matrix{\buildrel{\nu}\over\longrightarrow\cr\buildrel{\rho}\over\longleftarrow\cr} & GL_n(\R) & \longrightarrow & 0 \cr & & Id\uparrow & & \uparrow\iota & & \uparrow\iota & & & \cr 0 & \longrightarrow & \R^n & \buildrel{\mu\vert}\over\longrightarrow & {\cal P}_n & \matrix{\buildrel{\nu\vert}\over\longrightarrow\cr\buildrel{\rho\vert}\over\longleftarrow\cr} & {\cal L}_n & \longrightarrow & 0 \cr}$$ with $\mu(\xi)=\pmatrix{I_n & \xi \cr 0 & 1 \cr}$ and $\nu(\pmatrix{g & \lambda \cr 0 & 1})=g$. $\mu$ is 1-1, $\nu$ is onto, and $ker(\nu)=Im(\mu)=\{\pmatrix{I_n & \xi \cr 0 & 1 \cr}, \ \xi\in \R^n\}$. We have also restricted $\mu$ and $\nu$ (respectively $\mu\vert$ and $\nu\vert$) to the connected components of the Poincar\'e (${\cal P}_n$) and Lorentz (${\cal L}_n$) groups in $n$ dimensions. Both s.e.s.'s split, i.e. there exists the group homomorphism $\rho:GL_n(\R)\to GA_n(\R)$, $g\mapsto\rho(g)=\pmatrix{g & 0 \cr 0 & 1 \cr}$ and its restriction $\rho\vert$ to ${\cal L}_n$, such that $\nu\circ\rho=Id_{GL_n(\R)}$ and $\nu\vert\circ\rho\vert=Id_{{\cal L}_n}$. So $$GA_n(\R)=\R^n\odot GL_n(\R), \ {\cal P}_n=\R^n\odot{\cal L}_n$$ with composition law $$(\lambda^\prime,g^\prime)(\lambda,g)=(\lambda^\prime+g^\prime\lambda,g^\prime g).$$ As a consequence, the factorization of an element of $GA_n(\R)$ (${\cal P}_n$) in terms of elements of $\R^n$ and $GL_n(\R)$ (${\cal L}_n$) is unique: $\pmatrix{g & \xi \cr 0 & 1 \cr}=\mu(\xi)\rho(g)$ (or $\mu\vert(\xi)\rho\vert(g)$). The dimensions of $GA_n(\R)$, $GL_n(\R)$, ${\cal P}_n$ and ${\cal L}_n$ are, respectively, $n+n^2$, $n^2$, ${{n(n+1)}\over{2}}$ and ${{n(n-1)}\over{2}}$ (20, 16, 10 and 6 for $n=4$). 

\

The above s.e.s.'s pass to s.e.s.'s of the corresponding Lie algebras: $$\matrix{0 & \longrightarrow & \R^n & \buildrel{\tilde{\mu}}\over\longrightarrow & ga_n(\R) & \matrix{\buildrel{\tilde{\nu}}\over\longrightarrow\cr\buildrel{\tilde{\rho}}\over\longleftarrow\cr} & gl_n(\R) & \longrightarrow & 0 \cr & & Id\uparrow & & \uparrow\iota & & \uparrow\iota & & & \cr 0 & \longrightarrow & \R^n & \buildrel{\tilde{\mu}\vert}\over\longrightarrow & p_n & \matrix{\buildrel{\tilde{\nu}\vert}\over\longrightarrow\cr\buildrel{\tilde{\rho}\vert}\over\longleftarrow\cr} & l_n & \longrightarrow & 0 \cr} $$ with $gl_n(\R)=\R(n)$, $ga_n(\R)=\R^n\odot gl_n(\R)$ with Lie product $$(\lambda^\prime,R^\prime)(\lambda,R)=(R^\prime\lambda-R\lambda^\prime,[R^\prime,R]),$$ where $[R^\prime,R]$ is the Lie product in $gl_n(\R)$ and $[\lambda^\prime,\lambda]=0$ in $\R^n$, $\tilde{\mu}(\xi)=(\xi,0)$, $\tilde{\nu}(\xi,R)=R$, and $\tilde{\rho}(R)=(0,R)$. $\tilde{\mu}$, $\tilde{\nu}$  and $\tilde{\rho}$ (and their corresponding restrictions $\tilde{\mu}\vert$, $\tilde{\nu}\vert$ and $\tilde{\rho}\vert$) are Lie algebra homomorphisms, with $\tilde{\nu}\circ\tilde{\rho}=Id_{gl_n(\R)}$ and $\tilde{\nu}\vert\circ\tilde{\rho}\vert=Id_{l_n}$. The s.e.s.'s split only at the level of vector spaces i.e. if $(\lambda,R)\in ga_n(\R)$, then $(\lambda,R)=\tilde{\mu}(\lambda)+\tilde{\rho}(R)$, but $(\lambda,R)\neq \tilde{\mu}(\lambda)\tilde{\rho}(R)$.

\

Let us denote:

\

$M^n$: n-dimensional differentiable manifold

\

${\cal F}_{M^n}$: frame bundle of $M^n$: $GL_n\to FM^n\buildrel{\pi_F}\over\longrightarrow M^n$

\

${\cal A}_{M^n}$: affine frame bundle of $M^n$: $GA_n\to AM^n\buildrel{\pi_A}\over\longrightarrow M^n$

\

$GL_n$: general linear group in $n$ dimensions (section {\bf 24}), $dim_{\R}GL_n=n^2$

\

$GA_n$: general affine group in $n$ dimensions, $dim_{\R}GA_n=n+n^2$

\

${\cal F}^L_{M^n}$: bundle of Lorentz frames of $M^n$: ${\cal L}_n\to F^LM^n\buildrel{\pi_F\vert}\over\longrightarrow M^n$ (section {\bf 25})

\

${\cal F}^P_{M^n}$: bundle of Poincar\'e frames of $M^n$: ${\cal P}_n\to A^PM^n\buildrel{\pi_A\vert}\over\longrightarrow M^n$ 

\

One has the following diagram of bundle homomorphisms:

\

$$\matrix{AM^n\times GA_n & \buildrel{\iota\times\iota}\over\longleftarrow & A^PM^n\times{\cal P}_n & \matrix{\buildrel{\beta\vert\times\nu\vert}\over\longrightarrow\cr\buildrel{\gamma\vert\times\rho\vert}\over\longleftarrow\cr} & F^LM^n\times{\cal L}_n & \buildrel{\iota\times\iota}\over\longrightarrow & FM^n\times GL_n\cr
 \psi_A\downarrow & & \psi_A\vert\downarrow & & \downarrow\psi_F\vert & & \downarrow\psi_F \cr AM^n & \buildrel{\iota}\over\longleftarrow & A^PM^n & \matrix{\buildrel{\beta\vert}\over\longrightarrow\cr\buildrel{\gamma\vert}\over\longleftarrow\cr} & F^LM^n & \buildrel{\iota}\over\longrightarrow & FM^n\cr \pi_A\downarrow & & \pi_A\vert\downarrow & & \downarrow\pi_F\vert & & \downarrow\pi_F \cr M^n & \buildrel{Id}\over\longrightarrow & M^n & \buildrel{Id}\over\longrightarrow & M^n & \buildrel{Id}\over\longrightarrow & M^n\cr}$$ where $\beta$ is the bundle homomorphism 

\

$$\matrix{AM^n\times GA_n & \matrix{\buildrel{\beta\times\nu}\over\longrightarrow\cr\buildrel{\gamma\times\rho}\over\longleftarrow \cr} & FM^n\times GL_n\cr \psi_A\downarrow & & \downarrow\psi_F \cr AM^n & \matrix{\buildrel{\beta}\over\longrightarrow\cr\buildrel{\gamma}\over\longleftarrow\cr} & FM^n \cr \pi_A\downarrow & & \downarrow\pi_F \cr M^n & \buildrel{Id}\over\longrightarrow & M^n\cr}$$ between the bundle of affine frames and the bundle of linear frames over $M^n$, with $$\beta(x,(v_x,r_x))=(x,r_x), \ \gamma(x,r_x)=(x,(0_x,r_x)), \ 0\in T_xM^n,$$ $$AM^n=\cup_{x\in M^n}(\{x\}\times AM^n_x), \ AM^n_x=\{(v_x,r_x), \ v_x\in A_xM^n, \ r_x\in F_x\},$$ where $A_xM^n$ is the tangent space at $x$ considered as an affine space (Appendix E); and $$\psi_A((x,(v_x,r_x)),(\xi,g))=(x,(v_x+r_x\xi,r_xg))$$ is the action of $GA_n$ on $AM^n$, with $r_x\xi=\sum_{a=1}^n e_{ax}\xi^a$. $\pi|_F,\dots,\psi|_A,\dots,\beta|,\dots$, etc. are restrictions; in particular $\beta|(a)=\beta(a)$ and $\gamma|(e)=\gamma(e)$. 

\

A {\it general affine connection} (g.a.c.) on $M^n$ is a connection in the bundle of affine frames ${\cal A}_{M^n}$. If $\omega_A$ is the 1-form of the connection, then $$\omega_A\in \Gamma(T^*AM^n\otimes ga_n)$$ i.e. $$\omega_A:AM^n\to T^*AM^n\otimes ga_n, \ (v_x,r_x)\mapsto ((v_x,r_x),\omega_{A(v_x,r_x)}), \ \omega_{A(v_x,r_x)}:T_{(v_x,r_x)}AM^n\to ga_n,$$ $$V_{(v_x,r_x)}\mapsto\omega_{A(v_x,r_x)}(V_{(v_x,r_x)})=(\lambda,R)\equiv\lambda\odot R\in \R^n\odot gl_n(\R).$$ Obviously, $\omega_A$ obeys the usual axioms of connections. 

\

From the smoothness of $\gamma$, the pull-back $\gamma^*(\omega_A)$ is a $ga_n$-valued 1-form on $FM^n$: $$\gamma^*(\omega_A)=\varphi\odot\omega_F,$$ where $\omega_F$ is a connection on $FM^n$, and $\varphi$ is an $\R^n$-valued 1-form. There is a 1-1 correspondence between g.a.c.'s on $AM^n$ and pairs $(\omega_F,\varphi)$ on $FM^n$: $$\{\omega_A\}_{g.a.c.}\longleftrightarrow\{(\omega_F,\varphi)\}.$$ 

\

$\omega_A$ is an {\it affine connection} (a.c.) on $M^n$ if $\varphi$ is the soldering (canonical) form $\theta$ on $FM^n$ (section {\bf 26}). Then, if $\omega_A$ is an a.c. on $AM^n$, $$\gamma^*(\omega_A)=\theta\odot\omega_F$$ where $\omega_F$ is a connection on $FM^n$. There is then a 1-1 correspondence $$\{\omega_A\}_{a.c.}\longleftrightarrow\{\omega_F\},$$ since $\theta$ is fixed. Also, if $\Omega_A$ is the curvature of $\omega_A$, then $$\gamma^*(\Omega_A)=D_{\omega_F}\theta\odot\Omega_F.$$ But $D_{\omega_F}\theta =T_F$: the {\it torsion} of the connection $\omega_F$ on $FM^n$: in fact, from section {\bf 26}, $\theta^\mu={(X^{-1})^\mu}_\nu dx^\nu$, then $\theta^a={e_\mu}^a\theta^\mu={e_\mu}^a{(X^{-1})^\mu}_\nu dx^\nu={(X^{-1})^a}_\nu dx^\nu={e_\nu}^a dx^\nu=e^a$, so $D_{\omega_F}\theta^a=d\theta^a+{\omega_F^a}_be^b=T_F^a$. Therefore, $$\gamma^*(\Omega_A)=T_F\odot\Omega_F.$$ The facts that $A^PM^n$ is a subbundle of $AM^n$ and $F^LM^n$ is a subbundle of $FM^n$, with structure groups and Lie algebras the corresponding subgroups and sub-Lie algebras, and the existence of the restrictions $\beta|:A^PM^n\to F^LM^n$ and $\gamma|:F^LM^n\to A^PM^n$, allow us to obtain similar conclusions for the relations between affine connections on the Poincar\'e bundle and linear connections on the Lorentz bundle: 

\

There is a 1-1 correspondence between affine Poincar\'e connections $\omega_P$ on $F^PM^n$ and Lorentz connections on $F^LM^n$: $$\{\omega_P\}\longleftrightarrow \{\omega_L\}$$ with $$\gamma\vert^*(\omega_P)=\theta_L\odot\omega_L$$ where $\theta_L=\theta_{FM^n}\vert_{F^LM^n}$ is the canonical form on $F^LM^n$. Also, $$\gamma\vert^*(\Omega_P)=D_{\omega_L}\theta_L\odot\Omega_L=T_L\odot\Omega_L.$$ So, there is a 1-1 correspondence between curvatures of affine connections on $F^PM^n$ and torsion and curvature pairs on $F^LM^n$: $$\{\Omega_P\}\longleftrightarrow \{(T_L,\Omega_L)\}.$$ For pure gravity governed by the Einstein-Hilbert action, $T_L=0$, as it was shown in section {\bf 32}. 

\

The Poincar\'e gauge invariance of G.R. and E-C theory has been discussed by several authors (Hayashi and Shirafuji, 1980; Ali et al, 2009; Gronwald and Hehl, 1996; Hehl, 1998). To explicitly prove it, we have to consider as gauge transformations both the Lorentz part, already studied in the previous section, and the translational part. This has to be done using the bundle of Poincar\'e frames ${\cal F}_4:{\cal P}_4\to A^PM^4\buildrel{\pi_P}\over\longrightarrow M^4 \ (\pi_P=\pi_A\vert)$, (D'Olivo and Socolovsky, 2011). The action of ${\cal P}_4$ over on $A^PM^4$ is given by $$\psi_P:A^PM^4\times{\cal P}_4\to A^PM^4, \ (\psi_P=\psi_A\vert), \ \psi_P((x,(v_x,r_x)),(\xi,h))\equiv(x,(v_x,r_x))(\xi,h)=(x,(v_x+r_x\xi,r_xh))$$ $=(x,(v^\prime_x,r^\prime_x))$, where $r_x=(e_{ax}), \ a=1,2,3,4,$ is a Lorentz frame, $h\in{\cal L}_4$, and $\xi\in\R^4\cong\R^{1,3}$ is a Poincar\'e gauge translation. For a pure translation, $h=I_L$ i.e. ${h_a}^b={\delta_a}^b$ and therefore $$(x,(v_x,r_x))(\xi,I_L)=(x,(v_x+r_x\xi,r_xI_L))=(x,(v_x+r_x\xi,r_x))$$ i.e. $$r^\prime_x=r_x.$$ Therefore $e^\prime_{ax}=e_{ax}, \ a=1,2,3,4,$ and then, from ($c$) or ($c^\prime$) in section {\bf 28}.9, $$\omega^{\prime a}_{\mu b}=\omega^a_{\mu b}$$ since $\Gamma^\mu_{\nu\rho}=(\Gamma_{LC})^\mu_{\nu\rho}+K^\mu_{\nu\rho}$ remains unchanged (in the case of pure gravity $K^\mu_{\nu\rho}=0$), and so the coordinate Ricci scalar $R$ of section {\bf 32}.A is also a gauge scalar and $S_G$ is invariant.

\

The Poincar\'e bundle extends the symmetry group of GR and E-C theory to the semidirect sum $$G_{GR}={\cal P}_4\odot{\cal D}$$ cf. {\bf 28}.11, with composition law $$((\xi^\prime,h^\prime),g^\prime)((\xi,h),g)=((\xi^\prime,h^\prime)(g^\prime(\xi,h)g^{\prime -1}),g^\prime g).$$ The left action of ${\cal D}$ on ${\cal P}_4$ is given by the commutative diagram $$\matrix{A^PM^4 & \buildrel (\xi,h)\over\longrightarrow & A^PM^4 \cr g\downarrow & & \downarrow g \cr A^PM^4 & \buildrel (\xi^\prime,h^\prime)\over\longrightarrow \ & A^PM^4 \cr}$$ with $$g:A^PM^4\to A^PM^4, \ (x,(v_x^\mu{{\partial}\over{\partial x^\mu}}\vert_x,({e_{ax}}^\nu{{\partial}\over{\partial x^\nu}}\vert_x)))\mapsto (x,(v^{\prime\mu}_x{{\partial}\over{\partial x^{\prime\mu}}}\vert_x,({e_{ax}}^{\prime\nu}{{\partial}\over{\partial x^{\prime\nu}}}))),$$ where $v_x^{\prime\mu}={{\partial x^{\prime\mu}}\over{\partial x^\alpha}}\vert_xv_x^\alpha$ and ${e_{ax}}^{\prime\nu}={{\partial x^{\prime\nu}}\over{\partial x^\beta}}\vert_x{e_{ax}}^\beta.$

\

In section {\bf 29}, following Hehl (Hehl, 1985; Hehl et al, 1976; Hammond, 2002), we called ${e^a}'s$ the translational gravitational gauge potentials. This is not, however, strictly correct, since the ${e^a}'s$ or their duals, the tetrad fields $e_a={e_a}^\mu\partial_\mu$, are not connections, but tensors in both their world ($\mu$) and internal ($a$) indices (Hayashi, 1977; Leston, 2008; Leston and Socolovsky, 2011). The {\it translational potentials} ${B_a}^\mu$ (in fact their inverses ${B_\mu}^a$) are given by the 1-form fields locally defined as follows (Hayashi and Nakano, 1967; Aldrovandi and Pereira, 2007): $${B_\mu}^a={e_\mu}^a-{{\partial v_x^a}\over{\partial x^\mu}} \ \ or \ \ B^a=e^a-dv_x^a,$$ where $v_x=\sum_{a=0}^3v_x^ae_{ax}\in A_xM^4$; the $v_x^a$'s are here considered the coordinates of the tangent space at $x$. 

\

The transformation properties of the ${B_\mu}^a$'s are the following:

\

Internal Lorentz: $${B_\mu}^{\prime a}={h_a}^b{B_\mu}^b-\partial_\mu({h_b}^a)v_x^b \ \ or \ \ B^{\prime a}={h_b}^aB^b-(d{h_b}^a)v_x^b.$$

\

{\it Proof}: \ \ ${B_\mu}^{\prime a}={e_\mu}^{\prime a}-{{\partial v^{\prime a}_x}\over{\partial x^\mu}}$, with ${e_\mu}^{\prime a}={h_b}^a{e_\mu}^b$ and $v^{\prime a}_x={h_b}^av^b_x$, then ${B_\mu}^{\prime a}={h_b}^a{e_\mu}^b-{{\partial({h_b}^av^b_x)}\over{\partial x^\mu}}$. 

\

General coordinate transformations: $${B_\mu}^{\prime a}={{\partial x^\nu}\over{\partial x^{\prime\mu}}}{B_\nu}^a.$$

\

{\it Proof}: \ \ ${e_\mu}^{\prime a}={{\partial x^\nu}\over{\partial x^{\prime\mu}}}{e_\nu}^a$ i.e. ${e_\mu}^a$ is a 1-form, and ${{\partial}\over{\partial x^\mu}}={{\partial x^{\prime\nu}}\over{\partial x^\mu}}{{\partial}\over{\partial x^{\prime\nu}}}$.

\

Internal translations: $${B_\mu}^{\prime a}={B_\mu}^a-\partial_\mu \xi^a  \ \ or \ \ B^{\prime a}=B^a-d\xi^a.$$ 

\

{\it Proof}: \ \ ${B_\mu}^{\prime a}={e_\mu}^{\prime a}-{{\partial v^{\prime a}_x}\over{\partial x^\mu}}={e_\mu}^a-{{\partial v^{\prime a}_x}\over{\partial x^\mu}}={B_\mu}^a-{{\partial (v^{\prime a}_x-v^a_x)}\over{\partial x^\mu}}={B_\mu}^a-{{\partial \xi^a}\over{\partial x^\mu}}$. 

\

Then, $B=B_\mu dx^\mu={B_\mu}^adx^\mu b_a$, where $b_a$, $a=0,1,2,3$, is the canonical basis of $\R^4$, is the connection 1-form corresponding to the translations. 

\

{\it Remark}: The local ($x^\mu$) dependence of the internal Lorentz and translational transformations is a consequence of the general definition of a gauge transformation in fibre bundle theory (Appendix F).

\

In terms of the ${B_\mu}^a$ fields and the spin connection, the Ricci scalar in section {\bf 32} is given by $$R=({{\partial v_x^a}\over{\partial x^\mu}}{{\partial v_x^b}\over{\partial x^\nu}}+{{\partial v_x^a}\over{\partial x^\mu}}{B_\nu}^b+{{\partial v_x^b}\over{\partial x^\nu}}{B_\mu}^a+{B_\mu}^a{B_\nu}^b)(\partial^\mu\omega^\nu_{ab}-\partial^\nu\omega^\mu_{ab}+\omega^\mu_{ac}\omega^{\nu c}_b-\omega^\nu_{ac}\omega^{\mu c}_b). \eqno{(41)}$$ If one intends to use this Lagrangian density as describing a $({B_\mu}^a,\omega^\nu_{bc})$ (or $({e_\mu}^a,\omega^\nu_{bc})$) interaction (Randono, 2010), then immediately faces the problem that the ${B_\mu}^a$ (or ${e_\mu}^a$) does not have a free part (in particular a kinematical part), since all its powers are multiplied by $\omega$'s or $\partial\omega$'s. So an interpretation in terms of fields interaction seems difficult, and may be, impossible. 

\

{\bf 35. Torsion and gauge invariance}

\

It is well known the problem of the violation, in the presence of torsion, of the local gauge invariance of theories like Maxwell and Yang-Mills due to the straightforward application of the minimal coupling procedure to introduce the interaction with the gauge fields: the ``comma goes to semicolon'' rule. In fact, as we shall show below, the ``prohibition'' of this procedure should only be applied to the definition of the field strengths $F$, as emphasized by Hammond (Hammond, 2002). For simplicity of the presentation we shall restrict to the abelian case. 

\

In special relativity, for the field strength in terms of the gauge potential one has $F=dA=d(A_\nu dx^\nu)=(\partial_\nu A_\nu)dx^\mu\wedge dx^\nu={{1}\over{2}}(\partial_\mu A_\nu-\partial_\nu A_\mu)dx^\mu\wedge dx^\nu=F_{\mu\nu}dx^\mu\wedge dx^\nu$, which is clearly gauge independent: $F(A)=F(A+d\lambda)$. Replacing $\partial_\mu$ by $D_\mu$ one obtains $$F_{\mu\nu}\to f_{\mu\nu}=D_\mu A_\nu-D_\nu A_\mu=(\partial_\mu-\Gamma^\rho_{\mu\nu}A_\rho)-(\partial_\nu A_\mu-\Gamma^\rho_{\nu\mu}A_\rho)=F_{\mu\nu}-2\Gamma^\rho_{[\mu\nu]}A_\rho=F_{\mu\nu}-2T^\rho_{\mu\nu}A_\rho.\eqno{(*)}$$ When torsion vanishes, $f_{\mu\nu}=F_{\mu\nu}$ i.e. $A_{\mu;\nu}-A_{\nu;\mu}=A_{\mu,\nu}-A_{\nu,\mu}$; when $T^\rho_{\mu\nu}\neq 0$, $f_{\mu\nu}\neq F_{\mu\nu}$ and moreover, $f_{\mu\nu}$ is {\it not} gauge invariant: if $A_\mu\to A_\mu+\partial_\mu\lambda$, then $f_{\mu\nu}\to f_{\mu\nu}^\prime$ with $$f_{\mu\nu}^\prime-f_{\mu\nu}\equiv\delta_{g.tr.}(f_{\mu\nu})=-2T^\rho_{\mu\nu}\partial_\rho\lambda. \eqno{(**)}$$ The equations $(*)$ and $(**)$ show that the acquired dependence on torsion of the classical electric and magnetic fields, also depends on the chosen gauge (by the presence of $\lambda$), what is inadmissible. 

\

At this point, we criticize the ``solution'' given by some authors (Hehl et al, 1976; de Sabbata, 1997), which consists in the assertion that {\it torsion does not couple to the gauge field}. This statement would have sense if also the Levi-Civita part of the connection would not couple, since both $\Gamma_{LC}$ and torsion ``come together'' in the sum $\Gamma=\Gamma_{LC}+K$ (appendix B), where torsion is the antisymmetric part of the contortion $K$. However, $\Gamma_{LC}$ does couple. Moreover, de Sabbata (de Sabbata, 1997) shows that at the microscopic quantum level photons couple to torsion in a gauge invariant way through virtual pairs $e^+e^-$ creation. There is no reason to expect that in the transit to the classical limit the coupling should disappear; though, as we shall see below, partly due to the absence of the intermediate fermion field, gauge invariance breaks down. 

\

A partial solution to this problem has been given by Benn, Dereli and Tucker (Benn et al, 1980), leaving, as we show below, $F=dA$ with $F^\prime=dA^\prime=F$ if $A^\prime=A+d\lambda$ in a completely natural way. 

\

Let $A=A_\mu dx^\mu=A_ae^a$ be the connection 1-form of the electromagnetic field, with $A_a={e_a}^\mu A_\mu$ and $e^a={e_\mu}^a dx^\mu$.  If ${\omega^a}_b=\omega^a_{\mu b}dx^\mu$ is the spin connection with $\omega_{cb}=\eta_{ca}{\omega^a}_b=-\omega_{bc}$, then the {\it exterior covariant derivative} of $A_a$ with respect to ${\omega^a}_b$ is given by $$DA_a=dA_a-{\omega^b}_aA_b\eqno{(***)}$$ with $dA_a=dx^\mu\partial_\mu A_a$ and ${\omega^b}_aA_b=A_b\omega^b_{\mu a}dx^\mu$. $(***)$ gives the {\it minimal coupling} of the electromagnetic connection with the space-time connection, i.e. $dA_a\buildrel{m.c.}\over\longrightarrow DA_a$. 

\

Exterior multiplication with $e^a$ gives $$DA_a\wedge e^a=(da_a-{\omega^b}_aA_b)\wedge e^a=(dA_a)\wedge e^a-A_b{\omega^b}_a\wedge e^a,$$ and using the expression for torsion $T^a=de^a+{\omega^a}_be^b$ (section {\bf 29}), we obtain $$DA_a\wedge e^a=(dA_a)\wedge e^a-A_b(T^b-de^b)=(dA_a)\wedge e^a-A_bT^b+A_bde^b$$ i.e. $$DA_a\wedge e^a+A_bT^b=(dA_a)\wedge e^a+A_ade^a=d(A_ae^a)=dA.$$ Then, $$F=dA=DA_a\wedge e^a+A_bT^b. \eqno{(****)}$$ $F$ is closed, $$dF=d^2A=0$$ and, most important, $U(1)$-gauge invariant: $$A\to A^\prime=A+d\lambda\Longrightarrow F\to F^\prime=F+d^2\lambda=F.$$ We notice however that $DA_a$ is not $U(1)$-gauge invariant: in fact, with $d_a={e_a}^\mu\partial_\mu$, $$DA_a\to (DA_a)^\prime=D(A_a+d_a\lambda)=d(A_a+d_a\lambda)-{\omega^b}_a(A_b+d_b\lambda)=DA_a+d(d_a\lambda)-{\omega^b}_ad_b\lambda$$

\

$=DA_a+({\delta^b}_a-{\omega^b}_a)d_b\lambda$ i.e. $$(DA_a)^\prime=DA_a+{D^b}_ad_b\lambda$$ with $${D^b}_a={\delta^b}_ad-{\omega^b}_a.$$

\
 
Nevertheless, even with a gauge invariant field strength, due to the non gauge invariance of the spin density tensor of the electromagnetic field, the solution of the Cartan equation gives a $U(1)$-gauge dependent torsion, which points to a difficult (if not impossible) to cure illness of the EC theory.
In fact, the Maxwell-Einstein action describing the interaction between the electromagnetic field and gravity is given by $$S_{M-E}=l\int d^4x \ eL_{M-E}=l\int d^4x \ e(-{{1}\over{4}}F_{\mu\nu}F^{\mu\nu})$$ with $l={{G}\over{c^4}}$. From the r.h.s. of ($****$) and using ($***$), we obtain the expression for $F_{\mu\nu}$: $$(dA_a-{\omega^b}_a)\wedge e^a=((\partial_\mu A_a){e_\nu}^a-\omega^b_{\mu a}A_b{e_\nu}^a)dx^\mu\wedge dx^\nu,$$ and $$A_bT^b=A_bT^b_{\mu\nu}dx^\mu\wedge dx^\nu,$$ then $$F=((\partial_\mu A_a-\omega^b_{\mu a}A_b){e_\nu}^a+A_bT^b_{\mu\nu})dx^\mu\wedge dx^\nu$$ and so $$F_{\mu\nu}=2((\partial_{[\mu}A_a-\omega^b_{[\mu a}A_b){e_{\nu]}}^a+A_bT^b_{\mu\nu}).$$ The {\it Cartan equation} results from the variation with respect to the spin connection of the total action $S_G+S_{M-E}$: $$0=\delta_\omega S_G+\delta_\omega S_{M-E}.$$ The first term was obtained in section {\bf 32}.A.1.; for the second term, using the expression for $T^b_{\mu\nu}$ in section {\bf 29}, $$-{{1}\over{4}}\delta_\omega(F^{\mu\nu}F_{\mu\nu})=-{{1}\over{2}}F^{\mu\nu}\delta_\omega F_{\mu\nu}=-F^{\mu\nu}\delta_\omega({{1}\over{2}}((\partial_\mu A_a-\omega^b_{\mu a}A_b){e_\nu}^a-(\partial_\nu A_a-\omega^b_{\nu a}A_b){e_\mu}^a)$$ $$+A_b(\partial_\mu{e_\nu}^b-\partial_\nu{e_\mu}^b+\omega^b_{\mu a}{e_\nu}^a-\omega^b_{\nu a}{e_\mu}^a))$$ $$=-F^{\mu\nu}(-{{1}\over{2}}(\delta\omega^b_{\mu a})A_b{e_\nu}^a+{{1}\over{2}}(\delta\omega^b_{\nu a})A_b{e_\mu}^a+A_b((\delta\omega^b_{\mu a}){e_\nu}^a-(\delta\omega^b_{\nu a}){e_\mu}^a))$$ $$=F^{\mu\nu}(\delta\omega^b_{\mu a})A_b{e_\nu}^a-2F^{\mu\nu}A_b(\delta\omega^b_{\mu a}){e_\nu}^a=-F^{\mu\nu}A_b{e_\nu}^a(\delta\omega^b_{\mu a})=-F^{\mu\nu}A^b{e_\nu}^a\delta\omega_{\mu ba}=-F^{\mu\nu}A^{[b}{e_\nu}^{a]}\delta\omega_{\mu ba}$$ where we used the antisymmetry $\omega_{\mu ba}=-\omega_{\mu ab}$. Then $$0=-{e_a}^\mu T_b+{e_b}^\mu T_a+T^\mu_{ba}-l{F^\mu}_\nu A_{[b}{e_{a]}}^\nu.$$ Multiplying by ${e_\rho}^b{e_\sigma}^a$ we obtain $$T^\mu_{\rho\sigma}+\delta^\mu_\rho T_\sigma-\delta^\mu_\sigma T_\rho=lF^\mu_{[\sigma}A_{\rho ]}={{l}\over{2}}({F^\mu}_\sigma A_\rho-{F^\mu}_\rho A_\sigma)=-{{l}\over{2}}S^\mu_{\rho\sigma}$$ where $$S^\mu_{\rho\sigma}={F^\mu}_\rho A_\sigma-{F^\mu}_\sigma A_\rho$$ is the {\it canonical spin density tensor of the electromagnetic field} obtained from the gauge invariant Lagrangian density $L_{M-E}=-{{1}\over{4}}F^{\mu\nu}F_{\mu\nu}$ through the Noether theorem (Bogoliubov and Shirkov, 1980). $S^\mu_{\rho\sigma}$ is antisymmetric in its lower indices but it is {\it not} gauge invariant: $$\delta_{g.tr.}(S^\mu_{\rho\sigma})=2{F^\mu}_{[\rho}\partial_{\sigma ]}\lambda$$ if $\delta_{g.tr.}(A_\mu)=\partial_\mu\lambda$. In contradistinction with the density of energy-momentum $T_{\mu\nu}$ of any matter field, which always can be made gauge invariant (and symmetric), there is no known way to construct a gauge invariant spin density tensor for the electromagnetic field. However, at least in the special relativistic classical and quantum field theory context, after space integration of $S^0_{\rho\sigma}$, all the results for the conserved spin angular momentum tensor of the electromagnetic field are physical (light polarization, helicity states, etc.), and therefore gauge independent. This means that $S^\mu_{\rho\sigma}$ is not directly observable, and then the same could be concluded for the non gauge invariant torsion tensor produced by the electromagnetic spin.

\

By the same method of section {\bf 32} applied to the Dirac field, for the torsion one obtains $$T^\mu_{\rho\sigma}={{l}\over{2}}(S^\mu_{\rho\sigma}+{{1}\over{2}}(\delta^\mu_\sigma S_\rho-\delta^\mu_\rho S_\sigma))$$ with $S_\rho=S^\mu_{\rho\mu}={F^\mu}_\rho A_\mu$. 

\

{\bf Acknowledgements} 

\

This work was partially supported by the projects PAPIIT IN 118609, 113607-2, and 101711-2, DGAPA-UNAM, M\'exico. The author thanks for the hospitality at the IAFE-UBA-CONICET, Argentina.

\

{\bf References}

\

Aldrovandi, R. and Pereira, J. G. (2007). {\it An Introduction to Teleparallel Gravity}, Instituto de F\'\i sica Te\'orica, UNESP, Sao Paulo, Brazil.

\

Ali, S. A., Cafaro, C., Capozziello, S., and Corda, Ch. (2009). On the Poincar\'e Gauge Theory of Gravitation, {\it International Journal of Theoretical Physics} {\bf 48}, 3426-3448; arXiv: gr-qc/0907.0934.

\

Benn, I. M., Dereli, T., and Tucker, R. W. (1980). Gauge Field Interactions in Spaces with Arbitrary Torsion, {\it Physics Letters} {\bf 96}B, 100-104.

\

Bogoliubov, N. N. and Shirkov, D. V. (1980). {\it Introduction to the Theory of Quantized Fields}, Wiley, 50-51.

\

Carroll, S. (2004). {\it Spacetime and Geometry. An Introduction to General Relativity}, Addison Wesley, San Francisco. 

\

Cartan, E. (1922). Sur une g\'en\'eralization de la notion de courbure de Riemann et les espaces  $\grave{a}$ torsion, {\it Comptes Rendus Acad. Sci.}, Paris, {\bf 174}, 593-595.

\
  
Cheng, T.-P. (2010).{\it Relativity, Gravitation and Cosmology}, 2nd. ed., Oxford University Press; pp. 307-308.

\

de Sabbata, V. (1997). Evidence for torsion in gravity?, in {\it Spin Gravity: Is it possible to give an experimental basis to torsion?}, International School of Cosmology and Gravitation, XV Course, Erice, Italy, eds. P. G. Bergmann et al, World Scientific, 52-85.

\

Dirac, P. A. M. (1975). {\it General Theory of Relativity}, Princeton University Press, Princeton, New Jersey, 

\

1996; pp. 15, 25.

\

D'Olivo, G. and Socolovsky, M. (2011). Poincar\'e gauge invariance of general relativity and Einstein-Cartan theory; arXiv: gr-qc/1104.1657.

\

Einstein, A. (1956). {\it The Meaning of Relativity}, Chapman and Hall, p. 96.

\

Fabbri, L. (2010). Private communication.

\

Fabbri, L. (2011). On the Principle of Equivalence, in {\it Einstein and Hilbert: Dark Matter}, editor V. V. Dvoeglazov, Nova Publishers; arXiv: gr-qc/0905.2541.

\

Gronwald, F., and Hehl, F. W. (1996). On the Gauge Aspects of Gravity, in {\it Proceedings of the International School of Cosmology and Gravitation: 14th Course: Quantum Gravity}, Erice, Italy, 1995; P. G. Bergmann et al eds., World Scientific, Singapore; arXiv: gr-qc/9602013.

\

Hammond, R. T. (2002). Torsion gravity, {\it Reports on Progress in Physics} {\bf 65}, 599-649.

\

Hartley, D. (1995). Normal frames for non-Riemannian connections, {\it Classical and Quantum Gravity} {\bf 12}, L103-L105.

\

Hayashi, K. (1977). The gauge theory of the translation group and underlying geometry, {\it Physics Letters} {\bf 69}B, 441-444.

\

Hayashi, K. and Nakano, T. (1967). Extended Translation Invariance and Associated Gauge Fields, {\it Progress of Theoretical Physics} {\bf 38}, 491-507.

\

Hayashi, K., and Shirafuji, T. (1980). Gravity from Poincar\'e Gauge Theory of the Fundamental Particles. I, {\it Progress of Theoretical Physics} {\bf 64}, 866-882.

\

Hehl, F. W. (1985). On the Kinematics of the Torsion of Space-Time, {\it Foundations of Physics} {\bf 15}, 451-471.

\

Hehl, F. W. (1998). Alternative Gravitational Theories in Four Dimensions, in {\it Proceedings 8th M. Grossmann Meeting}, ed. T. Piran, World Scientific, Singapore; arXiv: gr-qc/9712096.

\

Hehl, F. W., von der Heyde, P., Kerlick, G. D., and Nester, J. M. (1976). General relativity with spin and torsion: Foundations and prospects, {\it Review of Modern Physics} {\bf 48}, 393-416.

\

Kobayashi, S. and Nomizu, K. (1963). {\it Foundations of Differential Geometry}, Volume I, J. Wiley, New York.

\

Leston, M. (2008). {\it Simetr\'\i as Infinito-Dimensionales en Teor\'\i as de Gravedad}, Tesis doctoral, FCEN-UBA, Argentina.

\

Leston, M., and Socolovsky, M. (2011). The Status of Gravity as a Gauge Theory, in {\it Einstein and Hilbert: Dark Matter}, ed. V. V. Dvoeglazov, Nova Scie. Pub, pp. 141-156.

\

O'Raifearteigh, L. (1997). {\it The Dawning of Gauge Theory}, Princeton University Press, Princeton.

\

Randono, A. 2010. Gauge Gravity; a forward-looking introduction, arXiv: gr-qc/1010.5822.

\

Schroedinger, E. (1950). {\it Space-Time Structure}, Cambridge University Press, Cambridge.

\

Socolovsky, M. (2010). Locally inertial coordinates with totally antisymmetric torsion, arXiv: gr-qc/1009.3979.

\

Trautman, A. (1973). On the structure of Einstein-Cartan equations, in {\it Differential Geometry, Symp. Mathematics}, Vol. {\bf 12}, Academic Press, London, 139-162.

\

Utiyama, R. (1956). Invariant Theoretical Interpretation of Interaction, {\it Physical Review} {\bf 101}, 1597-1607.

\

Weyl, H. (1918). Gravitation und Elektrizit$\ddot{a}$t, {\it Sitzungsber. d. Preuss.}, Akad. d. Wissensch., 465. (English translation in O'Raifearteigh (1997), 24-37.) 

\

{\bf Appendix A}

\

{\bf Fundamental theorem of riemannian geometry}

\

{\it Definition}: Let $(M,g,\nabla)$ be a riemannian manifold with a linear connection $\nabla$. $\nabla$ and $g$ are {\it compatible} or, equivalently, $\nabla$ is a {\it metric connection} in $M$, if for any smooth path $c:(a,b)\to M$, $\lambda\mapsto c(\lambda)$, and any pair of parallel vector fields $V$ and $V^\prime$ along $c$ (i.e. ${{DV}\over{d\lambda}}={{DV^\prime}\over{d\lambda}}=0$), then $g(V,V^\prime)\equiv <V,V^\prime>$ is constant along $c$. 

\

{\it Theorem}: In any riemannian manifold $(M,g)$ there exists a unique symmetric linear connection $\nabla$, i.e. a connection in the tangent bundle of $M$, which is compatible with the metric.

\

Proof: 

\

Let $X,Y,Z\in \Gamma(TM)$, and let $\nabla$ be a metric and symmetric linear connection in $M$. Then:

\

$<\nabla_XY,Z>=X<Y,Z>-<Y,\nabla_XZ>=X<Y,Z>=X<Y,Z>-<Y,\nabla_ZX+[X,Z]>$

\

$=X<Y,Z>-<Y,\nabla_ZX>-<Y,[X,Z]>$;

\

$Z<Y,X>=<\nabla_ZY,X>+<Y,\nabla_Z,X>$ implies $<Y,\nabla_ZX>=Z<Y,X>-<\nabla_ZY,X>$

\

$=Z<Y,X>-<\nabla_YZ+[Z,Y],X>$; then:

\

$<\nabla_XY,Z>=X<Y,Z>-Z<Y,X>+<\nabla_YZ+[Z,Y],X>-<Y,[X,Z]>$; 

\

$Y<Z,X>=<\nabla_YZ,X>+<Z,\nabla_YX>$ implies $<\nabla_YZ,X>=Y<Z,X>-<Z,\nabla_YX>$

\

$=Y<Z,X>-<Z,\nabla_XY+[Y,X]>=Y<Z,X>-<\nabla_XY,Z>-<Z,[Y,X]>$;

\

then:

\

$<\nabla_XY,Z>=X<Y,Z>-Z<Y,X>+Y<Z,X>-<\nabla_XY,Z>-<Z,[Y,X]>$

\

$+<[Z,Y],X>-<[X,Z],Y>$

\

and therefore

\

$<\nabla_XY,Z>={{1}\over{2}}(X<Y,Z>-Z<X,Y>+Y<Z,X>-<X,[Y,Z]>+<Y,[Z,X]+<Z,[X,Y])$

\

$={{1}\over{2}}(X<Y,Z>-<X,[Y,Z]>+Y<Z,X>+<Y,[Z,X]>-Z<X,Y>+<Z,[X,Y]),$

\

which gives an explicit expression for $<\nabla_XY,Z>$ in terms of $X,Y,Z$ and $< \ , \ >=g$.

\

In local coordinates, we choose $X=\partial _i$, $Y=\partial_j$, $Z=\partial_k$; then $[\partial_i,\partial_j]=0$ and therefore 

\

$<\nabla_{\partial_i}\partial_j,\partial_k>=<\Gamma^l_{ij}\partial_l,\partial_k>=\Gamma^l_{ij}<\partial_l,\partial_k>=\Gamma^l_{ij}g(\partial_l,\partial_k)
=\Gamma^l_{ij}g_{lk}={{1}\over{2}}(\partial_i<\partial_j,\partial_k>+\partial_j<\partial_k,\partial_i>$ 

\

$-\partial_k<\partial_i,\partial_j>)={{1}\over{2}}(\partial_ig_{jk}+\partial_jg_{ki}-\partial_kg_{ij})$; multiplying by $g^{mk}=(g^{-1})_{mk}$, $\Gamma^l_{ij}g^{mk}g_{lk}=\Gamma^l_{ij}\delta^m_l=\Gamma^m_{ij}$ 

\

and therefore $$\Gamma^m_{ij}={{1}\over{2}}g^{mk}(\partial_ig_{jk}+\partial_jg_{ik}-\partial_kg_{ij}).\eqno{(qed)}$$ 

\

Clearly, $\Gamma^m_{ij}=\Gamma^m_{ji}$. 

\

{\it Remark}: The theorem is also valid in pseudo-riemannian geometry; in particular for lorentzian manifolds.

\

{\it Note}: In $m$ dimensions, the number of independent components of the metric tensor and the Levi-Civita connection are $N(g_{\mu\nu};m)={{m(m+1)}\over{2}}$ and $N(\Gamma^\mu_{\nu\rho};m)={{m^2(m+1)}\over{2}}$. 

\

{\bf Appendix B}

\

{\bf General form of the local version of a non (necessarily) metric and non (necessarily) symmetric connection}

\

Given a linear connection in a manifold $M^n$ ($L^n$-space), with local Christoffel symbols $\Gamma^\mu _{\nu\rho}$, if in addition the manifold is riemannian or pseudo-riemannian $(M^n,g)$ ($V^n$-space), one has a $(L^n ,g)$-space. The {\it non-metricity} tensor is defined as minus the covariant derivative of the metric: $$Q_{\mu\nu\rho}=-D_\mu g_{\nu\rho}.$$ ($Q_{\mu\nu\rho}=Q_{\mu\rho\nu}$.) Using section {\bf 7}, by cyclic permutation of indices, one obtains $$\Gamma^\alpha_{\nu\mu}={(\Gamma_{LC})}^\alpha_{\nu\mu}+K^\alpha_{\nu\mu}+{{1}\over{2}}g^{\alpha\rho}(Q_{\mu\nu\rho}+Q_{\nu\rho\mu}-Q_{\rho\mu\nu})$$ where ${(\Gamma_{LC})}^\alpha_{\nu\mu}$ is the Levi-Civita connection of section {\bf 13}, and $$K^\alpha_{\nu\mu}=T^\alpha_{\nu\mu}+g^{\alpha\rho}T^\lambda_{\rho\mu}g_{\lambda\nu}+g^{\alpha\rho}T^\lambda_{\rho\nu}g_{\lambda\mu}$$ is the {\it contortion} tensor, with $$K^\alpha_{\nu\mu}={(K_A)}^\alpha_{\nu\mu}+{(K_S)}^\alpha_{\nu\mu},$$  $${(K_A)}^\alpha_{\nu\mu}=T^\alpha_{\nu\mu}=-{(K_A)}^\alpha_{\mu\nu},$$ $${(K_S)}^\alpha_{\nu\mu}={(K_S)}^\alpha_{\mu\nu}=g^{\alpha\rho}(T^\lambda_{\rho\mu}g_{\lambda\nu}+T^\lambda_{\rho\nu}g_{\lambda\mu}).$$ A {\it metric connection} is one in which $Q_{\mu\nu\rho}=0$ i.e. the connection is compatible with the metric, but non necessarily symmetric: $$\Gamma=\Gamma_{LC}+contortion.$$ ($U^n$-space.) In particular, scalar products and lengths of vectors are constant in parallel transport. (In fact, $\vert\vert V\vert\vert^2_{,\mu}=(g_{\alpha\beta}V_\alpha V_\beta)_{,\mu}=g_{\alpha\beta ,\mu}V^\alpha V^\beta +2g_{\alpha\beta}V^\alpha {V^\beta}_{,\mu}=2V_\beta({V^\beta}_{,\mu}+\Gamma^\beta_{\mu\alpha}V^\alpha)=2V_\beta {V^\beta}_{;\mu}=0$.) A physical case corresponds to the Einstein-Cartan theory of gravity. (Cartan, 1922.)

\

A {\it symmetric connection} is one in which torsion vanishes i.e. $T^\mu_{\nu\rho}=0$: $$\Gamma=\Gamma_{LC}+non-metricity.$$ A particular case is the Weyl connection (1918): $$non-metricity=Q_{\mu\nu\rho}=g_{\nu\rho}A_\mu$$ with $A=A_\mu dx^\mu$ a 1-form. (In natural units, $[A_\mu]=[mass]$ if $[x^\mu]=[length]$.)

\

{\bf Appendix C}

\

1. If $a=(a_{ij})$ is an invertible matrix with $a_{ij}=a_{ij}(x)$, then $\partial_\mu(det \ a)=a_{ij,\mu}(det \ a)a^{ij}$, where $a^{ij}=(a^{-1})_{ij}$. In particular, for the metric tensor, $$g_{,\rho}=g_{\mu\nu ,\rho}gg^{\mu\nu}$$ where $g=det g_{\mu\nu}$. Then, for the Levi-Civita connection, $$(\Gamma_{LC})^\mu_{\nu\mu}={{1}\over{2}}g^{\mu\lambda}(\partial_\nu g_{\mu\lambda}+\partial_\mu g_{\nu\lambda}-\partial_\lambda g_{\nu\mu})={{1}\over{2}}g^{\mu\lambda}g_{\mu\lambda ,\nu}={{1}\over{2}}g^{-1}(gg^{\mu\lambda}g_{\mu\lambda ,\nu})={{1}\over{2g}}g_{,\nu}={{1}\over{2(-g)}}(-g)_{,\nu}$$ $$={{1}\over{(-g)^{{1}\over{2}}}}\partial_\nu(-g)^{{1}\over{2}}=\surd^{-1}\partial_\nu\surd.$$

\

2. From 1., $\partial_\nu\surd={{\surd}\over{2}}g^{\mu\lambda}g_{\mu\lambda,\nu}$; then, $\delta\surd=(\partial_\nu\surd)\delta x^\nu={{\surd}\over{2}}g^{\mu\lambda}g_{\mu\lambda,\nu}\delta x^\nu={{\surd}\over{2}}g^{\mu\lambda}\delta g_{\mu\lambda}$. Using $g^{\rho\lambda}=\eta^{ab}{e_a}^\rho{e_b}^\lambda$ (section {\bf 28}.6), we obtain $g^{\mu\lambda}\delta g_{\mu\lambda}=2{e_d}^\lambda\delta{e_\lambda}^d$ and so $\delta e=e{e_d}^\lambda\delta{e_\lambda}^d$ with $e=\surd$. From ${e_d}^\lambda{e_\lambda}^f=\delta^f_d$, $\delta e=-e{e_\lambda}^d\delta{e_d}^\lambda$. 

\

{\bf Appendix D}

\

{\bf Lorentz gauge invariance of the Ricci scalar}

\

(This proof is due to G. D'Olivo.)

\

The Ricci scalar is given by $$R=\eta^{bd}{e_a}^\mu{e_d}^\nu(\partial_\mu\omega_{\nu b}^a-\partial_\nu\omega_{\mu b}^a+\omega_{\mu c}^a\omega_{\nu b}^c-\omega_{\nu c}^a\omega_{\mu b}^c)\equiv \eta^{bd}{e_a}^\mu{e_d}^\nu((3)-(4)+(1)-(2)).$$ Under the transformation $$\omega_{\mu c}^a={h_c}^l\omega_{\mu l}^{\prime r}{h^{-1}_r}^a+(\partial{h_c}^l){h^{-1}_l}^a$$ we have: 

\

$(1)=(a)+(b)+(c)+(d)$ with $$(a)={h_c}^l\omega_{\mu l}^{\prime r}{h^{-1}_r}^a{h_b}^g\omega_{\nu g}^{\prime s}{h^{-1}_s}^c, \ (b)={h_c}^l\omega_{\mu l}^{\prime r}{h^{-1}_r}^a(\partial_\nu{h_b}^g){h^{-1}_g}^c,$$ $$(c)={h_b}^g\omega_{\nu g}^{\prime s}{h^{-1}_s}^c(\partial_\mu{h_c}^l){h^{-1}_l}^a, \ (d)=(\partial_\mu{h_c}^l){h^{-1}_l}^a(\partial_\nu{h_b}^g){h^{-1}_g}^c;$$ 

\

$(2)=(e)+(f)+(g)+(h)$ with $$(e)={h_c}^g\omega_{\nu g}^{\prime s}{h^{-1}_s}^a{h_b}^l\omega_{\mu l}^{\prime r}{h^{-1}_r}^c, \ (f)={h_c}^g\omega_{\nu g}^{\prime s}{h^{-1}_s}^a(\partial_\mu{h_b}^l){h^{-1}_l}^c,$$ $$(g)={h_b}^l\omega_{\mu l}^{\prime r}{h^{-1}_r}^c(\partial_\nu{h_c}^g){h^{-1}_g}^a, \ (h)=(\partial_\nu{h_c}^l){h^{-1}_l}^a(\partial_\mu{h_b}^g){h^{-1}_g}^c;$$

\

$(3)=[1]+[2]+[3]+[4]$ with $$[1]={h_b}^n{h^{-1}_t}^a(\partial_\mu\omega_{\nu n}^{\prime t}), \ [2]=\omega_{\nu n}^{\prime t}\partial_\mu({h_b}^n{h^{-1}_t}^a), \ [3]=(\partial_\mu\partial_\nu{h_b}^n){h^{-1}_n}^a, \ [4]=(\partial_\nu{h_b}^n)(\partial_\mu{h^{-1}_n}^a);$$ and $(4)=[5]+[6]+[7]+[8]$ with $$[5]={h_b}^l{h^{-1}_s}^a(\partial_\nu\omega_{\mu l}^{\prime s}), \ [6]=\omega_{\mu l}^{\prime s}\partial_\nu({h_b}^l{h^{-1}_s}^a), \ [7]=(\partial_\nu\partial_\mu{h_b}^l){h^{-1}_l}^a, \ [8]=(\partial_\mu{h_b}^l)(\partial_\nu{h^{-1}_l}^a).$$ Now, 

\

$[3]-[7]=(\partial_\mu\partial_\nu{h_b}^n){h^{-1}_n}^a-(\partial_\nu\partial_\mu{h_b}^l){h^{-1}_l}^a=0,$

\

$(b)+(c)=\omega_{\mu l}^{\prime r}{h^{-1}_r}^a\partial_\nu{h_b}^l-\omega_{\nu g}^{\prime s}{h_b}^g\partial_\mu{h^{-1}_s}^a,$

\

$(f)+(g)=\omega_{\nu g}^{\prime s}{h^{-1}_s}^a\partial_\mu{h_b}^g-\omega_{\mu l}^{\prime r}{h_b}^l\partial_\nu{h^{-1}_r}^a;$

\

so

\

$((b)+(c))-((f)+(g))=\omega_{\mu l}^{\prime r}\partial_\nu({h^{-1}_r}^a{h_b}^l)-\omega_{\nu g}^{\prime s}\partial_\mu({h^{-1}_s}^a{h_b}^g);$

\

also,

\

$[2]-[6]=\omega_{\nu g}^{\prime s}\partial_\mu({h_b}^g{h^{-1}_s}^a)-\omega_{\mu l}^{\prime r}\partial_\nu({h_b}^l{h^{-1}_r}^a);$

\

then 

\

$((b)+(c))-((f)+(g))+([2]-[6])=0.$

\

Also,

\

$[4]-[8]=(\partial_\nu{h_b}^l)(\partial_\mu{h^{-1}_l}^a)-(\partial_\mu{h_b}^l)(\partial_\nu{h^{-1}_l}^a)$

\

and

\

$(d)-(h)=(\partial_\nu{h^{-1}_l}^a)(\partial_\mu{h_b}^l)-(\partial_\mu{h^{-1}_l}^a)(\partial_\nu{h_b}^l);$

\

so

\

$([4]-[8])+((d)-(h))=0.$

\

Finally, 

\

$[1]-[5]+(a)-(e)={h_b}^l{h^{-1}_s}^a(\partial_\mu\omega_{\nu l}^{\prime s}-\partial_\nu\omega_{\mu l}^{\prime s}+\omega_{\mu r}^{\prime s}\omega_{\nu l}^{\prime r}-\omega_{\nu r}^{\prime s}\omega_{\mu l}^{\prime r})$.

\

Therefore,

\

$R=\eta^{bd}{e_a}^\mu{e_d}^\nu{h_b}^l{h^{-1}_s}^a(\partial_\mu\omega_{\nu l}^{\prime s}-\partial_\nu\omega_{\mu l}^{\prime s}+\omega_{\mu r}^{\prime s}\omega_{\nu l}^{\prime r}-\omega_{\nu r}^{\prime s}\omega_{\mu l}^{\prime r})=\eta^{lt}{e_s}^{\prime\mu}{e_t}^{\prime\nu}(\partial_\mu\omega_{\nu l}^{\prime s}-\partial_\nu\omega_{\mu l}^{\prime s}+\omega_{\mu r}^{\prime s}\omega_{\nu l}^{\prime r}-\omega_{\nu r}^{\prime s}\omega_{\mu l}^{\prime r})$

\

$=R^\prime. \ \ \ \ \ \ \ \ \ \ \ \ \ \ \ \ \ \ \ \  (qed)$

\

{\bf Appendix E}

\

{\bf Affine spaces}

\

An {\it affine space} is a triple $(V,\varphi,A)$ where $V$ is a vector space, $A$ is a set, and $\varphi$ is a free and transitive left action of $V$ as an additive group on $A$: $$\varphi:V\times A\to A, \ (v,a)\mapsto v+a,$$ with $$0+a=a \ and \ (v_1+v_2)+a=v_1+(v_2+a), \ for \ all \ a\in A \ and \ all \ v_1,v_2\in V.$$ Then, given $a,a^\prime\in A$, there exists a unique $v\in V$ such that $a^\prime=v+a$. Also, if $v_0$ is fixed in $V$, $\varphi_{v_0}:A\to A$, $\varphi_{v_0}(a)=\varphi(v_0,a)$ is a bijection.

\

Example. $A=V$: The vector space itself is considered as the set on which $V$ acts. In particular, when $V=T_xM^n$ and $A=T_xM^n$, the tangent space is called {\it affine tangent space} and denoted by $A_xM^n$. The points $``a"$ of $A_xM^n$ are the tangent vectors at $x$.

\

It is clear that to define an action of $GA_n(\R)$ on $A_xM^n$, we need a frame at $x$ i.e. we have to consider the bundle of affine frames $AM^n$ so that $(v_x,r_x)(\xi,g)=(v_x+r_x\xi,r_xg)$. 

\

{\bf Appendix F}

\

{\bf Gauge transformations in $G$-bundles}

\

A {\it gauge transformation} or {\it vertical automorphism} of a (smooth) principal $G$-bundle $\xi:G\to P\buildrel{\pi}\over\longrightarrow B$ is a diffeomorphism $\alpha:P\to P$ such that the following diagram commutes: $$\matrix{P\times G & \buildrel{\alpha\times Id_G}\over\longrightarrow & P\times G\cr \psi\downarrow & & \downarrow\psi\cr P & \buildrel{\alpha}\over\longrightarrow & P \cr \pi\downarrow & & \downarrow\pi \cr B & \buildrel{Id_B}\over\longrightarrow & B \cr}$$ ($\psi$ is the action of $G$ on $P$.) That is: $$\alpha\circ\psi=\psi\circ(\alpha\times Id_G) \ \ i.e. \ \ \alpha(pg)=\alpha(p)g$$ and $$\pi\circ\alpha=\pi \ \ i.e. \ \ \pi(\alpha(p))=\pi(p).$$ So, $\alpha(p)=ph$ with $h\in G$.

\

The set of gauge transformations of $\xi$, ${\cal G}(\xi)$, is called {\it the gauge group of the bundle}.

\

{\it Local form of $\alpha$}

\

A local trivialization of $\xi$ is given by the commutative diagram $$\matrix{P_U & \buildrel{\Phi_U}\over\longrightarrow & U\times G\cr \pi\downarrow & & \downarrow\pi_1 \cr U & \buildrel{Id_U}\over\longrightarrow & U \cr}$$ i.e. $\pi_1\circ\Phi_U=\pi$, where $P_U=\pi^{-1}(U)$, $\Phi_U$ is a diffeomorphism, $U$ is an open subset of $B$, and $\pi_1(p,g)=p$. 

\

$\Phi_U$ defines the local section of $\xi$, $\sigma_U:U\to P_U$, $\sigma_U(b)=\Phi^{-1}_U(b,e)$ where $e$ is the identity in $G$. Then there exists the smooth function $$\alpha_U:U\to G, \ \ b\mapsto\alpha_U(b)$$ which {\it determines} $\alpha$ {\it for all} $p\in P_U$. In fact, let $p=\sigma_U(b)$; then $$\alpha(\sigma_U(b))=\sigma_U(b)g\in P_b=\pi^{-1}(\{b\}),$$ and so $$ \alpha_U(b)=g$$ with $g$ unique since $\psi$ acts freely on $P$ and transitively on fibers. If $p^\prime\in P_b$, then $p^\prime=\sigma_U(b)h$ and $\alpha(p^\prime)=\alpha(\sigma_U(b)h)=\alpha(\sigma_U(b))h=(\sigma_U(b)\alpha_U(b))h=\sigma_U(b)(\alpha_U(b)h)$. This holds for all $b\in U$.

\

\

\

\

\

\

\

\

e-mail: socolovs@nucleares.unam.mx

\

\end

\
To explicitly illustrate the effect of the translation connection $B^a={B_\mu}^a dx^\mu$ and the spin connection $\omega_{abc}=\omega_{\mu bc}dx^\mu$ on matter fields, we rewrite the covariant derivative of the Dirac field in section {\bf 32}B: $$D_a\psi=({\delta_a}^\mu+{B_a}^\mu)(\partial_\mu-{{i}\over{4}}\omega_{\mu bc}\sigma^{bc})\psi=({\delta_a}^\mu\partial_\mu-{{i}\over{4}}\omega_{abc}\sigma^{bc}+{B_a}^\mu\partial_\mu-{{i}\over{4}}{B_a}^\mu\omega_{\mu bc}\sigma^{bc})\psi$$ $$=({\delta_a}^\mu\partial_\mu-{{i}\over{4}}(\omega_{abc}+C_{abc})\sigma^{bc}+{B_a}^\mu\partial_\mu)\psi,$$ where $C_{abc}={B_a}^\mu\omega_{\mu bc}.$

\

$\omega_{abc}$ and ${B_a}^\mu$ are multiplied by the generators of the Lorentz ($\sigma^{bc}$) and translation ($\partial_\mu$) Lie algebras; the appearance of the crossed term $C_{abc}$ is a consequence of the fact that the Poincar\'e group is the semi-direct product of the translation group and the Lorentz group.

\

\

\

\

\

\

\

\

{\bf 2. Leggett's inequalities} 

\

Let a source $S$ emit pairs of photons with polarization directions represented by unit vectors $\vec{u}$ and $\vec{v}$, towards corresponding analizers  1 and 2, with orientations given by unit vectors $\vec{a}$ and $\vec{b}$. When each photon of each pair is detected, the results of the polarization measurements are given by functions $A_{\vec{u}}(\vec{a}, \vec{b}; \lambda)$ and $B_{\vec{v}}(\vec{b},\vec{a};\lambda)$ which, at detection, take the values  +1 or -1. (Analizer 1 corresponds to $A$ and analyzer 2 corresponds to $B$.) $\lambda$ is a supplementary (``hidden'') variable taking values in a real domain $\Lambda$, such that for the subensemble of photons with polarizations $\vec{u}$ and $\vec{v}$, has a probability distribution $\rho_{\vec{u},\vec{v}}(\lambda)$ obeying $$\rho_{\vec{u},\vec{v}}(\lambda)\geq 0 \  \  \ and \  \  \ \int_\Lambda d\lambda \rho_{\vec{u},\vec{v}}(\lambda)=1. \eqno{(2.1)}$$ Then one has the following three average values over the subensemble: $$av.(A)=\int_\Lambda d\lambda \rho_{\vec{u},\vec{v}}(\lambda)A_{\vec{u}}(\vec{a},\vec{b};\lambda), \eqno{(2.2)}$$ $$av.(B)=\int_\Lambda d\lambda \rho_{\vec{u},\vec{b}}(\lambda)B_{\vec{v}}(\vec{b},\vec{a};\lambda), \eqno{(2.3)}$$ and $$av.(AB)=\int_\Lambda d\lambda \rho_{\vec{u},\vec{v}}(\lambda)A_{\vec{u}}(\vec{a},\vec{b};\lambda)B_{\vec{v}}(\vec{b},\vec{a};\lambda). \eqno{(2.4)}$$ In principle, $av.(A)$, $av.(B)$ and $av.(AB)$ depend on the set of variables $\{\vec{u},\vec{a},\vec{v},\vec{b}\}$. Non locality is allowed by the posible dependence of $A$ on $\vec{b}$ and of $B$ on $\vec{a}$, and realism is represented by the variable $\lambda$.

\
 
In fact, the source emits pairs of photons with polarizations in directions $(\vec{u},\vec{v})$, $(\vec{u}^\prime,\vec{v}^\prime)$, $(\vec{u}^{\prime\prime},\vec{v}^{\prime\prime})$,... (it depends on the decaying state, see next section). However, within each subensemble $(\vec{u}^{(n)},\vec{v}^{(n)})$, the formulae (2.2)-(2.4) for the average values of $A$, $B$, and $AB$ are clearly given in the same spirit of Bell in his original introduction of the classical, unknown, and undetectable hidden variables $\lambda$ in ref. 3. 

\

The quantities $A$ and $B$ obey the conditions listed in the Appendix. Then one has the inequalities $$1-\vert\int_\Lambda d\lambda \rho_{\vec{u},\vec{v}}(\lambda)(A_{\vec{u}}(\vec{a},\vec{b};\lambda)-B_{\vec{v}}(\vec{b},\vec{a};\lambda))\vert \geq \int_\Lambda d\lambda \rho_{\vec{u},\vec{v}}(\lambda)A_{\vec{u}}(\vec{a},\vec{b};\lambda)B_{\vec{v}}(\vec{b},\vec{a};\lambda)$$ $$\geq -1+\vert\int_\Lambda d\lambda \rho_{\vec{u},\vec{v}}(\lambda)(A_{\vec{u}}(\vec{a},\vec{b};\lambda)+B_{\vec{v}}(\vec{b},\vec{a};\lambda))\vert, \eqno{(2.5)}$$ known as Leggett's inequalities. $^1$

\

{\bf 3. Violation of the inequalities by quantum mechanics}

\

We shall now show that quantum mechanics violates the Leggett's inequalities. Coordinates are chosen such that the photons propagate in the $z$-direction, and that polarizations and analyzers are in the $xy$-plane. We shall consider the following two cases: 

\

(i). Positive parity state: $$\psi_+={{1}\over {\sqrt{2}}}(\vec{x}_1\otimes\vec{x}_2+\vec{y}_1\otimes \vec{y}_2), \eqno{(3.1)}$$ product of a $J=0\to J=1\to J=0$ radiative cascade decay of calcium $^{14}$. In this case, $\vec{u}=\vec{v}=\vec{x}$ or $\vec{u}=\vec{v}=\vec{y}$.

\

(ii) Negative parity state: $$\psi_-={{1}\over {\sqrt{2}}}(\vec{x}_1\otimes \vec{y}_2-\vec{y}_1\otimes\vec{x}_2), \eqno{(3.2)}$$ product of the ground state  positronium decay $e^+e^-\to \gamma\gamma$. $^{15}$ In this case, $\vec{u}=\vec{x}$ and $\vec{v}=\vec{y}$ or $\vec{u}=\vec{y}$ and $\vec{v}=\vec{x}$.

\

The quantum probability that a photon with polarization $\vec{p}$ passes through an analyzer in direction $\vec{c}$ (or to go through the ordinary ray in a calcite crystal) is given by $(\vec{p}\cdot \vec{c})^2$ (Malus law), while the probability to be absorbed (or to go through the extraordinary ray) is $1-(\vec{p}\cdot \vec{c})^2$. The quantum average $av.(C)_q(\vec{p},\vec{c})$ is obtained by assigning a +1 to the first possibility and a -1 to the second one, namely $$av.(C)_q (\vec{p},\vec{c})=(+1)(\vec{p}\cdot \vec{c})^2+(-1)(1-(\vec{p}\cdot\vec{c})^2)=2(\vec{p}\cdot \vec{c})^2-1. \eqno{(3.3)}$$ Then $av.(A)_q(\vec{u},\vec{a})=2(\vec{u}\cdot \vec{a})^2-1$ and $av.(B)_q(\vec{v}, \vec{b})=2(\vec{v}\cdot\vec{b})^2-1$. If the quantum averages obey the classical inequalities (a.7) (or (2.5)), one should have $$1-2\vert (\vec{u}\cdot \vec{a})^2-(\vec{v}\cdot \vec{b})^2\vert \geq av.(AB)_{q\pm}\geq -1+2\vert (\vec{u}\cdot\vec{a})^2+(\vec{v}\cdot\vec{b})^2-1 \vert , \eqno{(3.4)}$$ where the + and - signs respectively correspond to the cases (i) and (ii). From the expressions of the joint (coincidence) quantum probabilities $^{16}$ $$P_{12,q+}(\vec{a},\vec{b})={{1}\over{2}}(\vec{a}\cdot\vec{b})^2 \ \ \ and \ \ \ P_{12,q-}(\vec{a},\vec{b})={{1}\over{2}}(1-(\vec{a}\cdot\vec{b})^2), \eqno{(3.5)}$$ one easily obtains $$av.(AB)_{q+}=-av.(AB)_{q-}={{1}\over{2}}(2(\vec{a}\cdot \vec{b})^2-1). \eqno{(3.6)}$$ From (3.4) and (3.6) one then obtains $$1-2\vert(\vec{x}\cdot \vec{a})^2-\vec{x}\cdot\vec{b})^2\vert \geq {{1}\over{2}}(2(\vec{a}\cdot\vec{b})^2-1) \geq -1+2\vert(\vec{x}\cdot\vec{a})^2+(\vec{x}\cdot\vec{b})^2-1\vert \eqno{(3.7)}$$ for case (i), and $$1-2\vert(\vec{x}\cdot\vec{a})^2-(\vec{y}\cdot\vec{b})^2\vert \geq {{1}\over{2}}(1-2(\vec{a}\cdot\vec{b})^2) \geq -1+2\vert (\vec{x}\cdot\vec{a})^2+(\vec{y}\cdot\vec{b})^2-1\vert \eqno{(3.8)}$$ for case (ii). 

\

From these expressions it can be easily shown the following:

\

Case (i): For $\vec{b}=\pm\vec{a}$, quantum mechanics violates the r.h.s. of (3.7) for $\vec{x}\cdot \vec{a}=cos \Phi_{\vec{x},\vec{a}}>{{1}\over{2}}\sqrt{{{7}\over{2}}}$ {\it i.e.} $\Phi_{\vec{x},\vec{a}}<cos^{-1}({{1}\over{2}}\sqrt{{{7}\over{2}}})$ but preserves its l.h.s.; and the other way around when $\vec{a}\cdot\vec{b}=0$ for $\Phi_{\vec{x},\vec{a}}\in [0, {{\pi}\over{2}}]$ but $\Phi_{\vec{x},\vec{a}}\neq {{1}\over{2}}cos^{-1}({{1}\over{4}})$.

\

Case (ii): For $\vec{b}=\pm \vec{a}$, quantum mechanics violates the l.h.s. of (3.8) for ${{1}\over{2}}cos^{-1}({{3}\over{4}})>\Phi_{\vec{x},\vec{a}} \geq 0$ and ${{\pi}\over{2}}\geq \Phi_{\vec{x},\vec{a}}>cos^{-1}(-{{3}\over{4}})$ but preserves its r.h.s.; and the other way around when $\vec{a}\cdot\vec{b}=0$ for the same angles.

\

{\bf 4. Case of spin ${{1}\over{2}}$}

\

For completeness, and with the purpose of giving another example of the violation of the Leggett's inequalities by quantum mechanics at the level of pure ensambles, we consider the case of the EPR-Bohm $^{3,17}$ experiment with two spin ${{1}\over{2}}$ particles $A$ and $B$ (``electrons'') in the singlet state $$\psi={{1}\over{\sqrt{2}}}((\uparrow_A\otimes \downarrow_B-\downarrow_A\otimes\uparrow_B). \eqno{(4.1)}$$ The quantum probabilities that an electron $A$ with spin vector $\vec{u}$ will be deviated by a Stern-Gerlach (SG) apparatus in directions $\vec{a}$ and $-\vec{a}$ are respectively given by $$pr_{q;A}(\vec{u},\vec{a})=cos^2{{\theta_{\vec{u},\vec{a}}}\over{2}} \ \ \ \ and \ \ \ \ pr_{q;A}(\vec{u},-\vec{a})=sin^2{{\theta_{\vec{u},\vec{a}}}\over {2}}, \eqno{(4.2)}$$ where $\theta_{\vec{u},\vec{a}}$ is the angle between $\vec{u}$ and $\vec{a}$. Then the quantum average of the spin projetions for particle $A$ is $$av.(A)_q(\vec{u},\vec{a})=(+{{1}\over{2}})pr_{q;A}(\vec{u},\vec{a})+(-{{1}\over{2}})pr_{q;A}(\vec{u},-\vec{a})={{1}\over{2}}(cos^2{{\theta_{\vec{u},\vec{a}}}\over{2}}-sin^2{{\theta_{\vec{u},\vec{a}}}\over{2}})={{1}\over{2}}cos_{\vec{u},\vec{a}}={{1}\over{2}}\vec{u}\cdot\vec{a}. \eqno{(4.3)}$$ The analogous result for electron $B$ with spin vector $\vec{v}$ and SG in direction $\vec{b}$ is $$av.(B)_q(\vec{v},\vec{b})={{1}\over{2}}\vec{v}\cdot \vec{b}. \eqno{(4.4)}$$
When $\vec{v}=-\vec{u}$ (singlet state), the quantum formula for the correlation betwen the two spins is $^3$ $$av.(AB)_q(\vec{u},\vec{a};-\vec{u},\vec{b})=-\vec{a}\cdot\vec{b}. \eqno{(4.5)}$$ A non-local realistic (hidden variable) theory again leads to averages satisfying the formula (2.5), with $\vec{u}$ and $\vec{v}$ being spin directions instead of photon polarizations, and $\vec{a}$ and $\vec{b}$ being SG directions instead of photon analizers. For the case $\vec{v}=-\vec{u}$, the replacement of the ``classical'' averages appearing in the re-interpreted equation (2.5) by the quantum expressions (4.3), (4.4) and (4.5), leads to the inequalities $$1-{{1}\over{2}}\vert\vec{u}\cdot (\vec{a}+\vec{b})\vert\geq -\vec{a}\cdot\vec{b}\geq -1+{{1}\over{2}}\vert\vec{u}\cdot(\vec{a}-\vec{b})\vert .\eqno{(4.6)}$$ It is easy to see that the left (right) and right (left) hand sides of (4.6) are respectively violated (preserved) if $$\vec{u}={{\vec{a}+\vec{b}}\over{\vert \vec{a}+\vec{b}\vert}} \ , \ \ and \ \ {{2}\over{3}}\pi<\theta_{\vec{a},\vec{b}}<\pi, \eqno{(4.7)}$$ and $$\vec{u}={{\vec{a}-\vec{b}}\over {\vert\vec{a}-\vec{b}\vert}} \ , \ \ and \ \ 0<\theta_{\vec{a},\vec{b}}<{{\pi}\over{3}}. \eqno{(4.8)}$$

\

{\bf 5. Summary and conclusions}

\

Leggett in 2003 $^1$ and Gr\"oblacher {\it et al} in 2007 $^2$ have derived, in a context of non local realism, new Bell type inequalities for pairs of polarized photons in a mixture state of subensembles, each consisting of a coherent superposition of states with definite polarization for each photon. These inequalities were shown to be violated by quantum mechanics and, in ref. 2, also by the experiment. Both Leggett and Gr\"oblacher {\it et al} considered the hypotesis leading to the inequalities as  ``reasonable'' and/or ``plausible'', limiting the non local theories to the so called {\it crypto non local theories}. In this note, however, working at the level of subensembles, what is enough for our purposes, we argue that, within the context of realistic or hidden variable theories {\it \`a la} Bell, the choice of non locality made by these authors {\it is the most general one}. We derive the inequalities at the subensemble level, and show that they are violated by quantum mechanics for certain angles between analyzers, and between photon polarizations and analyzers. Then we conclude that {\it non local realism \`a la Bell is not possible}; and moreover, since local realism has also been shown to be wrong both theoretically and experimentally by extensive work during the last decades, {\it it is the whole proposal of the existence of hidden variables for completing quantum mechanics that comes to an end. Quantum mechanics is a complete theory.}

\

{\bf Acknowledgements} 

\

This work was partially supported by the project PAPIIT IN113607, DGAPA-UNAM, M\'exico. M. S. thanks for hospitality at the Facultad de Astronom\'\i a y Astrof\'\i sica de la Universidad de Valencia, Spain, where part of this work was performed.

\

{\bf References}

\

1. A. J. Leggett, Nonlocal Hidden-Variable Theories and Quantum Mechanics: An Incompatibility Theorem, {\it Foundations of Physics} {\bf 33}, 1469-1493 (2003).

\

2. S. Gr\"oblacher, T. Paterek, R. Kaltenbaek, C. Brukner, M. Zukowski, M. Aspelmayer, and A. Zeilinger, An experimental test of non-local realism, {\it Nature} {\bf 446}, 871-875 (2007).

\

3. J. S. Bell, On the Einstein-Podolsky-Rosen paradox, {\it Physics} {\bf 1}, 195-200 (1965).

\

4. J. F. Clauser, M. A. Horne, A. Shimony, and P. A. Hold, Proposed experiment to test local hidden variable theories, {\it Physical Review Letters} {\bf 23}, 880-884 (1969).

\

5. A. Aspect, J. Dalibard, and G. Roger, Experimental test of Bell's inequalities using time-varying analyzers, {\it Physical Review Letters} {\bf 49}, 1804-1807 (1982).

\

6. G. Weihs, T. Jennewein, C. Simon, H. Weinfurter, and A. Zeilinger, A violation of Bell's inequality under strict locality conditions, {\it Physical Review Letters} {\bf 81}, 5039-5043 (1998).

\

7. D. Bohm, A suggested interpretation of the quantum theory in terms of ``hidden variables'' (I and II), {\it Physical Review} {\bf 85}, 166-193 (1952).

\

8. N. Bohr, {\it Atomic Physics and Human Knowledge}, Wiley, New York (1958).

\

9. W. Heisenberg, {\it The Physical Principles of the Quantum Theory}, Dover, New York (1930).

\

10. J. J. Sakurai, {\it Modern Quantum Mechanics}, Addison-Wesley (1994).

\

11. M. Socolovsky, EPR, Bell, GHZ, and Hardy theorems, and quantum mechanics, {\it Revista Cubana de F\'\i sica} {\bf 22}, 104-118 (2005).

\

12. R. Lapiedra, {\it Las Carencias de la Realidad. La conciencia, el Universo y la mec\'anica cu\'antica}, Tusquets, Barcelona (2008).

\

13. E. Schr\"odinger, Discussion of probability relations between separated systems, {\it Proc. Camb. Phil. Soc.} {\bf 31}, 555-563 (1935).

\

14. A. Aspect, P. Gangier, and G. Rogers, Experimental Tests of Realistic Local Theories via Bell's Theorem, {\it Physical Review Letters} {\bf 47}, 460-463  (1981); A. Peres, {\it Quantum Theory, Concepts and Methods}, Kluwer, The Netherlands (1993).

\

15. L. R. Kasday, J. D. Ullman, and C. S. Wu, Angular Correlation of Compton-Scattered Annihilation Photons and Hidden Variables, {\it Nuovo Cimento B} {\bf 25}, 633-661 (1975); G. Faraci, D. Gutkowski, S. Nottarigo, and A. R. Pennisi, An Experimental Test of the EPR Paradox, {\it Lettere Nuovo Cimento} {\bf 9}, 607-611 (1974); A. Peres, {\it Quantum Theory, Concepts and Methods}, Kluwer, The Netherlands (1993).

\

16. L. E. Ballentine, {\it Quantum Mechanics}, World Scientific, pp. 597-598 (2001).

\

17. D. Bohm, {\it Quantum Theory}, Dover, p. 611 (1989).

\

{\bf Appendix}

\

Let $A$ and $B$ be two quantities which take the values +1 and -1. Then it holds $$1-\vert A-B \vert=AB=-1+\vert A+B \vert \eqno{(a.1)}$$ where $\vert \ \ \vert$ denotes the absolute value. Suppose both $A$ and $B$ depend on the variables $\{\lambda, \vec{a}, \vec{b},..., \vec{w} \}$ where $\lambda$ lies in a real domain $\Lambda$ and $\vec{a}$, $\vec{b}$,...,$\vec{w}$ are certain unit vectors in ordinary space. The variable $\lambda$ has a weight given by a classical probability distribution $\rho(\lambda)$ with $\rho(\lambda)\geq 0$ and normalized according to $$\int_\Lambda d\lambda \rho(\lambda)=1. \eqno{(a.2)}$$ From (a.1) and (a.2) $$1-\int_\Lambda d\lambda \rho(\lambda)\vert A-B \vert =\int_\Lambda d\lambda \rho(\lambda) AB =-1+\int_\Lambda d\lambda \rho(\lambda)\vert A+B \vert \eqno{(a.3)}$$ {\it i.e.} $$1-av.(\vert A-B \vert)=av.(AB)=-1+av.(\vert A+B \vert) \eqno{(a.4)}$$ where $$av.(X)=\int_\Lambda d\lambda \rho(\lambda)X(\lambda,\vec{a},\vec{b},...,\vec{w}) \eqno{(a.5)}$$ is the {\it classical average value} of $X$. 

\

Since for any quantity $X$, the average of its absolute value is greater than or equal to the absolute value of its average {\it i.e.} $$av.(\vert X \vert )\geq \vert av.(X) \vert, \eqno{(a.6)}$$ we obtain the inequalities $$1-\vert av.(A)-av.(B)\vert \geq av.(AB) \geq -1+\vert av. (A)+av.(B) \vert \eqno{(a.7)}$$ since $av.(A\pm B)=av.(A)\pm av.(B)$.

\

\

e-mails:

Ramon.Lapiedra@uv.es, socolovs@nucleares.unam.mx, socolovs@iafe.uba.ar

\

\end